\documentclass[useAMS,usenatbib]{mn2e}

\usepackage{graphicx}
\usepackage{amsmath}
\usepackage{amssymb}
\usepackage{bm}
\usepackage{ulem}
\usepackage{enumerate}
\usepackage{times}

\defcitealias{Chamandy+13a}{CSS13a}
\defcitealias{Chamandy+13b}{CSS13b}
\defcitealias{Comparetta+Quillen12}{CQ12}

\newcommand{\Exp}[1]{{\rm e}^{#1}}

\newcommand{\del}{\partial}
\newcommand{\Del}{{\nabla}}

\newcommand{\bmDel}{\bm{\nabla}}

\newcommand{\eps}{\epsilon}

\newcommand{\Emf}{\bm{\mathcal{E}}}

\newcommand{\Flux}{\bm{\mathcal{F}}}

\newcommand{\mean}[1]{\overline{#1}}
\newcommand{\meanv}[1]{\overline{\bm{#1}}}

\newcommand{\D}{_\mathrm{D}}						
\newcommand{\eq}{_\mathrm{eq}}						
\newcommand{\f}{_\mathrm{0}}					   	
\newcommand{\1}{_\mathrm{1}}					   	
\newcommand{\2}{_\mathrm{2}}					   	
\newcommand{\kin}{_\mathrm{k}}			   		
\newcommand{\magn}{_\mathrm{m}}			   		
\newcommand{\turb}{_\mathrm{t}}			   		
\newcommand{\const}{\mathrm{const}}			   		
\newcommand{\ma}{_\mathrm{max}}			   		

\newcommand{\cro}{\times}

\newcommand{\mbr}{\mean{B}_r}
\newcommand{\mbp}{\mean{B}_\phi}

\newcommand{\mup}{\mean{U}_\phi}
\newcommand{\muz}{\mean{U}_z}

\newcommand{\alphatilde}{\widetilde{\alpha}}

\newcommand{\rci}{r_{\mathrm{c},i}}

\newcommand\bgreek[1]{ \mathchoice
    {\hbox{\boldmath$\displaystyle{#1}$\unboldmath}}%
    {\hbox{\boldmath$\textstyle{#1}$\unboldmath}}%
    {\hbox{\boldmath$\scriptstyle{#1}$\unboldmath}}%
    {\hbox{\boldmath$\scriptscriptstyle{#1}$\unboldmath}}}

  \newcommand{\kms}{\,{\rm km\,s^{-1}}}
  \newcommand{\kmskpc}{\,{\rm km\,s^{-1}\,kpc^{-1}}}

  \newcommand{\kpc}{\,{\rm kpc}}

  \newcommand{\Myr}{\,{\rm Myr}}
  \newcommand{\Gyr}{\,{\rm Gyr}}

\title[Magnetic arms generated by multiple patterns]{Magnetic arms generated by multiple interfering galactic spiral patterns}
\author[L. Chamandy, K. Subramanian, A. Quillen]{Luke Chamandy$^{1}$\thanks{E-mail: luke@iucaa.ernet.in}, Kandaswamy Subramanian$^{1}$
\& Alice Quillen$^{2}$\\
$^{1}$Inter-University Centre for Astronomy and Astrophysics, Post Bag 4, Ganeshkhind, Pune 411007, India\\
$^{2}$Department of Physics and Astronomy, University of Rochester, Rochester, NY 14627, USA}

\begin{document}


\pagerange{\pageref{firstpage}--\pageref{lastpage}} \pubyear{2012}

\maketitle

\label{firstpage}

\begin{abstract}
Interfering two- and three-arm spiral patterns have previously been inferred to exist in many galaxies and also in numerical simulations,
and invoked to explain important dynamical properties,
such as lack of symmetry, kinks in spiral arms, and star formation in armlets.
The non-axisymmetric galactic mean-field dynamo model of \citet{Chamandy+13a} is generalized 
to allow for such \textit{multiple} co-existing spiral patterns in the kinetic $\alpha\kin$ effect, 
leading to the existence of magnetic spiral arms in the large-scale magnetic field with several new properties.
The large-scale magnetic field produced by an evolving superposition of two- and three-arm (or two- and four-arm) patterns 
evolves with time along with the superposition. 
Magnetic arms can be stronger and more extended in radius and in azimuth when produced by two interfering patterns rather than by one pattern acting alone.
Transient morphological features arise in the magnetic arms, 
including bifurcations, disconnected armlets, and temporal and spatial variation in arm strength and pitch angles.
Pitch angles of the large-scale magnetic field and magnetic arm structures (ridges)
 are smaller than those typically inferred from observations of spiral galaxies for model parameters of \citet{Chamandy+13a}, 
but can become comparable to typically inferred values for certain (still realistic) parameters.
The magnetic field is sometimes strongest in between the $\alpha\kin$-arms, 
unlike in standard models with a single pattern, where it is strongest within the $\alpha\kin$-arms.
Moreover, for models with a two- and three-arm pattern, 
some amount of $m=1$ azimuthal symmetry is found to be present in the magnetic field, 
which is generally not the case for forcing by single two- or three-arm patterns.
Many of these results are reminiscent of observed features in the regular magnetic fields of nearby spiral galaxies,
like NGC~6946, which has previously been inferred to have significant two- and three-arm spiral patterns,
and IC~342, which has been reported to contain an inner two-arm and outer four-arm pattern.
\end{abstract}

\section{Introduction}
\label{sec:introduction}
Magnetic arms, akin to the familiar stellar or gaseous spiral arms, 
are here defined as spiral-shaped enhancements of the regular (i.e. mean, or large-scale) magnetic field in galaxies.\footnote{In 
the observational literature, magnetic arms usually refer to narrow arm-like structures of polarized radio emission 
located in between the optical arms, that, moreover, 
have about the same winding (pitch) angle as the optical arms, 
and have magnetic field vectors roughly aligned with both types of arms.}
They have been identified through polarized synchrotron emission and its Faraday rotation in a handful of galaxies
\citep{Fletcher10, Beck+Wielebinski13}.
On the theoretical side, 
mean-field dynamo models have successfully produced strongly non-axisymmetric magnetic fields by modulating a dynamo parameter,
such as $\alpha$, the turbulent diffusivity $\eta\turb$, or the seed large-scale magnetic field,
along a rotating spiral (typically presumed to be co-spatial with the gaseous spiral pattern of the galaxy) 
(\citealt{Chiba+Tosa90, Mestel+Subramanian91, Subramanian+Mestel93, Moss98, Shukurov98, Rohde+99}, \citealt{Chamandy+13a}, \citealt{Chamandy+13b}
(hereafter \citetalias{Chamandy+13a} and \citetalias{Chamandy+13b}), \citealt{Moss+13}).

Certain properties of magnetic arms seem to be rather generic: 
(i) large radial and azimuthal extents, comparable to those of the gaseous spiral arms \citep[e.g.][]{Krause+89,Frick+00,Chyzy08,Fletcher+11}, 
(ii) pitch angles of the magnetic arms (ridges) comparable to those of the spiral arms \citep[e.g.][]{Frick+00},
(iii) pitch angles of the field within the magnetic arms again comparable to those of the gaseous arms
\citep[e.g.][and references therein]{Fletcher10,Beck+Wielebinski13}, and 
(iv) cases where magnetic arms are concentrated in interarm regions or, 
in any case, not confined to the gaseous arms \citep{Krause+89,Ehle+96,Beck+Hoernes96,Beck07,Beck+05,Chyzy08,Fletcher+11}.

None of the above properties have been easy to explain using mean-field dynamo models.
Properties (i) and (ii) do not arise naturally when the dynamo is forced by a steady rigidly rotating spiral
because magnetic arms tend to be localized to within only $\sim$1-3 $\kpc$ of the corotation radius
of the spiral pattern invoked to force the dynamo,
and to be rather tightly wound, crossing the gaseous arms (\citealt{Mestel+Subramanian91},\citetalias{Chamandy+13a,Chamandy+13b}).
However, these two assumptions, typical of most models, viz. that the spiral pattern is both steady and rigidly rotating,
are likely unrealistic \citep{Sellwood11,Grand+12a,Quillen+11,Donghia+13}.
Indeed, spiral arms that do not rotate rigidly but rather follow the galactic rotation curve 
and hence rapidly wind up as `corotating' arms \citep{Wada+11, Grand+12a} can resolve both of these
issues, although the resulting magnetic arms are both relatively weak and short-lived as compared 
to those produced by a steady rigidly rotating spiral pattern
\citepalias{Chamandy+13a}.

Property (iii) is difficult to reproduce in dynamo models for two separate reasons: 
firstly, standard galactic dynamo parameter values yield pitch angles $\sim-(8^\circ-20^\circ)$ for the magnetic field $\meanv{B}$, 
which is reasonably consistent with the range of pitch angles of the large-scale magnetic field inferred from observations \citep{Fletcher10}.
However, there is at least one notable exception: the pitch angle in M33 is inferred from observations to be $\sim-40^\circ$ \citep{Tabatabaei+08}.
Secondly, it is not clear how the magnetic field could get aligned to the spiral arms (with pitch angles often in the range $\sim30-40^\circ$),
although large-scale streaming motions and shocks may play a role.

Finally, while many interesting ideas have been put forward to explain property (iv), 
(\citealt{Shukurov98, Moss98, Rohde+99}, \citetalias{Chamandy+13a}, \citetalias{Chamandy+13b}, \citealt{Moss+13}), 
arguably, none of them are yet conclusive.

It should be noted, however, that the above properties are not strictly universal.
For example, at least one magnetic arm (ridge) of IC~342 seems to cross an optical arm \citep{Krause+89}.
Typical pitch angles of the regular magnetic field in M33, though large in magnitude compared to other galaxies,
are significantly smaller in magnitude than those of the optical arms \citep{Fletcher10}.
In both NGC~6946 and M51 pitch angles of the polarized component of the magnetic field match quite well 
with those of the magnetic arm structure (ridge) in one of the main arms, 
but the correspondence is much poorer in the other arm \citep{Beck07, Patrikeev+06}.

To complicate matters, polarized intensity is produced by a combination of regular field and anisotropic random field
(e.g. due to compression in spiral shocks and shearing motions),
and these two contributions must be disentangled using Faraday rotation measures, 
which depends on modelling \citep[e.g.][]{Fletcher+11}.
Given that compression in shocks would tend to align the magnetic field with the spiral arms \citep{Patrikeev+06},
polarization angles would be expected to be larger than pitch angles of the regular magnetic field within spiral arms,
especially for galaxies with strong density waves.
All in all, the above four properties can serve to guide the present investigation, 
but should not be taken as absolute.

As mentioned above, the structure and dynamics of the gaseous spiral can strongly affect the strength, 
morphology, and time evolution of the resulting magnetic arms \citepalias{Chamandy+13a}.
Therefore, it is important to explore what effect different types 
of spiral structure and dynamics have on the mean-field dynamo.
There is a rapidly growing body of literature that argues for alternate theories of spiral structure and dynamics.
Aside from the steady rigidly rotating spiral pattern and corotating (maximally winding up) arms models mentioned above,
galactic spirals can also be modeled as resulting from multiple interfering transient patterns (density waves).
This latter possibility is explored in the present work.
Other models, such as density waves that wind up and propagate \citep{Binney+Tremaine08}, are left for future work.

Both observational and numerical studies of spiral morphology have provided justification 
for the existence of more than one spiral pattern in some disk galaxies.
On the observational side, \citet{Elmegreen+92}, \citet{Shetty+07} and \citet{Meidt+09} 
have found that galaxies can contain more than one spiral symmetry, with different pattern speeds and multiplicities.
On the numerical side \citet{Sellwood11}, \citet{Quillen+11} and \citet{Roskar+12} 
have detected multiple spiral patterns (that may overlap in time) in their $N$-body simulations.
Generally, these patterns do not survive for galactic lifetimes, 
but can survive (in a quasi-steady state) for several rotation periods.
Non-linear coupling between density waves at different locations in the disk \citep{Masset+Tagger97}
may even lead to (composite) spiral patterns that last for $\gtrsim10$ rotation periods \citep{Donghia+13}.
A criticism of many spiral structure studies, e.g. those pertaining to the galactic dynamo, 
is that evolving spiral patterns have not been considered.   
We begin to address this issue here with a study of time dependent phenomena due to the presence of multiple patterns.

The paper is organized as follows. 
In Section \ref{sec:model}, the non-axisymmetric galactic dynamo model is presented.
In Section \ref{sec:results}, numerical solutions are described,
and where necessary, explained in the light of existing theory.
Key results are summarized and discussed in Section \ref{sec:conclusion}.
The relevance of the possible transience of individual spiral patterns is taken up in Appendix~\ref{sec:transient}.

\section{The model}
\label{sec:model}
The galactic disk dynamo model and numerical code of \citetalias{Chamandy+13a} are mostly retained. 
This model is based on the $\alpha^2\omega$ dynamo model \citep{Ruzmaikin+88}, 
with the $\omega$ effect produced by the differential rotation, 
and the $\alpha$ effect presumed to result from helicity in the small-scale turbulence 
in a rotating and stratified medium \citep{Brandenburg+Subramanian05a}.
The dynamo equations are solved on a polar grid, linear in $r$. 
The numerical code makes use of the thin disk approximation \citep{Ruzmaikin+88} to avoid having to calculate $z$-components of the magnetic field 
and mean electromotive force.
It also uses the `no-$z$' approximation, 
which approximates $z$-derivatives as simple divisions by the disk thickness \citep{Subramanian+Mestel93, Moss95, Phillips01},
to reduce the problem to two dimensions.
The resolution is $n_r=300$, $n_\phi=150$, the outer radius is $R=20\kpc$, and vacuum boundary conditions are imposed at $r=0,R$.

\subsection{The mean-field dynamo}
\label{sec:dynamo}
The mean-field induction equation \citep{Moffatt78, Krause+Radler80},
\begin{equation}
\label{meaninduction}
\frac{\partial \meanv{B}}{\partial t} = \bmDel \times \left( \meanv{U} \times \meanv{B} + \bgreek{\Emf} -\eta \nabla \times \meanv{B}\right),
\end{equation}
is solved to obtain the mean magnetic field. 
Here overbar denotes mean quantities, 
$\bm{B}$ is the magnetic field, $\bm{U}$ is the velocity field, 
and both are assumed to be separable into mean and random parts,
\begin{equation}
\label{scaleseparation}
\bm{U}= \meanv{U} +\bm{u}, \quad \bm{B}= \meanv{B} +\bm{b}.
\end{equation}
In addition, $\Emf=\mean{\bm{u}\cro\bm{b}}$ is the mean turbulent electromotive force, 
and Ohmic terms are henceforth neglected since the microscopic diffusivity $\eta$ 
is much smaller than the turbulent diffusivity ($\eta\turb$ below) in galaxies.
The mean-field induction equation is solved together with two other dynamical equations.
The first such equation is for $\Emf$
\citep{Rogachevskii+Kleeorin00, Blackman+Field02, Brandenburg+Subramanian05a, Rheinhardt+Brandenburg12},
\begin{equation}
\label{minimaltau}
\frac{\partial \Emf}{\partial t}=\frac{1}{\tau}(\alpha\meanv{B}-\eta\turb \bmDel\times\meanv{B}-\Emf),
\end{equation}
where $\alpha$ and $\eta\turb$ depend on the mean properties of the turbulence,
and $\tau$ is a relaxation time assumed here to be equal to the correlation time of the turbulence at the energy-carrying scale.
In the limit $\tau=0$, Eqs.~\eqref{meaninduction} and \eqref{minimaltau} can be combined to give the standard mean-field dynamo equation,
\begin{equation}
\label{dynamo_FOSA}
\frac{\del\meanv{B}}{\del t} =\Del\cro(\meanv{U}\cro\meanv{B}+\alpha\meanv{B}-\eta\turb\Del\cro\meanv{B}) \quad (\tau=0).
\end{equation}

The $\alpha$ effect in Eq.~\eqref{meaninduction} or \eqref{minimaltau} can be represented as the sum of kinetic and magnetic contributions,
\begin{equation}
\alpha= \alpha\kin +\alpha\magn,
\end{equation}
where $\alpha\kin$ is introduced using an analytical expression, 
while $\alpha\magn$  is solved for dynamically.
$\alpha\magn$ becomes comparable to $\alpha\kin$ only upon saturation, 
when the energy in the mean magnetic field is comparable to that in the turbulence.
$\alpha\magn$ is obtained using the `dynamical quenching' formalism,
which is based on magnetic helicity conservation
\citep{Pouquet+76,Kleeorin+96,Blackman+Field02,Blackman+Brandenburg02,Brandenburg+Subramanian05a,Subramanian+Brandenburg06,Hubbard+Brandenburg10,Shapovalov+Vishniac11}.
This formalism leads to the third dynamical equation to be solved \citep{Shukurov+06},
\begin{equation}
\label{dalpha_mdt}
\frac{\del\alpha\magn}{\del t}
=-\frac{2\eta\turb\Emf\cdot\meanv{B}}{l^2B\eq^2}-\bmDel\cdot\Flux_\alpha,
\end{equation}
where $l$ is the energy-carrying scale of the turbulence, $B\eq\equiv\sqrt{4\pi\rho}u$, with $\rho$ the gas density, 
is the `equipartition' field strength,
and $\Flux_\alpha$ is a flux of $\alpha\magn$ (related to the flux of the mean small-scale magnetic helicity).
The Ohmic term has been dropped, 
which is justified if the time scales considered are short compared to the resistive time scale 
or if the Ohmic term is negligible compared to the flux term.
In fact, the latter condition must hold in order to avert the catastrophic quenching of the dynamo, 
and so a vertical advective flux of $\alpha\magn$, 
expected to be present because of galactic winds/fountain flow, is included
\citep{Shukurov+06,Heald12,Bernet+13}.\footnote{
In \citetalias{Chamandy+13a} it was shown that a (perhaps equally plausible) diffusive flux \citep{Mitra+10} has essentially the same effect,
though with resulting large-scale field somewhat stronger in the saturated state.}

The kinetic contribution to $\alpha$ has an amplitude
given by Krause's law \citep{Steenbeck+66,Krause+Radler80,Ruzmaikin+88,Brandenburg+Subramanian05a,Brandenburg+13},
\begin{equation}
\label{Krause}
\alpha\kin=\frac{l^2}{h}\omega\alphatilde,
\end{equation}
where $h$ is the local disk half-thickness, $\omega$ is the angular velocity of the gas, 
and $\alphatilde$ has an azimuthal mean of unity.
Non-axisymmetry in the disk is imposed by modulating $\alphatilde$, and hence $\alpha$, along a spiral,
assumed to be correlated in some way with the gaseous spiral of the galaxy.

\begin{table}
\begin{center}
\caption{Parameter values for the various models studied: pattern index $i$, 
         multiplicity $n_i$, winding parameter $\kappa_i$, $\alpha\kin$-spiral pitch angle $p_{\alpha,i}$,
         normalized spiral perturbation amplitude $\epsilon_i$, corotation radius $\rci$,
         activation and de-activation times for spiral forcing $t_{\mathrm{on},i}$ and $t_{\mathrm{off},i}$, respectively.
         All models were run with $\tau=0$ and $\tau=l/u$ (the latter designated with `$\tau$' after the model name).
         A2: Standard two-arm model. A3: Standard three-arm model.
         B2: Alternate two-arm model. B3: Alternate three-arm model. A23: Standard two- and three-arm model.
         A24: Two- and four-arm model.
         T23: Transient two- and three-arm model. L23: Large separation two- and three-arm model.
         S23: Small separation two- and three-arm model.
         X23: Alternate disk two- and three-arm model.
         Distances $\rci$ are in $\kpc$, 
         while times $t_{\mathrm{on},i}$ and $t_{\mathrm{off},i}$ are in $\Gyr$.}
\label{tab:models}
\begin{tabular}{lllllllllll}
\hline
Model          &$i$   &$n_i$    &$\kappa_i$  &$p_{\alpha,i}$  &$\epsilon_i$  &$\rci$   &$t_{\mathrm{on},i}$    &$t_{\mathrm{off},i}$\\
\hline                   
\hline                   
A2/A2$\tau$    &1     &2        &-3          &$-34^\circ$     &0.5           &6        &0                      &--\\
\hline
A3/A3$\tau$    &1     &3        &-6          &$-27^\circ$     &0.5           &7        &0                      &--\\
\hline                   
B2/B2$\tau$    &1     &2        &-6          &$-27^\circ$     &0.5           &7        &0                      &--\\
\hline
B3/B3$\tau$    &1     &3        &-3          &$-34^\circ$     &0.5           &6        &0                      &--\\
\hline
A23/A23$\tau$  &1     &2        &-3          &$-34^\circ$     &0.5           &6        &0                      &--\\
               &2     &3        &-6          &$-27^\circ$     &0.5           &7        &0                      &--\\
\hline
A24/A24$\tau$  &1     &2        &-3          &$-34^\circ$     &0.5           &6        &0                      &--\\
               &2     &4        &-6          &$-27^\circ$     &0.5           &7        &0                      &--\\
\hline
T23/T23$\tau$  &1     &2        &-3          &$-34^\circ$     &0.5           &6        &$4.5$                  &$5$\\
               &2     &3        &-6          &$-27^\circ$     &0.5           &7        &$4.5$                  &$5$\\
\hline
L23/L23$\tau$  &1     &2        &-3          &$-34^\circ$     &0.5           &5        &0                      &--\\
               &2     &3        &-6          &$-27^\circ$     &0.5           &7        &0                      &--\\
\hline                   
S23/S23$\tau$  &1     &2        &-3          &$-34^\circ$     &0.5           &6.5      &0                      &--\\
               &2     &3        &-6          &$-27^\circ$     &0.5           &7        &0                      &--\\
\hline                   
X23/X23$\tau$  &1     &2        &-3          &$-34^\circ$     &1.0           &6        &0                      &--\\
               &2     &3        &-6          &$-27^\circ$     &1.0           &7        &0                      &--\\
\hline
\end{tabular}
\end{center}
\end{table}

\subsection{The galactic disk}
\label{sec:disk}
For the galactic rotation curve, the Brandt profile
\begin{equation}
  \label{Brandt}
  \omega(r)=\frac{\omega_0}{\left[1+(r/r_\omega)^2\right]^{1/2}},
\end{equation}
is used.
Parameter values $\omega_0\simeq127\kmskpc$ and $r_\omega=2\kpc$ are chosen 
so that the circular velocity $\mup=r\omega=250\kmskpc$ at $r=10\kpc$ (for all models but one).
With this profile, $\omega\rightarrow\const$ as $r\rightarrow0$ (solid body rotation) and $\omega\propto 1/r$ 
for $r\gg r_\omega$ (flat rotation curve).
The disk half-thickness varies hyperbolically with radius \citep{Ruzmaikin+88},
\begin{equation}
\label{hyperboloid}
h(r)=h\D\left[1+(r/r\D)^2\right]^{1/2},
\end{equation}
where $h\D$ is the scale height at $r=0$ and $r\D=10\kpc$ controls the disk flaring rate.
The value of $h$ at $r=10\kpc$ is chosen to be $0.5\kpc$, which gives $h\D\simeq0.35\kpc$.
With this form, $h\rightarrow h_D={\rm constant}$ as $r\rightarrow0$ and $h\propto r$ for $r\gg r_D$.
The extent to which the ionized component of galactic disks is flared has not been firmly established, 
and is an assumption of the model.
Some models that use an unflared disk with constant $h$ have also been explored, 
and the differences in the results are very minor.
The equipartition magnetic field strength is taken as
\begin{equation}
\label{Beq}
B\eq=B\f\Exp{-r/R},
\end{equation}
where $B\f$ is an adjustable parameter and $R=20\kpc$ is the outer radius of the disk (numerical grid).
This corresponds to an exponential scale length of $10\kpc$ for the turbulent energy in the ionized component.
This is comparable to but slightly larger than $R\approx7\kpc$ that is determined for the total magnetic field in the galaxy NGC~6946
\citep{Beck07}.

\subsection{The $\alpha\kin$-spirals}
Non-axisymmetric forcing of the dynamo is modelled as an enhancement of the $\alpha$ effect within multiple spiral patterns,
which are situated at different radii and hence rotate with different pattern speeds according to the galactic rotation curve.
The individual spiral patterns are taken to be steady, rigidly rotating, and logarithmic in shape.
The prescription used is close to that of \citet[][hereafter \citetalias{Comparetta+Quillen12}]{Comparetta+Quillen12}; 
the analogous quantity in that paper is the surface density.
In polar coordinates $(r,\phi)$, 
\begin{equation}
\label{alphatilde}
\alphatilde(r,\phi,t)=1 +\epsilon_1 A_1(r) f_1(r,\phi,t) +\epsilon_2 A_2(r) f_2(r,\phi,t)
\end{equation}
for all models except one; see below.
Using $i=1,2$ to represent the two spiral patterns, 
the overall amplitude of each pattern is set by the parameter $\epsilon_i$, 
modulated by the envelope \citepalias{Comparetta+Quillen12} 
\begin{equation}
A_i=\exp\left\{-\frac{[\log_{10}(r/r_{\mathrm{0},i})]^2}{2w_i^2}\right\},
\end{equation}
where $w_i$ is a parameter that determines how sharply peaked in radius is the amplitude,
and $r_{\mathrm{0},i}$ is the radius at which the amplitude peaks.
The values $w_i=1$ and $r_{\mathrm{0},i}=r_{\mathrm{c},i}$ (Table~\ref{tab:models}) are used, 
corresponding to a fairly flat envelope function gently peaked at the corotation radius of the respective pattern.
Values $w_1=0.14$, $w_2=0.17$, $r_{\mathrm{0},1}=5\kpc$, $r_{\mathrm{0},2}=6\kpc$ were also tried for the standard model A23,
making the patterns more localized in radius and peaked slightly inward of corotation, 
and thus closer to the model of \citetalias{Comparetta+Quillen12} and the patterns seen in \citet{Quillen+11}.
However, this was found to affect the results only very slightly, 
showing that the model is insensitive to modest adjustments to the envelope function.
Values $w_i=1$ and $r_{\mathrm{0},i}=r_{\mathrm{c},i}$ are therefore retained
to keep the number of independent parameters to a minimum.
Non-axisymmetry is implemented using 
\begin{equation}
\label{f}
f_i=\cos(\chi_i)+0.25\cos(2\chi_i),
\end{equation}
where the second term makes the function more peaked at $\phi=0$ than a simple cosine \citepalias{Comparetta+Quillen12},
and the angular part
\begin{equation}
\chi_i=-n_i(\phi-\Omega_it)+\kappa_i\ln(r/R)
\end{equation}
leads to a logarithmic spiral.
The quantity $n_i$ designates the number of arms in the pattern, 
$\kappa_i$ ($<0$ for a trailing spiral) determines how tight is the spiral,
and $\Omega_i\equiv\omega(r_{\mathrm{c},i})$ is the pattern angular speed.
Using the standard definition for pitch angle,
\[
\mathrm{pitch\;angle}\equiv\cot^{-1}\left(\frac{\del\phi\ma}{\del \ln r}\right), 
\]
where $\phi\ma$ is the azimuth at which a given spiral perturbation ($\eps_iA_if_i$ in this case) is maximum, 
it can easily be shown that the pitch angle of each individual $\alpha\kin$ spiral pattern is
\[
p_{\alpha,i}= \tan^{-1}\left(\frac{n_i}{\kappa_i}\right),
\]
which gives $p_{\alpha,1}=-34^\circ$ and $p_{\alpha,2}=-27^\circ$ for the two-pattern models studied.

\begin{figure*}                     
  $                                 
  \begin{array}{c c c c}              
    \includegraphics[width=42mm]{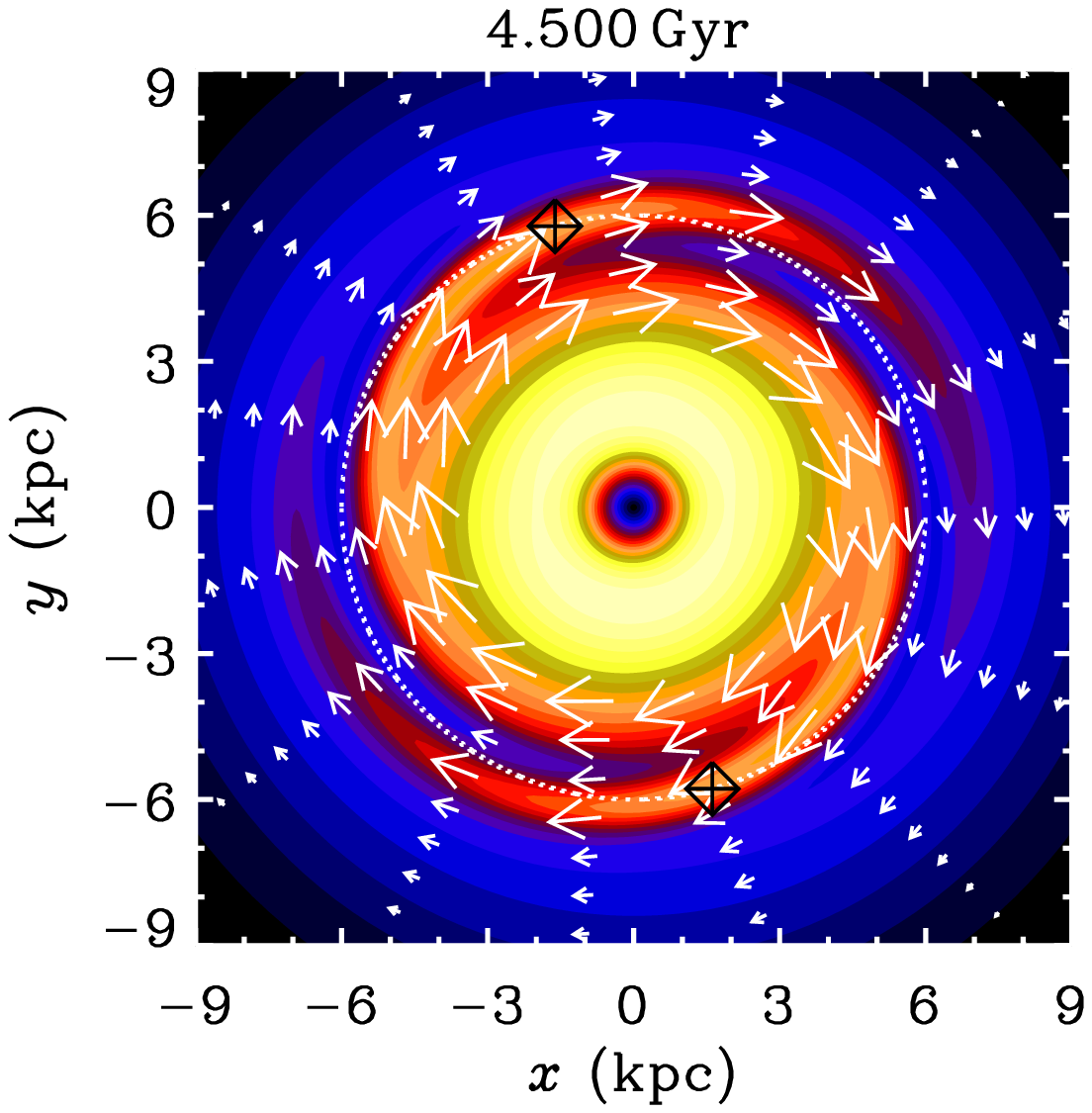}
    \includegraphics[width=42mm]{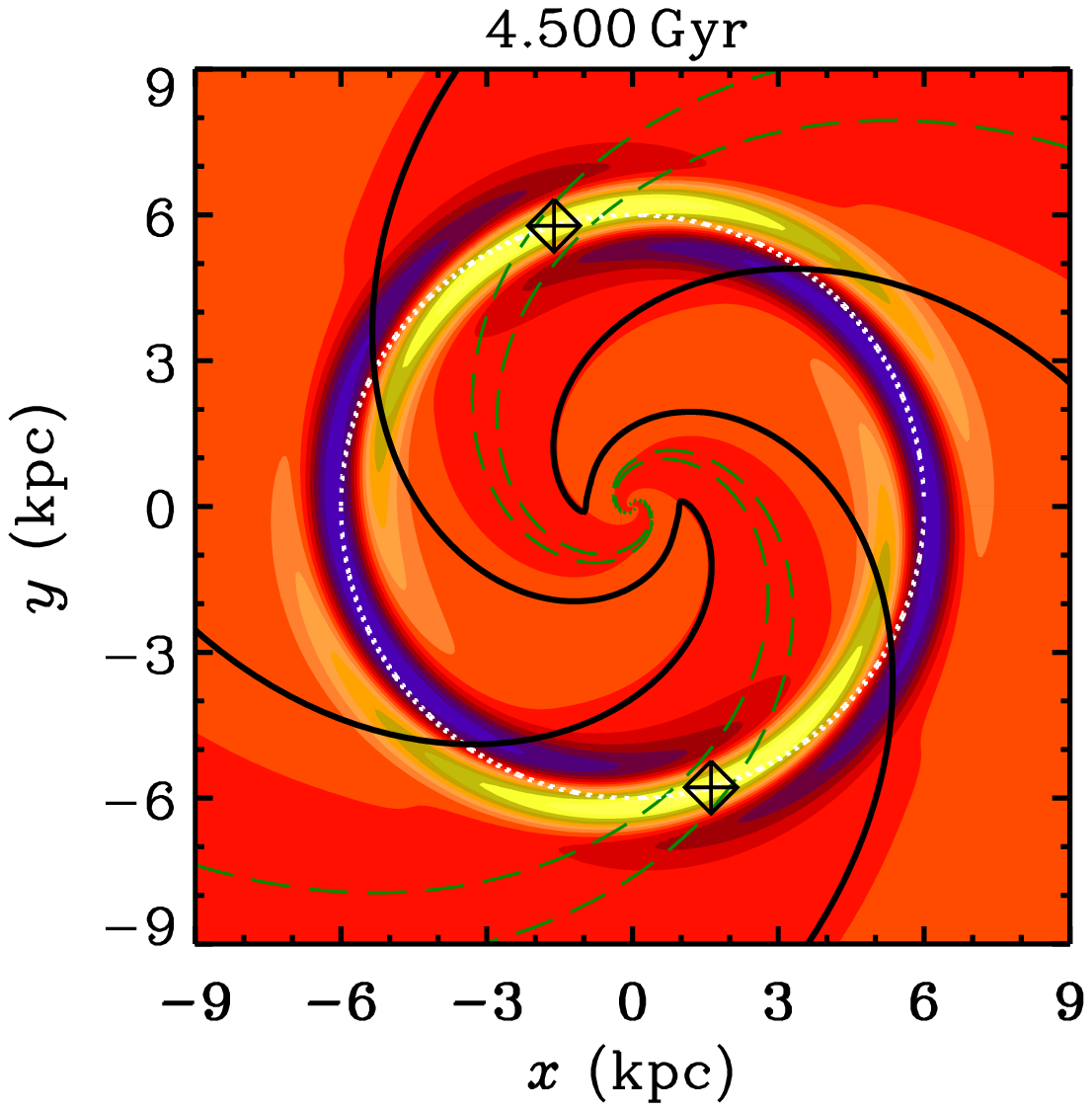}
    \includegraphics[width=42mm]{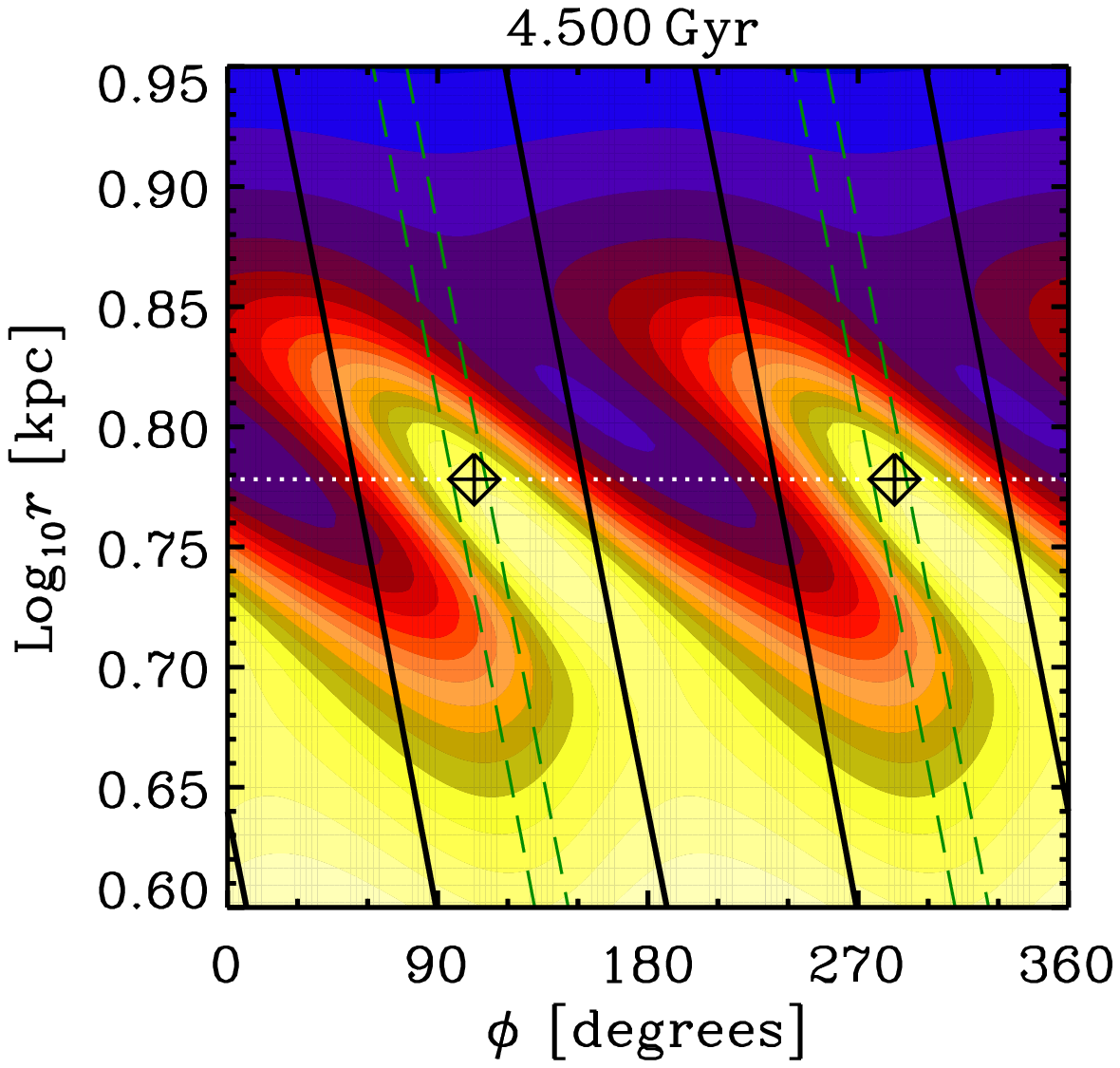}
    \includegraphics[width=42mm]{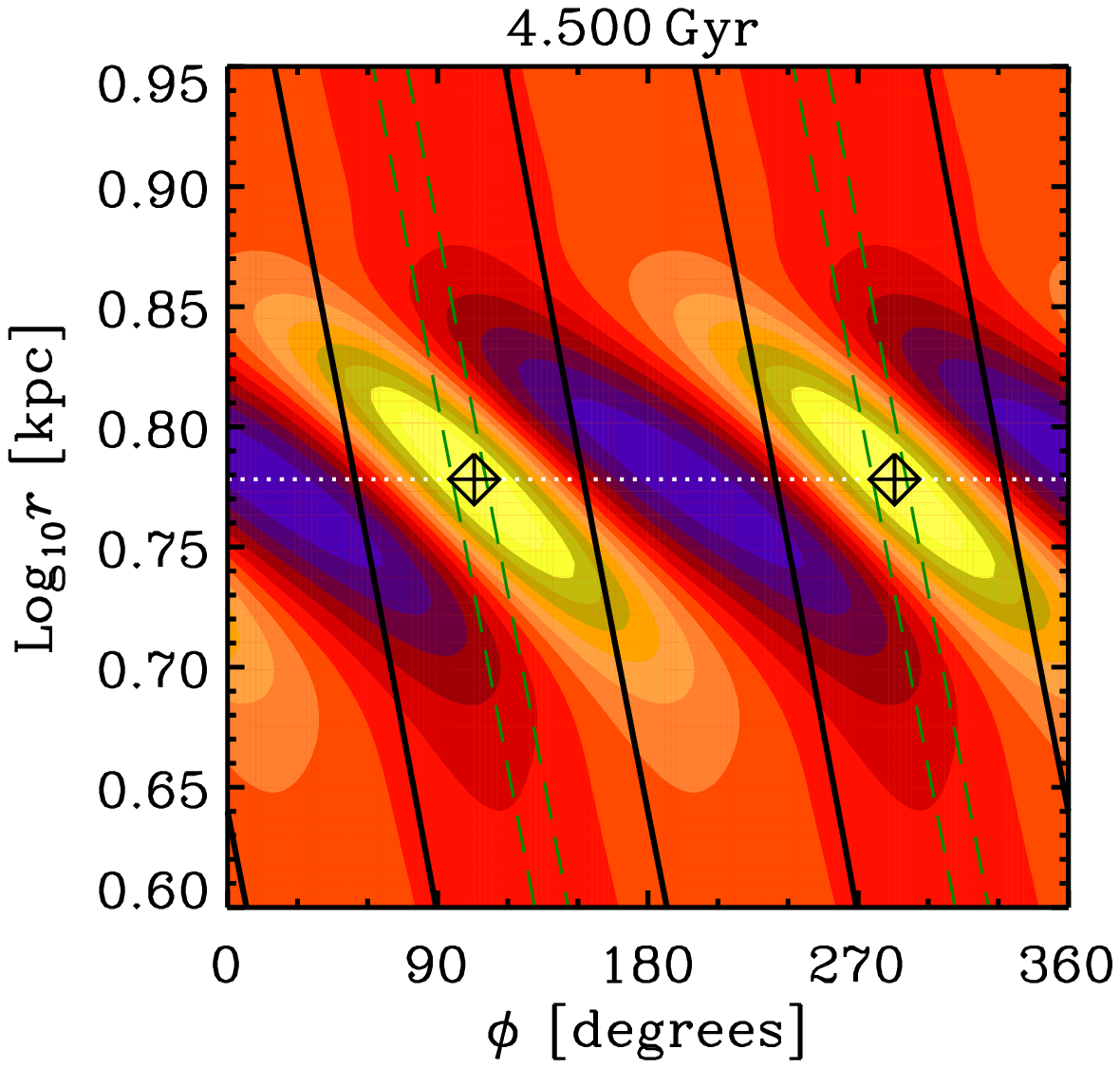}\\
    \includegraphics[width=42mm]{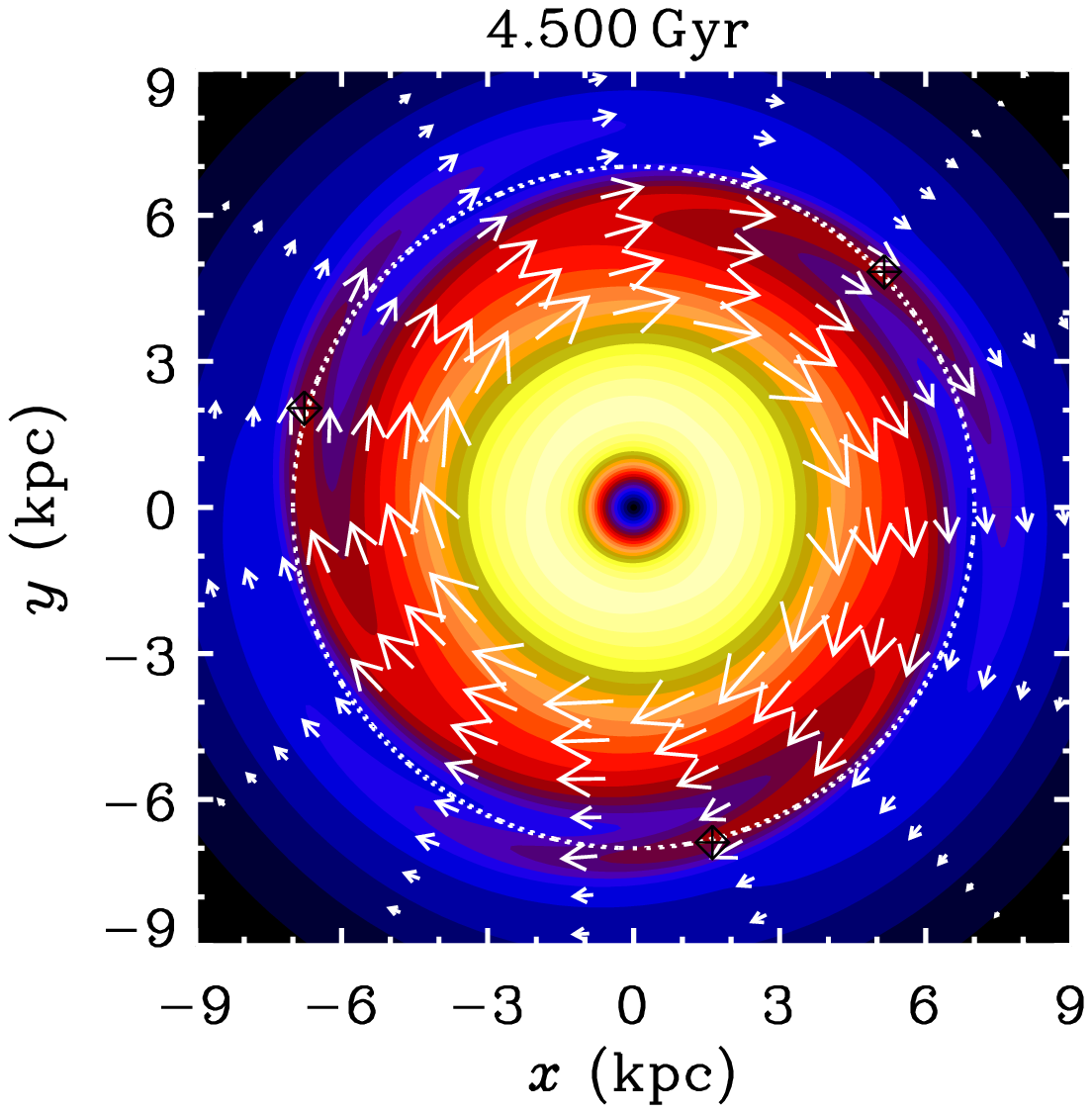}
    \includegraphics[width=42mm]{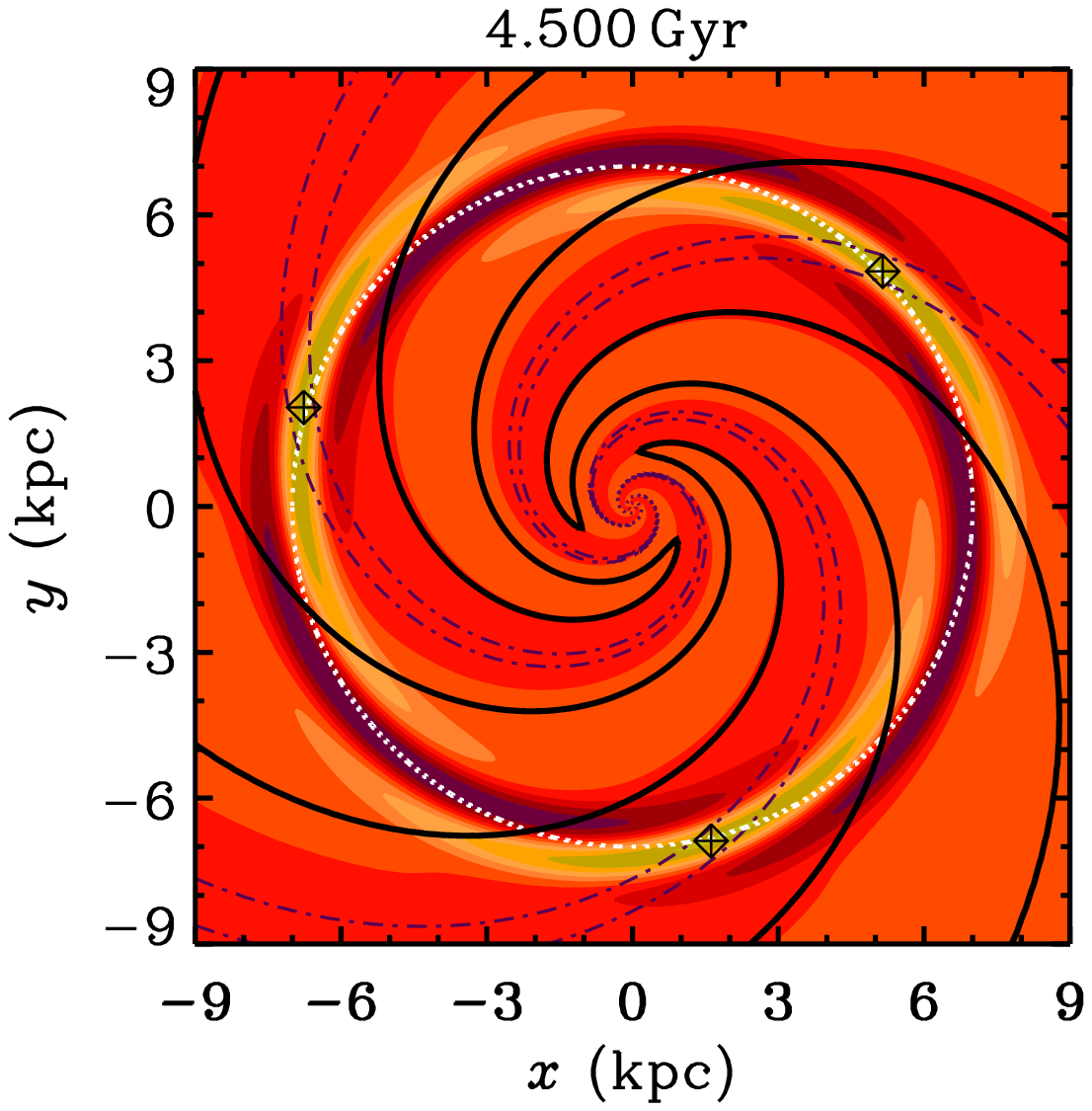}
    \includegraphics[width=42mm]{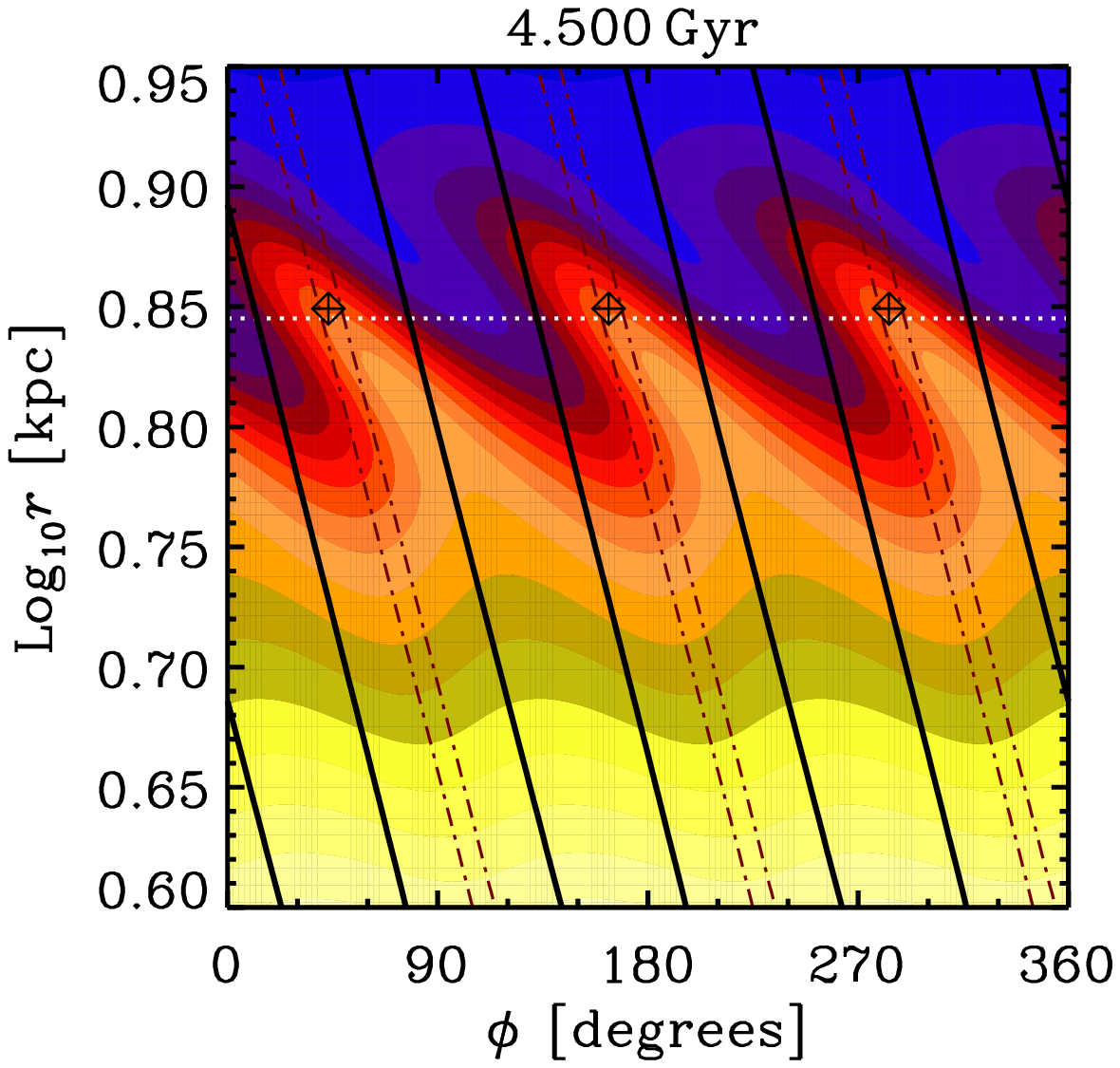}
    \includegraphics[width=42mm]{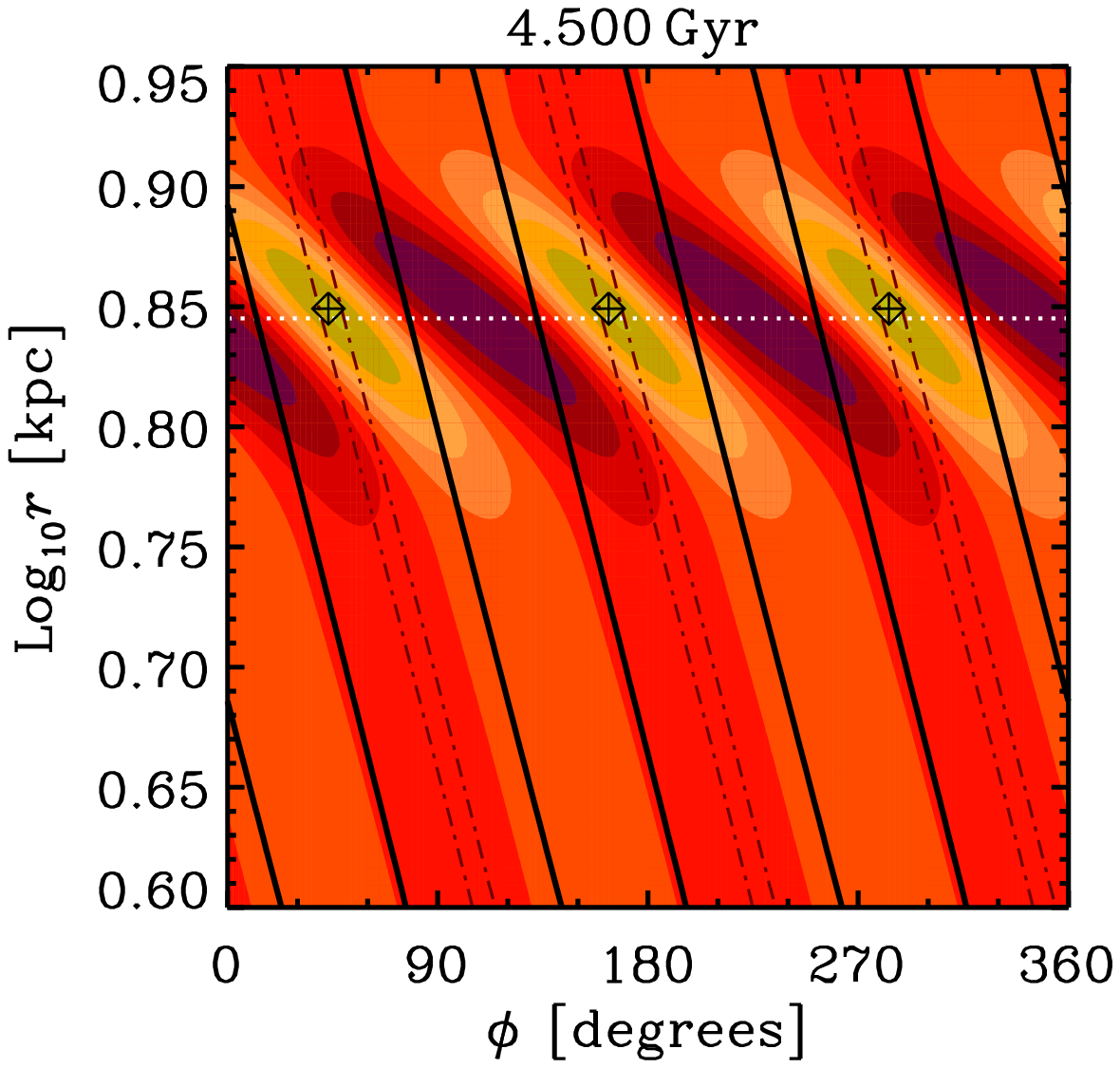}\\
    \includegraphics[width=42mm]{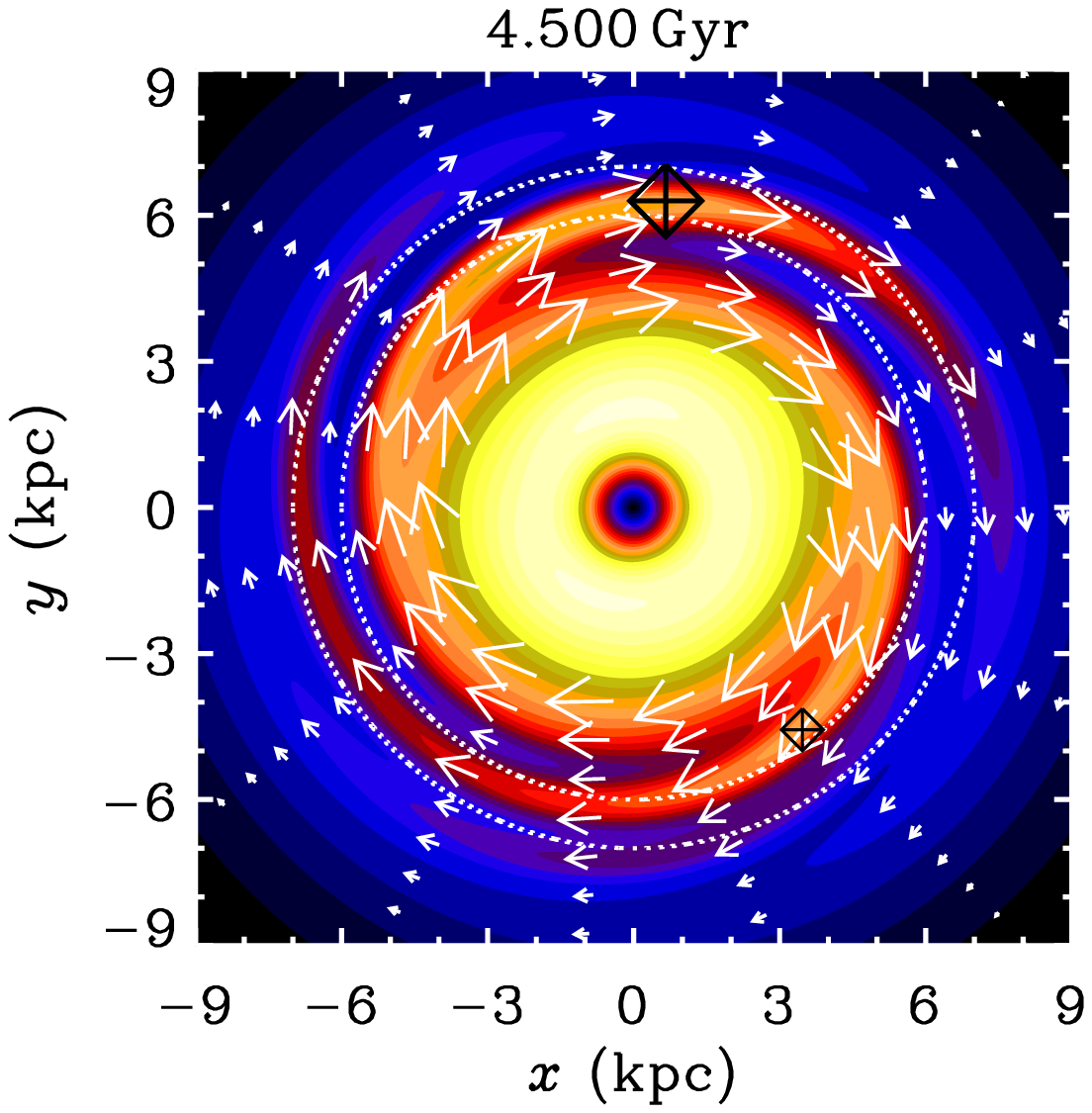}
    \includegraphics[width=42mm]{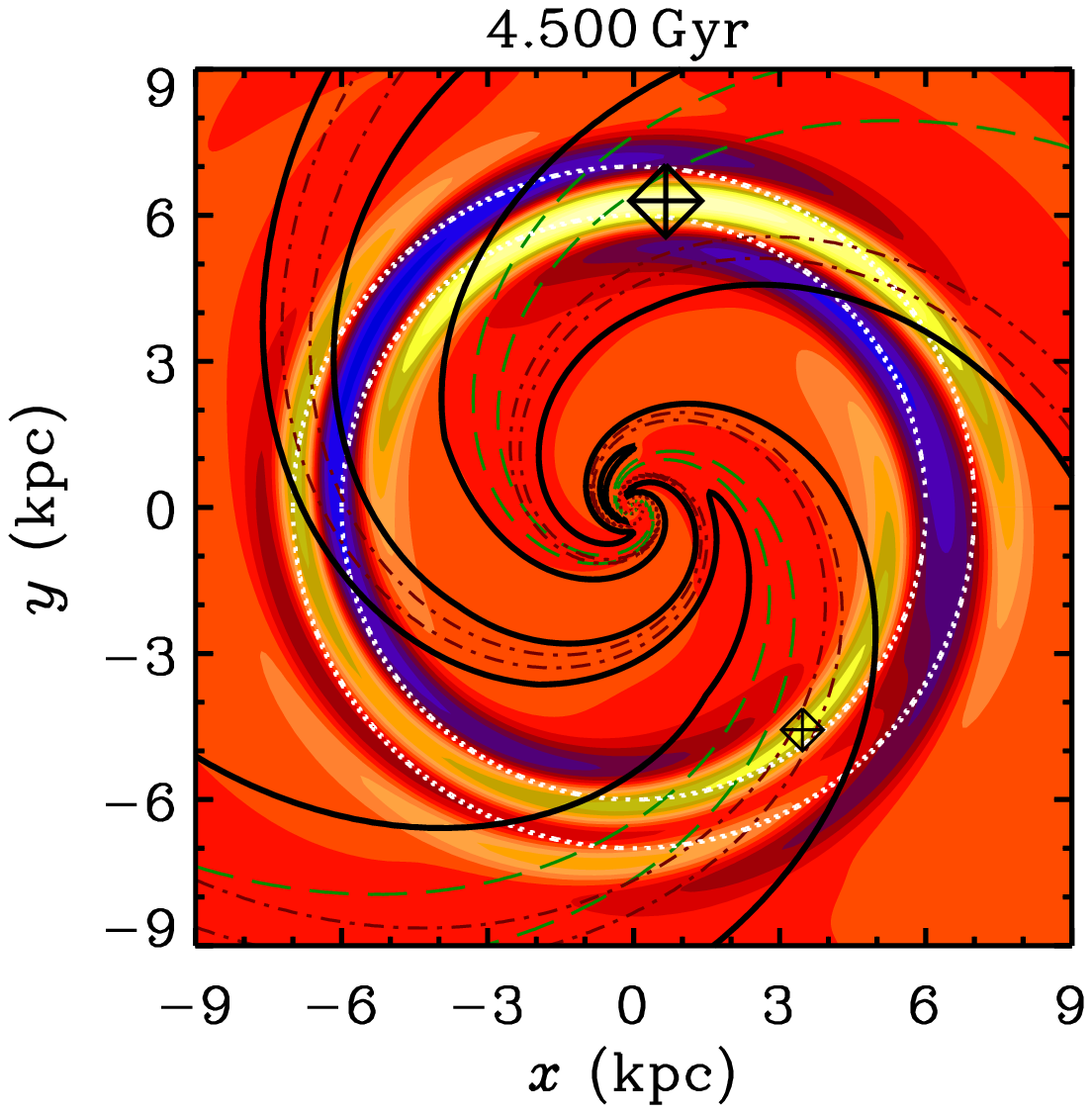}
    \includegraphics[width=42mm]{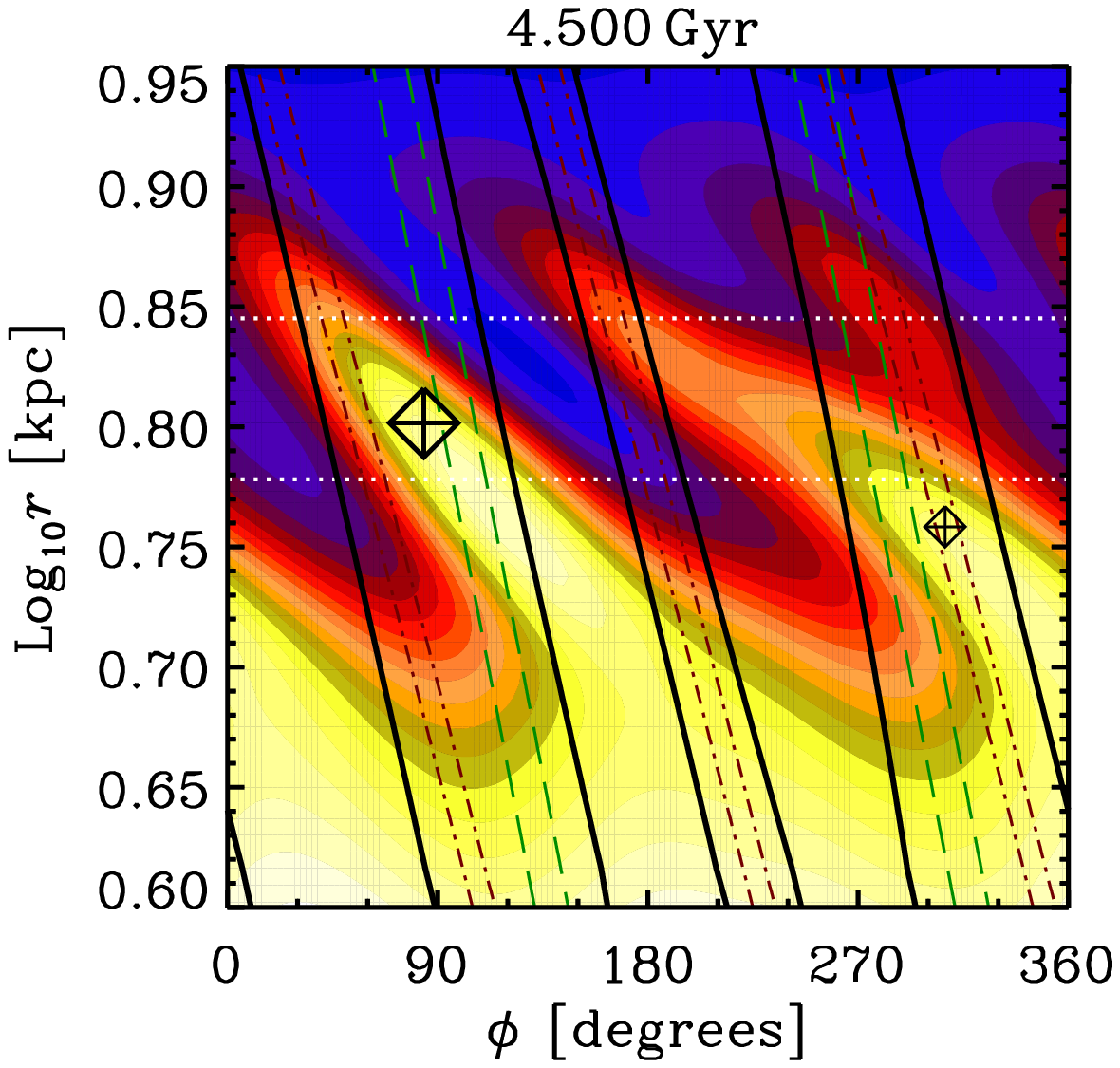}
    \includegraphics[width=42mm]{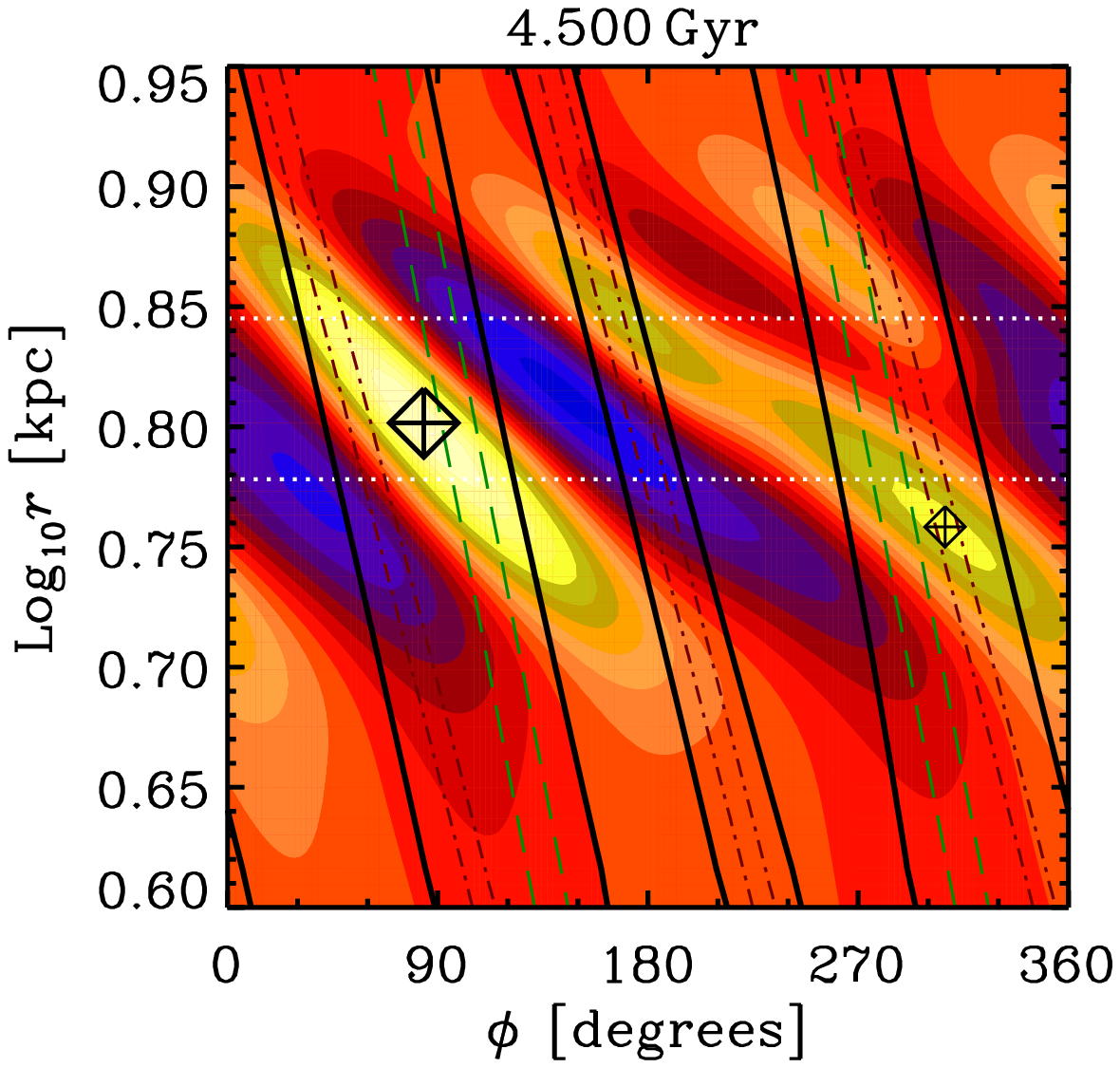}\\
    \includegraphics[width=42mm]{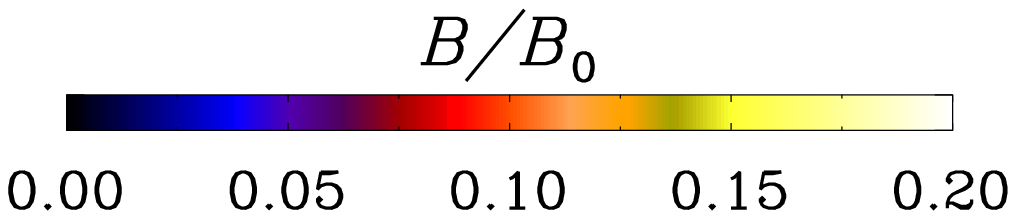}
    \includegraphics[width=42mm]{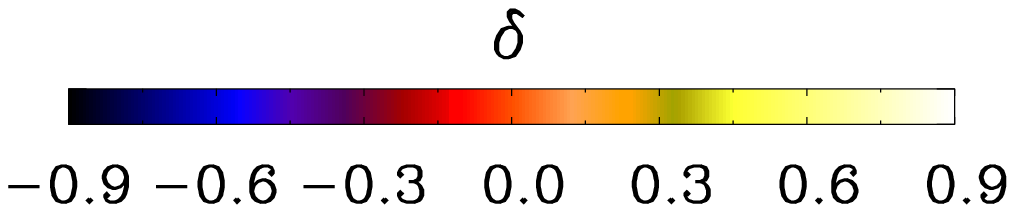}
    \includegraphics[width=42mm]{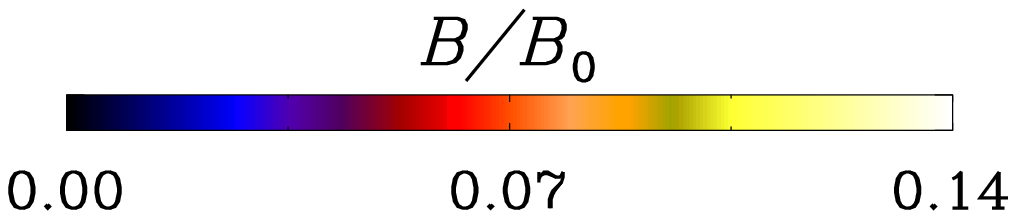}
    \includegraphics[width=42mm]{IS_018_colr_delta_nt180.eps}
  \end{array}                        
  $                                  
 \caption{Left column: magnitude of the total mean magnetic field, normalized to the equipartition field at $r=0$ (see \citetalias{Chamandy+13a})
          at $t=4.5\Gyr$ for Model~A2 with two-arm forcing (top), Model~A3 with three-arm forcing (middle) 
          and Model~A23 with both two- and three-arm forcing (bottom).
          Field vectors show the relative magnitude and pitch angle of the field. 
          Corotation circles for the two patterns are shown by white dotted lines.
          A cross within a diamond designates the locations of the two largest local maxima of $\delta$, defined in Eq.~\eqref{delta};
          the symbol size is proportional to the local value of $\delta$.
          Second column from left: the quantity $\delta$, defined in the text.
          Black contours identify the 50\% level of $\alpha\kin$, normalized to the azimuthal mean, 
          while dashed and dash-dotted lines are the 95\% contours of the two- and three-arm perturbations to $\alpha\kin$, respectively.
          Third column from left: the mean magnetic field strength in `unwound' $\log_{10}(r)$ vs. $\phi$ coordinates 
          (with $\phi$ starting at the positive $x$-axis and increasing in the counter-clockwise direction).
          Right column: $\delta$ in $\log_{10}(r)$ vs. $\phi$ coordinates.
          \label{fig:Bcompare}}            
\end{figure*}                       

\subsection{The parameter space studied}
\label{sec:parameter}
A list of models (runs), as well as the parameter values used for each, is given in Table~\ref{tab:models}.
For each set of parameter values in Table~\ref{tab:models}, the dynamo relaxation time $\tau$ is taken to be alternately equal to zero 
or the eddy turnover time $l/u=9.8\Myr$, where $l=0.1\kpc$ is the scale and $u=10\kms$ the rms velocity of the largest turbulent eddies. 
Those models suffixed with `$\tau$' have $\tau=l/u$ while those without this suffix have $\tau=0$.
Models A2, A3 and A23 are the fiducial two-, three-, and the combined two- and three-arm models, respectively.
Model B2 (B3) has the same parameters as A3 (A2), except for the number of arms,
to explore the effect of the number of arms on the solution.
Model L23 (S23) is the same as A23 but with the inner pattern moved inward (outward) 
to explore the effect of a larger (smaller) radial separation between the two- and three-arm patterns.
Model~A24 again has the same parameters as Model~A23 except that the outer pattern now contains four arms.
Model~T23 also has the same parameters, 
except now the transient nature of galactic spiral patterns are explored.
In this model, both patterns are imposed from the time $t=4.5\Gyr$ 
(magnetic field already in the saturated state),
and turned off at the time $t=5\Gyr$.  
Although this is admitedly an oversimplication, it is sufficient for its intended purpose, 
which is to test how rapidly and in what manner the magnetic field can adjust to the turning on and turning off of a transient spiral pattern,
in the saturated state.
For all other models, the spiral patterns are imposed from $t=0$, 
and their governing parameters are kept constant for the duration of the simulation ($\sim5\Gyr$).
The magnetic field is then studied at $t>4.5\Gyr$, which corresponds to the saturated state.
However, the field is studied for only a few rotation periods of the pattern (i.e. $\mathrm{few}\times0.15\Gyr$ for the innermost pattern in Model~A23), 
comparable to the typical lifetimes of such patterns seen in some simulations \citep[e.g.][]{Quillen+11}.
Model~X23 uses disk and dynamo parameters that, while still realistic, 
are found to be better suited to producing the observed properties of Sect.~\ref{sec:introduction}:
$\mup(10\kpc)=180\kms$, $\epsilon_1=\epsilon_2=1$, $h\D\simeq0.32\kpc$ and $\muz=2\kms$ at $r=0$, 
decreasing smoothly to $0.2\kms$ at $r=10\kpc$.
Moreover, $f_i$ in equation \eqref{alphatilde} is replaced by $(0.75 +f_i)$ 
so that $l^2\omega/h$ (Krause's law) now represents the \textit{minimum} of $\alpha\kin$ [see equations \eqref{Krause} and \eqref{f}].

\section{Results}
\label{sec:results}
The magnetic arm morphology that arises when two patterns force the dynamo is substantially more complex 
than that which arises from forcing by a single pattern.
This can be seen from Figure~\ref{fig:Bcompare}, part of which shows the large-scale magnetic field strength $B\equiv|\meanv{B}|$,
normalized to the equipartition field strength at $r=0$, $B\f\equiv B\eq(0)$, in the leftmost and third from left columns.
All snapshots are taken at the same time, $4.5\Gyr$ after the start of the simulation, 
when the magnetic field is in the saturated (almost steady) state.
In the left two columns Cartesian coordinates are used, 
while in the right two columns, $\log_{10}(r)$ vs. $\phi$ coordinates are used.
In the second from left and rightmost columns, the quantity illustrated is
\begin{equation}
\label{delta}
\delta\equiv\frac{\mbp-\mbp^{(0)}}{\mbp^{(0)}},
\end{equation}
where $\mbp^{(0)}$ is the axisymmetric part of $\mbp$.
($\mbp$ is used, rather than $\mbr$, because $\mbp$ is the dominant component.
The quantity $\delta$ is, in some sense, more useful than $|\meanv{B}|$ when discussing magnetic arms
because it is not biased to small radii, where the axisymmetric field itself is stronger.
For this reason, it represents quite faithfully the degree of non-axisymmetry as a function of $r$.)
The top row shows the result of forcing by the two-arm pattern alone (Model A2), 
the middle row shows the same for the three-arm pattern alone (Model A3),
and the bottom row shows the result obtained when both patterns are included (Model A23).
In all plots showing $|\meanv{B}|$ or $\delta$, 
a cross enclosed within a diamond has been plotted at the two largest local maxima of $\delta$,
with symbol size proportional to the local $\delta$.

\subsection{Relative strength of two- and three-arm magnetic field patterns}
\label{sec:multiplicity}
\begin{figure}                     
  $                                 
  \begin{array}{c c}              
    \includegraphics[width=42mm]{IS_020_deltacorotate_nt180.eps}		
    \includegraphics[width=42mm]{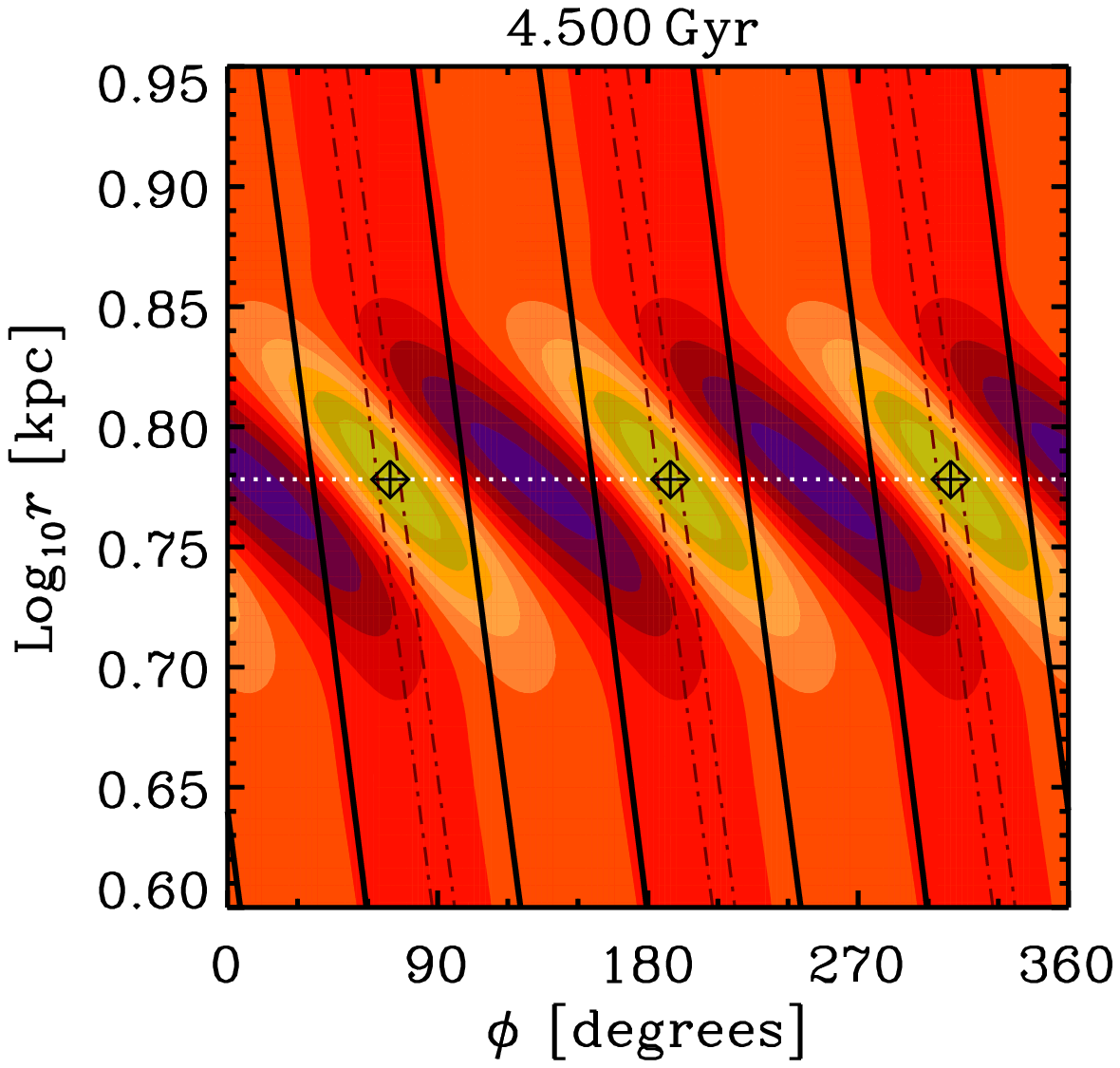}\\		
    \includegraphics[width=42mm]{IS_021_deltacorotate_nt180.eps}		
    \includegraphics[width=42mm]{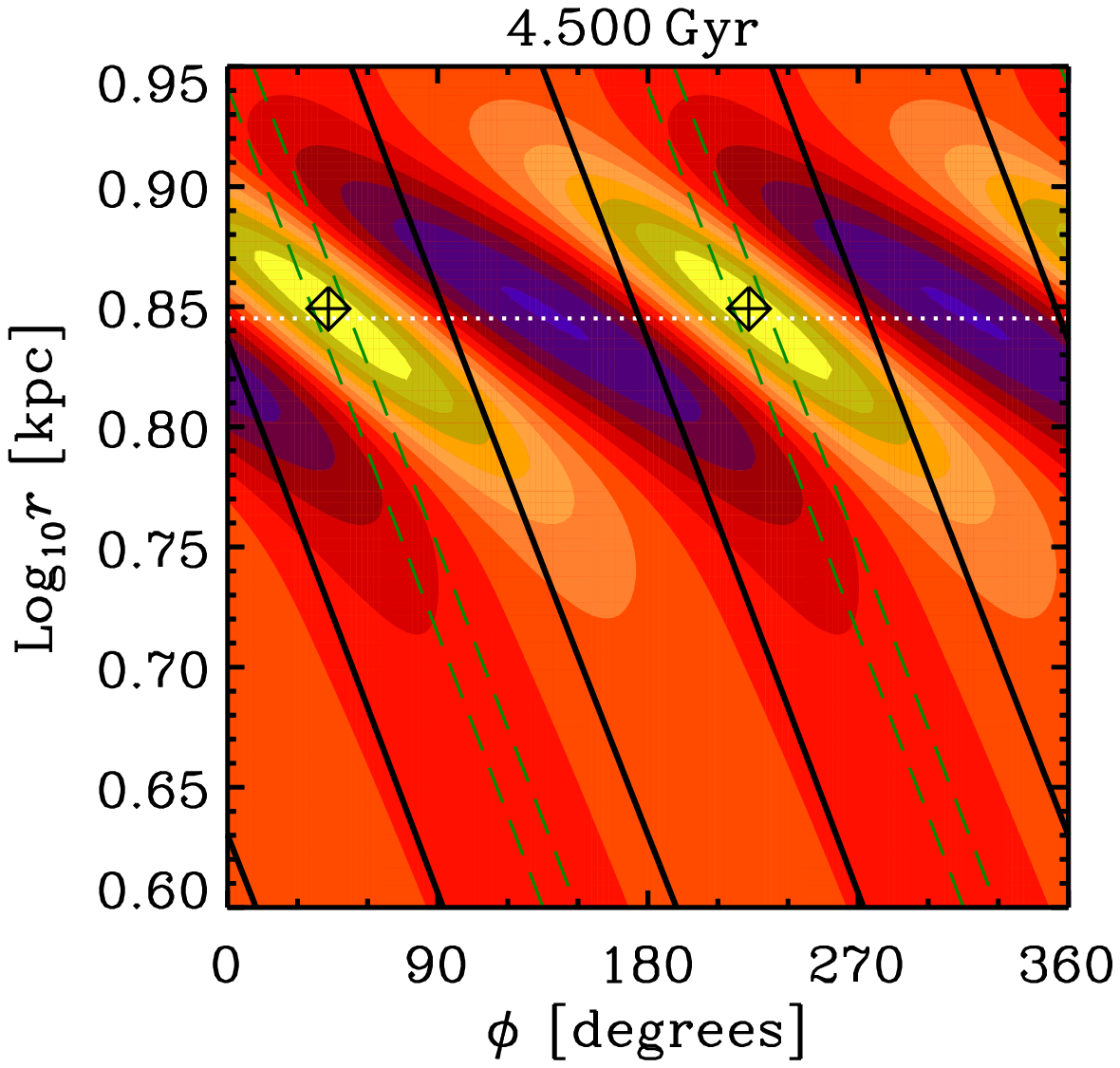}		
  \end{array}                        
  $                                  
  \caption{Left column: the same as the upper panels of the rightmost column of Figure~\ref{fig:Bcompare},
           showing $\delta$ for Model~A2 (top) and A3 (bottom).
           Right column: the same but now for Model~B3 (top) and B2 (bottom). 
           \label{fig:deltacompare_n}
          }            
\end{figure}

The effect of the three-arm $\alpha\kin$-pattern, whose corotation radius is $r_{\mathrm{c},2}=7\kpc$, on the magnetic field,
is somewhat weaker than that of the two-arm $\alpha\kin$-pattern ($r_{\mathrm{c},1}=6\kpc$).
This can be seen by comparing the strength of the large-scale field within the magnetic arms for Models A2 and A3 in, e.g., 
the third column from the left of Figure~\ref{fig:Bcompare}, 
or by noting that the two-arm ($m=2$) symmetry of the magnetic field dominates over the three-arm ($m=3$) symmetry for Model~A23.
As it turns out, $\delta$ also attains larger values for Model~A2 than for Model~A3.
This is mainly due to the number of arms being smaller in the former case, 
rather than to the location of the corotation radius.
In going from Model~A2 (A3) to B3 (B2),
only the multiplicity $n$ has been altered, resulting in a weaker (stronger) $\delta$ profile. 
This is shown in Figure~\ref{fig:deltacompare_n}, 
where the $\delta$ profile for Model~A2 in the top left panel can be compared with that of B3 in the top right,
and that of A3 in the bottom left can be compared with that of B2 in the bottom right.
Therefore, the amplitude of $\delta$ is not greatly affected by the location of the corotation radius,
but it is inversely related to the pattern multiplicity $n$.
This result was not previously appreciated in studies of forcing by non-axisymmetric $\alpha\kin$.

\subsection{Strength and spatial extent of magnetic arms}
\label{sec:morphology}
Magnetic arms generated in Model~A23 can be stronger, as well as more radially and azimuthally extended 
than those generated in Models~A2 or A3.
However, arms that are weaker and less extended than those of Models~A2 or A3 can co-exist with these more extended arms.
These attributes can be seen most clearly by comparing profiles of $\delta$ for Models~A2 (top right) and A3 (middle right) of Figure~\ref{fig:Bcompare}, 
with those of panel A23 (bottom right).
Peaks of $\delta$ can be significantly larger when more than one spiral is invoked (about $1.4$ times as large in Model~A23 compared to Model~A2).
Moreover, the contours corresponding to the regions of high $\delta$, diagonally oriented yellow/orange `fingers' in the panels, 
can clearly be more elongated for Model~A23 than for the models with only a single pattern.
In Models~A2 and A3, magnetic arms are centred and have maximum amplitude near the corotation radius,
whilst in Model~A23, magnetic arms can extend from inside the inner corotation to outside the outer corotation,
and be peaked either in between the two corotation circles, or inside the inner corotation circle.
The extent of a magnetic arm may be defined as a contiguous region for which $\delta>0.1$. 
Then the extents of the magnetic arms in Models~A2 are approximately $(\Delta r$, $\Delta\phi)=(2.7\kpc$, $210^\circ)$,
while those of A3 are approximately $(2.2\kpc$, $140^\circ)$.
On the other hand, the extents of the magnetic arms in Model~A23 at $t=4.5\Gyr$
are approximately ($3.5\kpc$, $230^\circ$), ($3.5\kpc$, $290^\circ$), ($1.3\kpc$, $90^\circ$) 
(for arms I, II and III in the order that they intersect the $n=3$ corotation radius when going from small to large $\phi$).
The extents for magnetic arms for this model at other times are generally similar to these.
It is therefore seen that the two main magnetic arms in Model~A23 
have significantly larger radial and azimuthal extents than magnetic arms from Models~A2 and A3,
while the third, less prominent arm (or armlet), 
has smaller radial and azimuthal extent than the other two arms or the arms of Models~A2 and A3.
At the same time, the strengths of the arms in Model~A23 are quite drastically different,
with arm I having a peak strength [$\max(\delta)=0.83$] of about $1.7$ times that of arm II $[\max(\delta)=0.49$] 
and $3.2$ times that of arm III [$\max(\delta)=0.26$].
For comparison, $\max(\delta)=0.60$ for Model~A2 and $\max(\delta)=0.38$ for Model~A3.
Therefore, arm I has greater strength than arms of both the two- and three-arm models, 
arm II has comparable, though slightly smaller strength than arms of Model~A2, 
while arm III has even smaller strength than the arms of Model~A3.
It should be emphasized, however, that the cutoff at $\delta=0.1$ is arbitrary, and a higher cutoff would imply more magnetic arms.
This is because while magnetic arms may be relatively thick and uniform in $\delta$ (e.g. arm I), 
they may also be thin and non-uniform,
with an inner maximum in $\delta$, followed by a clear minimum and another maximum as one moves out along the arm (e.g. arm II).

It is clear from Figure~\ref{fig:Bcompare} that the $\delta$-pattern in Model~A23 
can loosely be thought of as a superposition of those from Models A2 and A3.
Where these two patterns overlap, there is effectively constructive and destructive interference.
A magnetic arm `segment' from the two-arm pattern can become `merged' with one from the three-arm pattern,
with the region between the two being `filled in' due to effective constructive interference,
resulting in a more extended magnetic arm.
The remarkably high accuracy of the superposition approximation can be understood, qualitatively, 
as due to the superposition of the effective potentials, when the dynamo model is mapped to an eigenvalue problem
\citepalias{Chamandy+13b}.
This approximation works less well as the radial separation between the two patterns is decreased.

The effect of changing the radial separation between the two spiral patterns has also been explored.
In Model~L23 (`large separation'), the two sets of magnetic segments (i.e. the two-arm and three-arm sets) are more disjoint,
i.e. undergo less effective interference.
The magnetic field evolves more rapidly with time, 
as the beat frequency between the two patterns is greater due to their increased separation.
In Model~S23 (`small separation'), magnetic arms are less elongated than those in Model~A23,
and the field configuration evolves less rapidly than in Model~A23. 
It should be noted, however, that for real galaxies, the rotation curve may vary more rapidly with radius than the Brandt curve used here,
and therefore, a small separation in corotation radii need not imply a small separation in pattern speed.

\subsection{Effect of a finite dynamo relaxation time $\tau$}
\label{sec:tau}
\begin{figure}                     
  $                                 
  \begin{array}{c c}              
    \includegraphics[width=42mm]{IS_020_deltacorotate_nt180.eps}		
    \includegraphics[width=42mm]{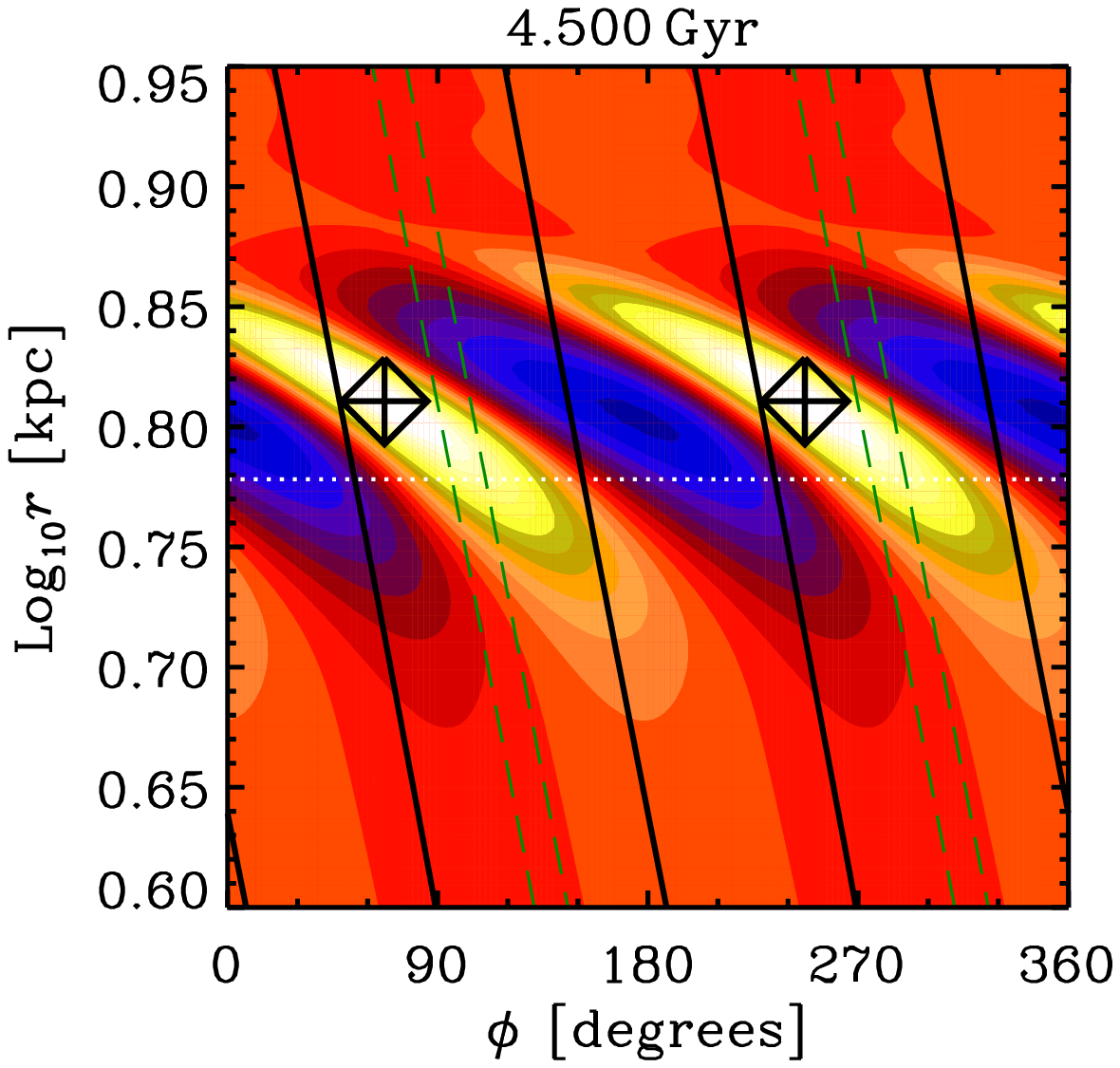}\\		
    \includegraphics[width=42mm]{IS_021_deltacorotate_nt180.eps}		
    \includegraphics[width=42mm]{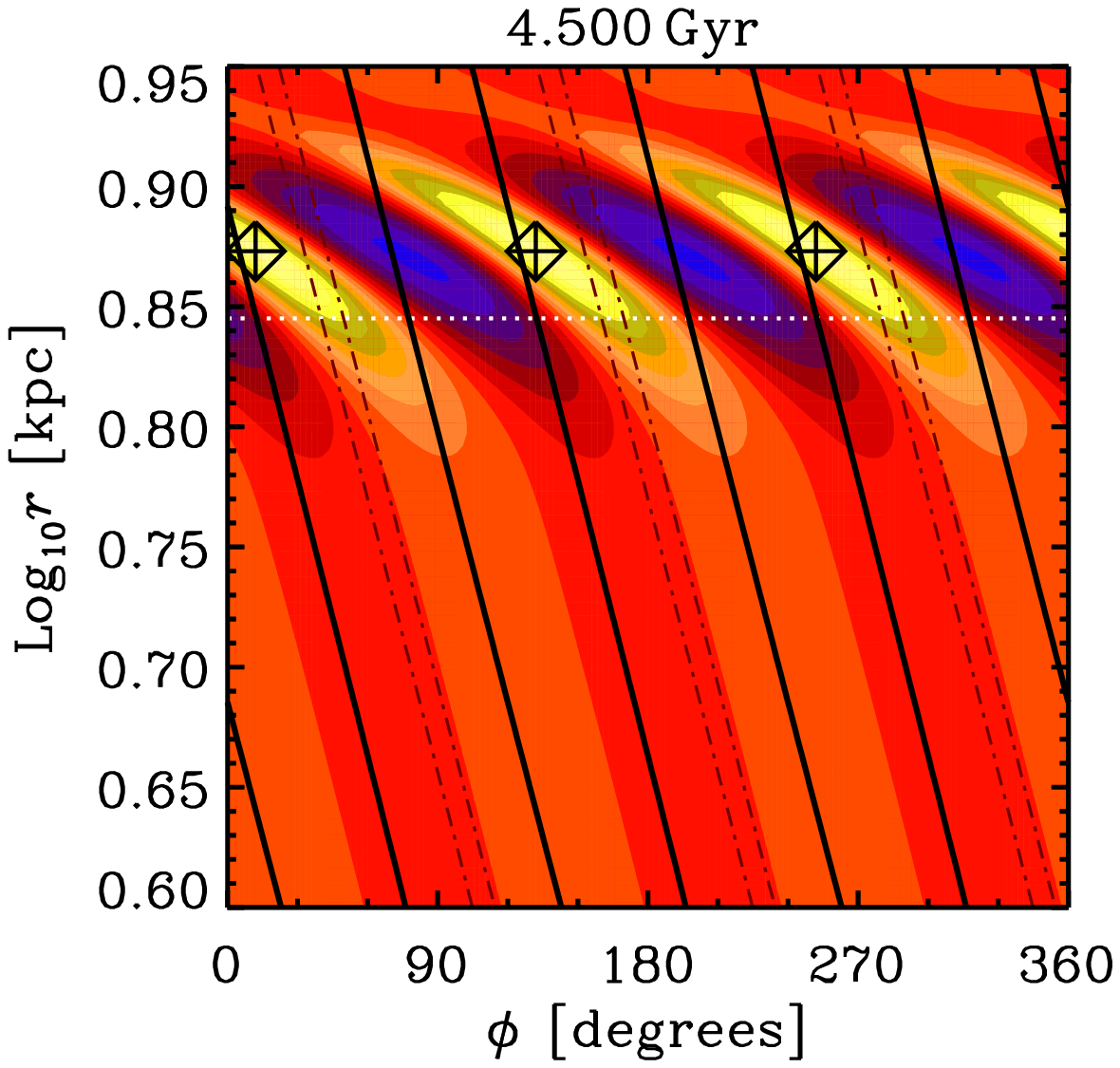}\\		
    \includegraphics[width=42mm]{IS_018_deltacorotate_nt180.eps}		
    \includegraphics[width=42mm]{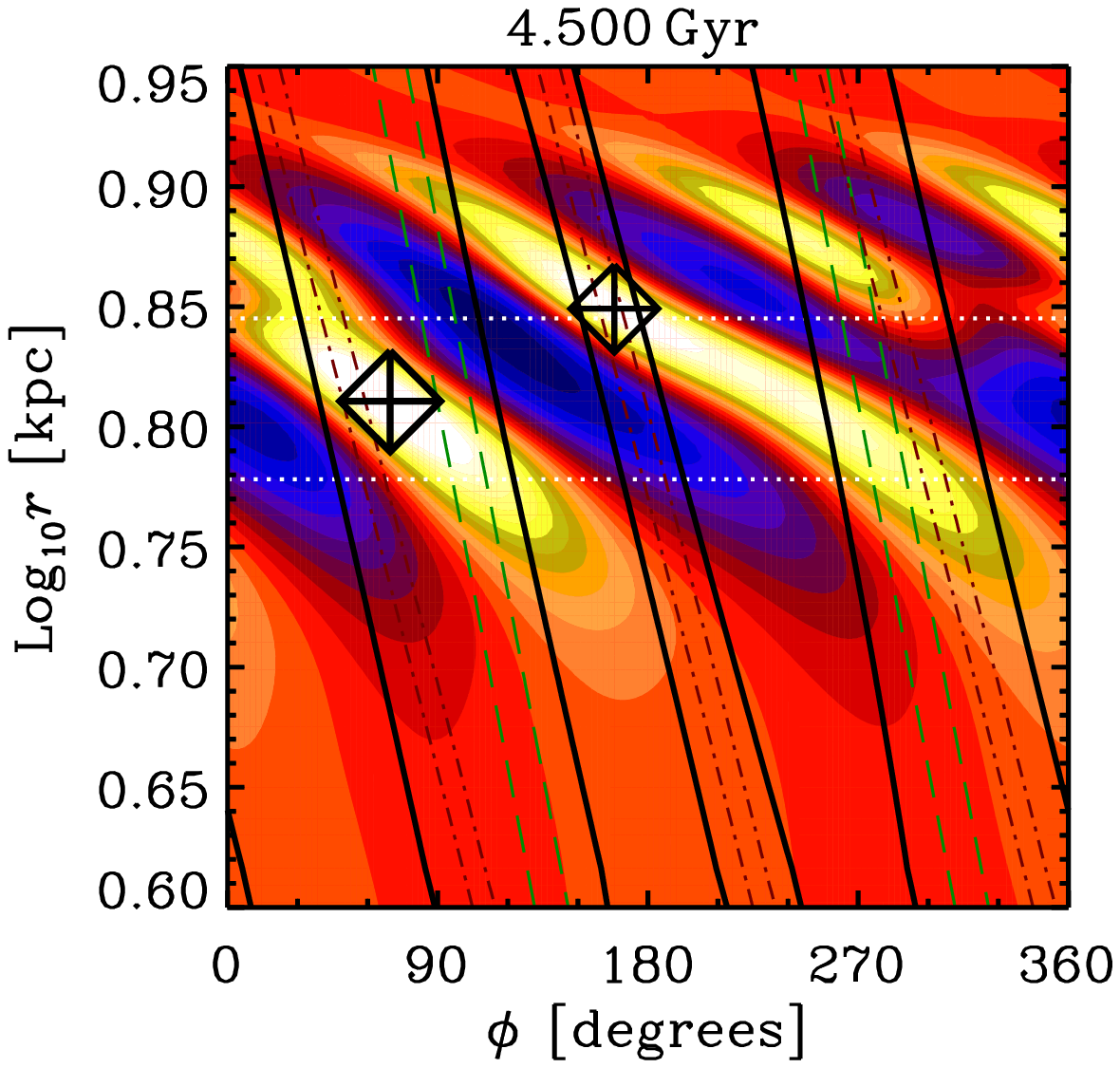}		
  \end{array}                        
  $                                  
  \caption{Comparison of $\tau=0$ and $\tau=l/u$ cases.
           Left column: the same as the rightmost column of Figure~\ref{fig:Bcompare} (Models A2, A3 and A23).
           Right column: shows the same plots, but for Models A2$\tau$, A3$\tau$ and A23$\tau$.
           For $\tau=l/u$, the colour table has been clipped at $\delta=0.9$, though $\delta$ exceeds this value in some regions.
           \label{fig:deltacompare_tau}
           }            
\end{figure}

As discussed in \citetalias{Chamandy+13a,Chamandy+13b}, the effects of a finite relaxation time $\tau$
are to strengthen the magnetic arms and cause them to be more tightly wound, 
as well as to shift their peaks outward in radius, 
and backward in azimuth (relative to the sense of the galactic rotation).
These effects are evident in Figure~\ref{fig:deltacompare_tau}, 
where a copy of the rightmost column of Figure~\ref{fig:Bcompare} ($\tau=0$) is shown on the left,
and the corresponding plots for the $\tau=l/u$ case are shown on the right.
The bottom row illustrates $\delta$ for $t=4.5\Gyr$ in Model~A23 on the left and for Model~A23$\tau$ on the right.
Magnetic arms are significantly stronger in Model~A23$\tau$ than in Model~A23.
Magnetic arms produced for $\tau=0$ are found to be close in shape to logarithmic spirals in the single pattern case,
as they appear as straight lines in the plots for Models~A2 and A3,
while those produced for $\tau=l/u$ in Models~A2$\tau$ and A3$\tau$ show obvious deviations from logarithmic spirals.
Qualitatively, the effects, discussed throughout this work, that are caused by the presence of two $\alpha\kin$ patterns rather than one,
occur in both $\tau=0$ and $\tau=l/u$ cases.
In order to separate the effects of multiple interfering spiral patterns from those of a finite $\tau$,
the focus is placed on the $\tau=0$ case in what follows.

\begin{figure*}                     
  $                                 
  \begin{array}{l l l l}              
    \includegraphics[width=42mm]{IS_018_deltacorotate_nt180.eps}
    \includegraphics[width=42mm]{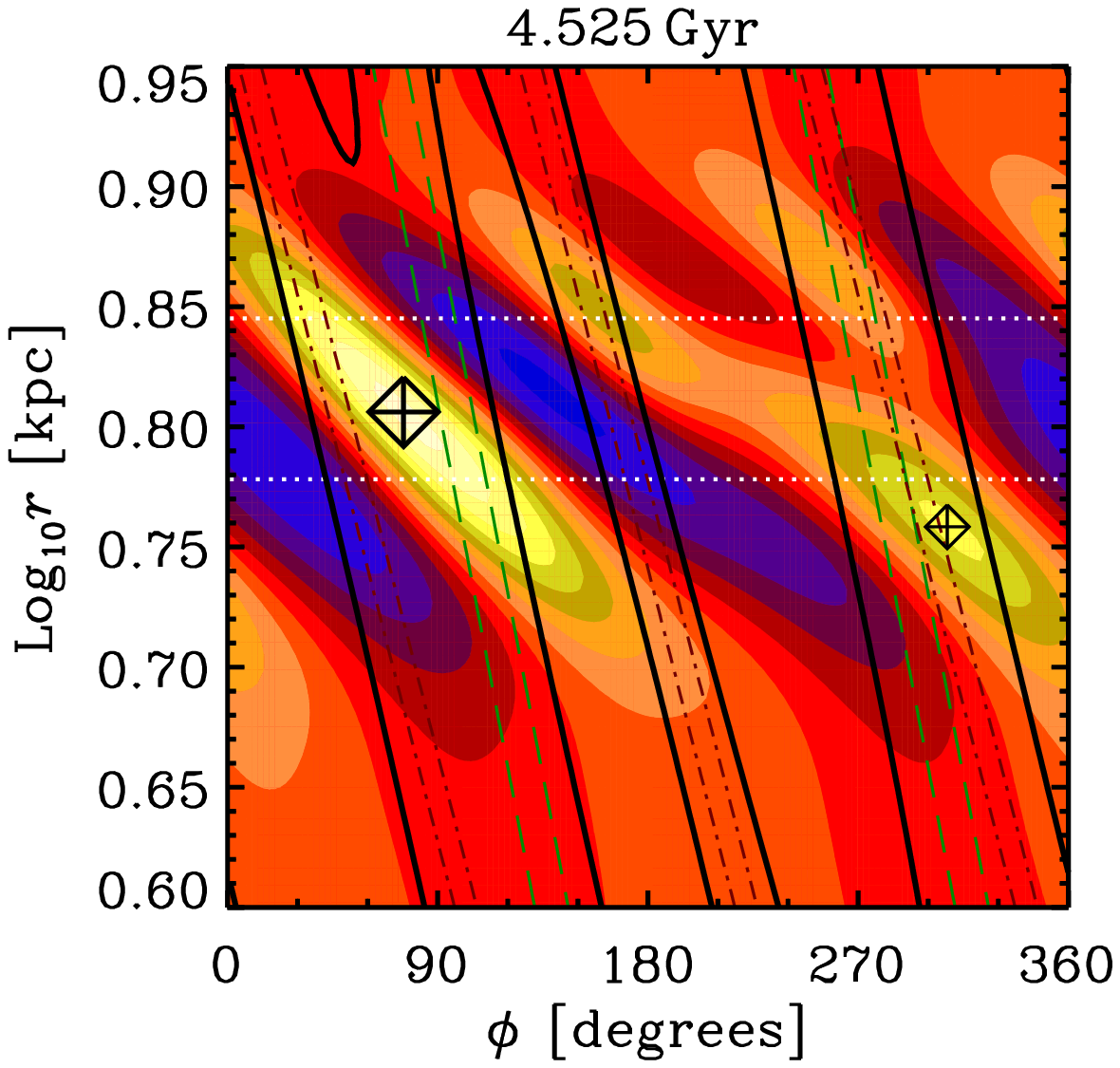}
    \includegraphics[width=42mm]{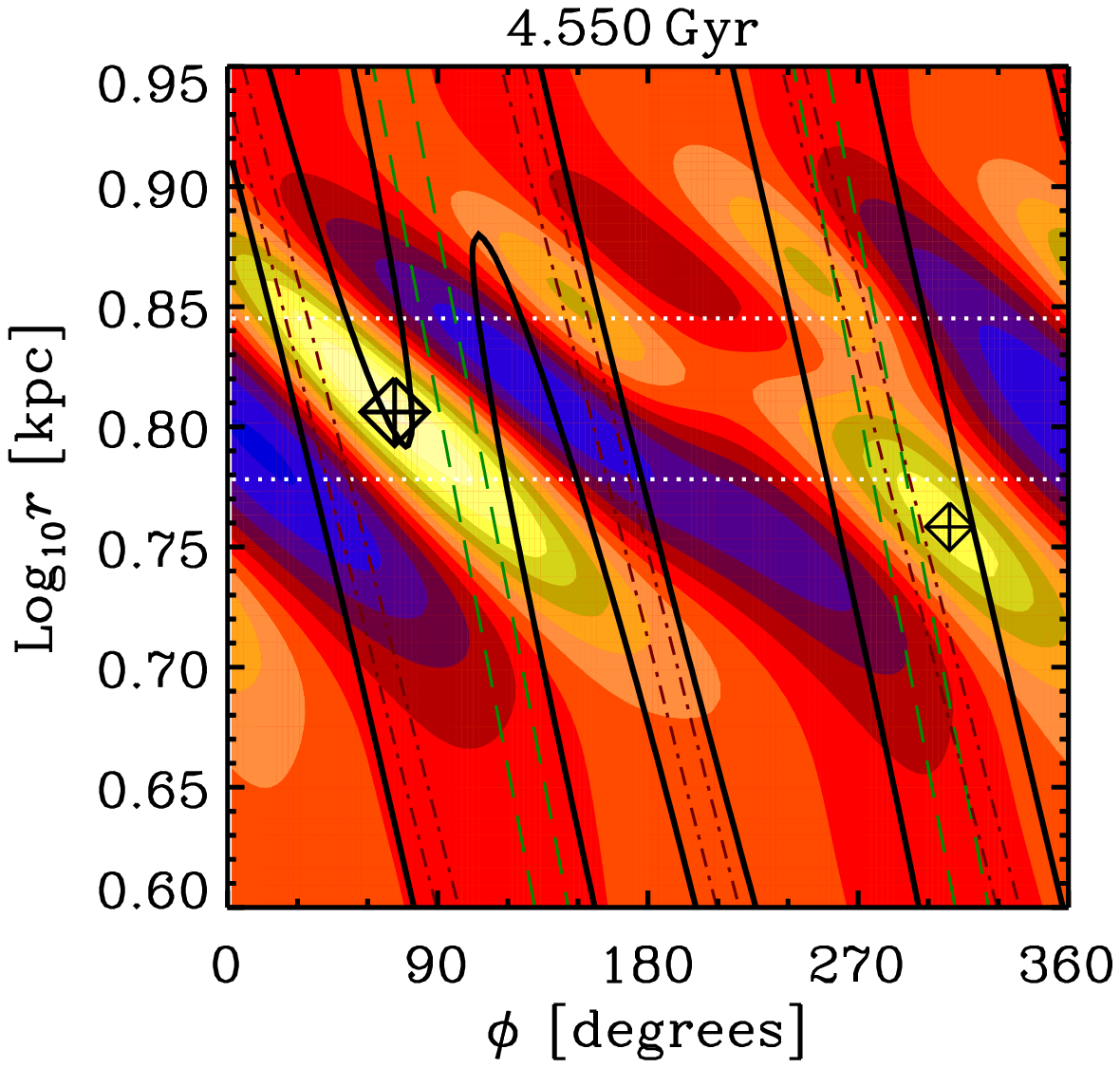}
    \includegraphics[width=42mm]{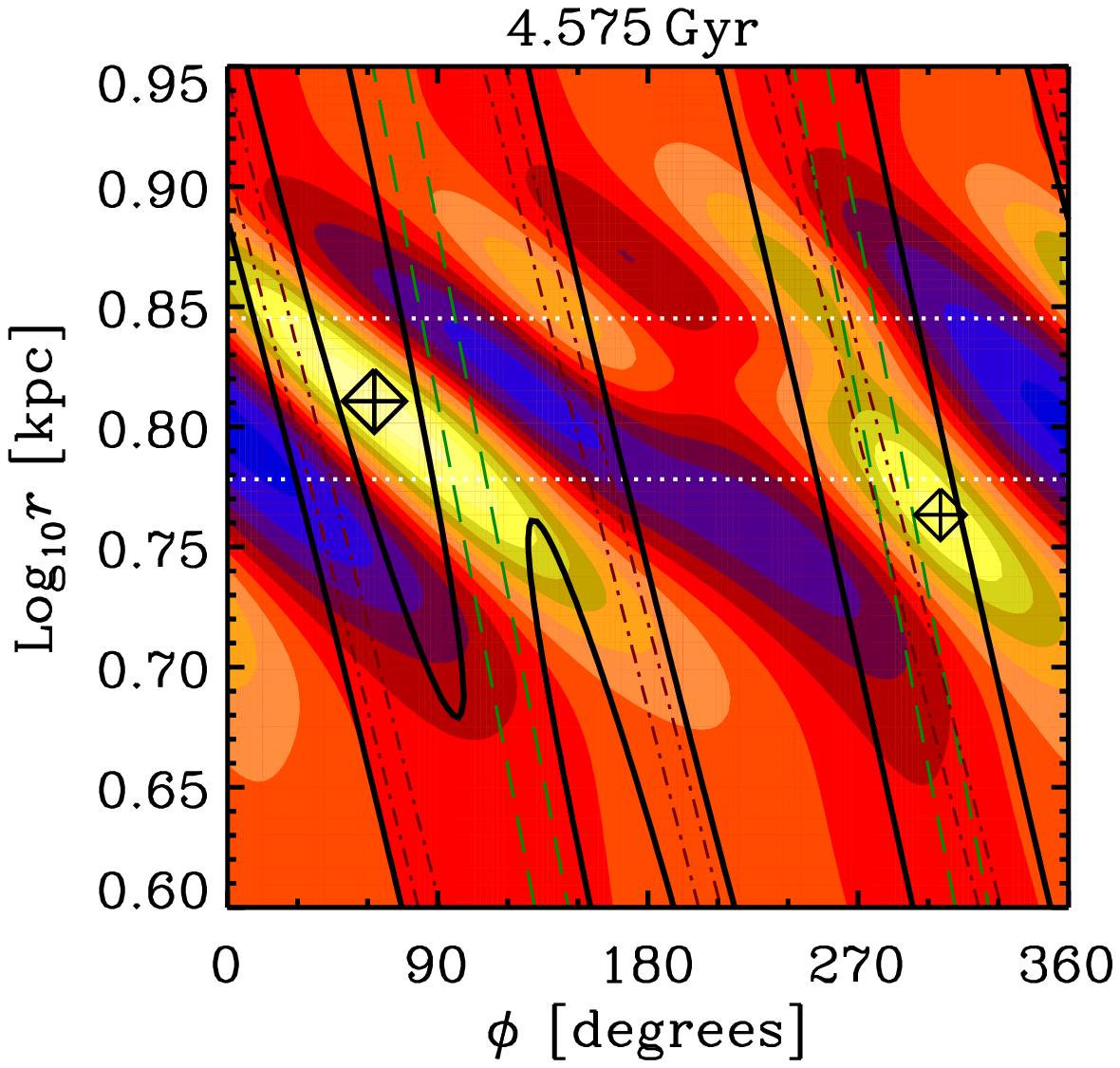}\\
    \includegraphics[width=42mm]{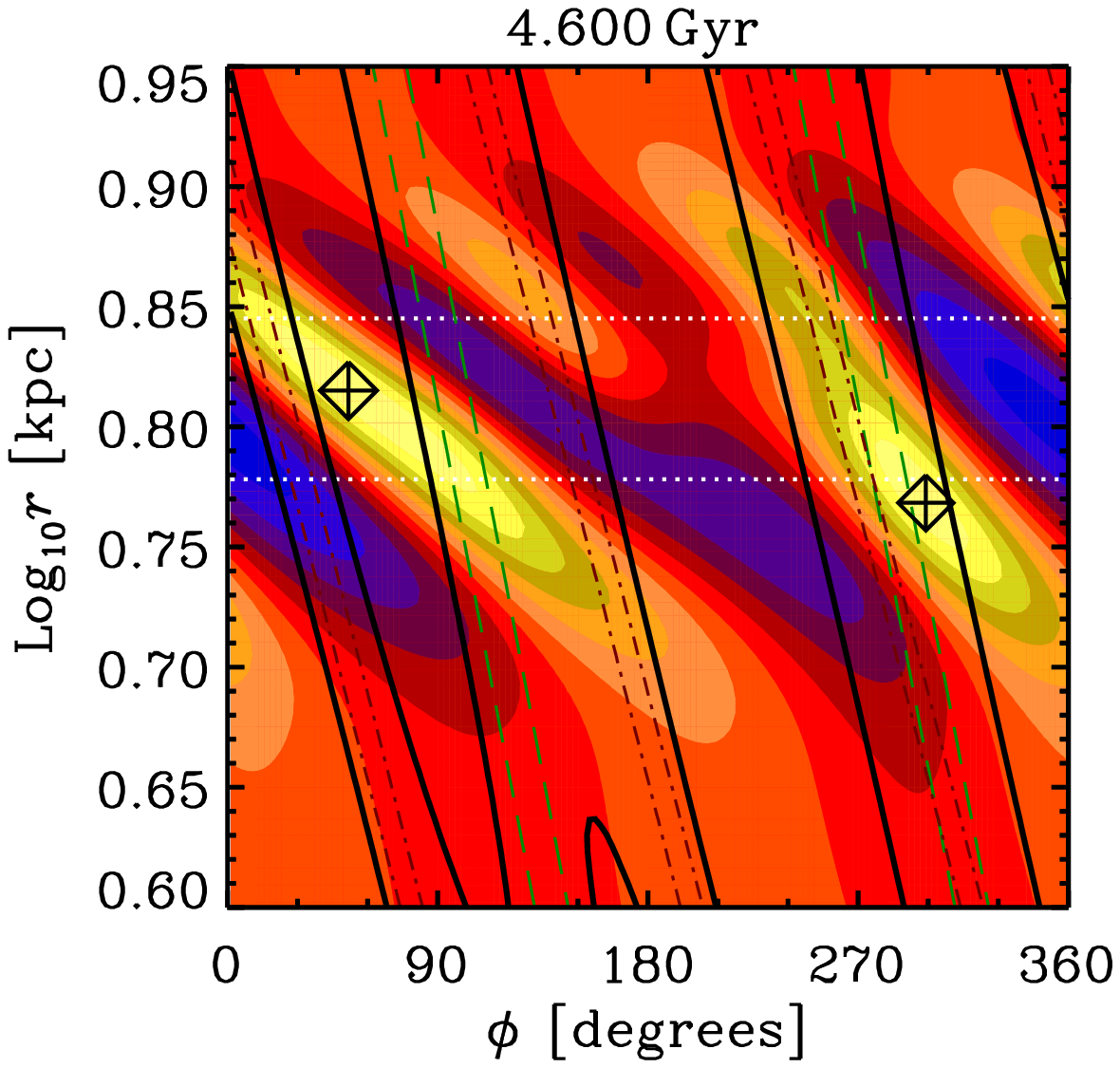}
    \includegraphics[width=42mm]{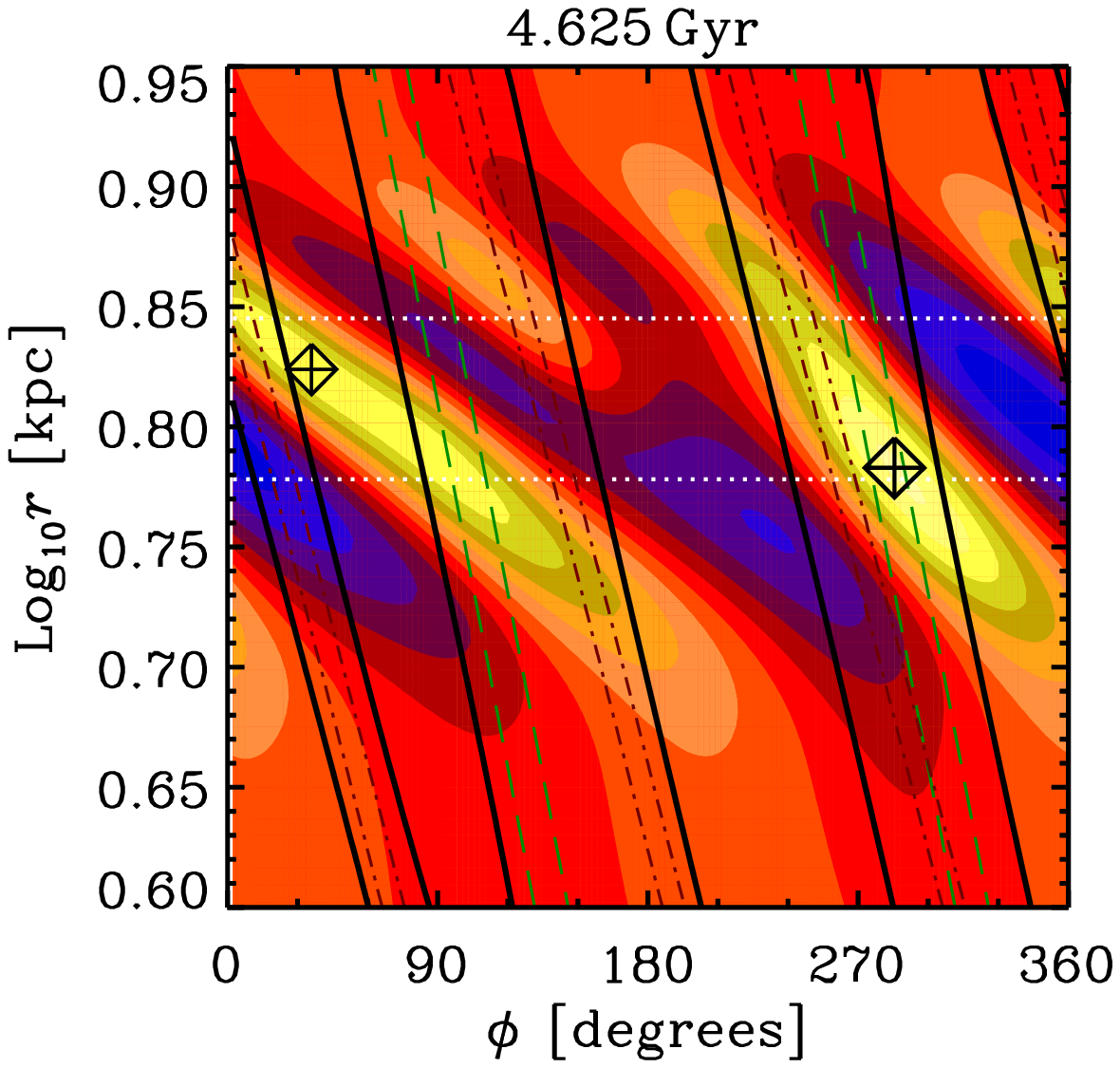}
    \includegraphics[width=42mm]{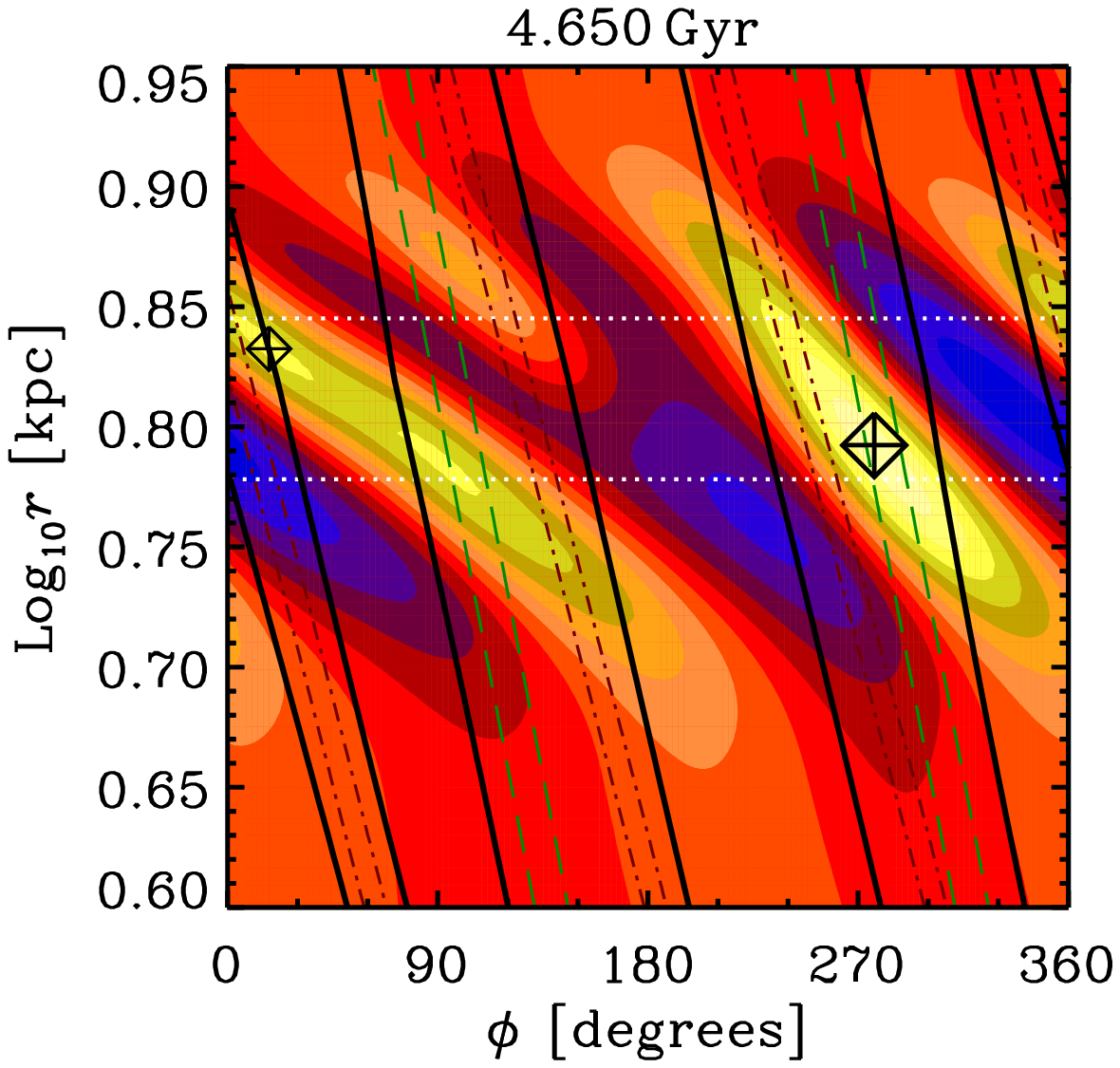}
    \includegraphics[width=42mm]{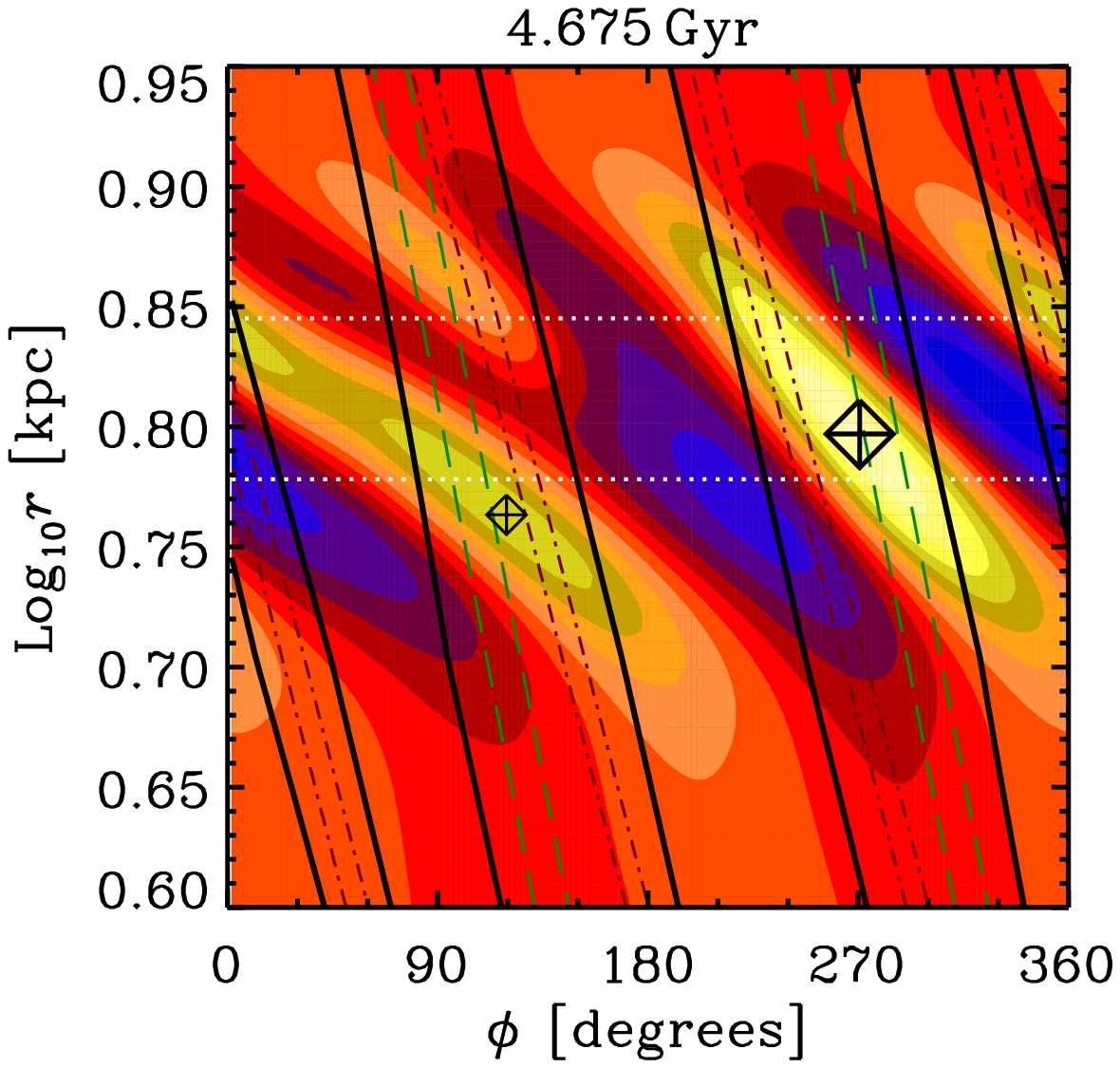}
  \end{array}                       
  $
  \caption{Time sequence showing the ratio $\delta$ of the non-axisymmetric to axisymmetric part of $\mbp$ for Model A23,
           with times shown on panels.
           Colours, contours, corotation radii, and symbols are as in Figure~\ref{fig:Bcompare}.
           \label{fig:delta}}
\end{figure*}                       

\begin{figure*}                     
  $                                 
  \begin{array}{l l l l}              
    \includegraphics[width=42mm]{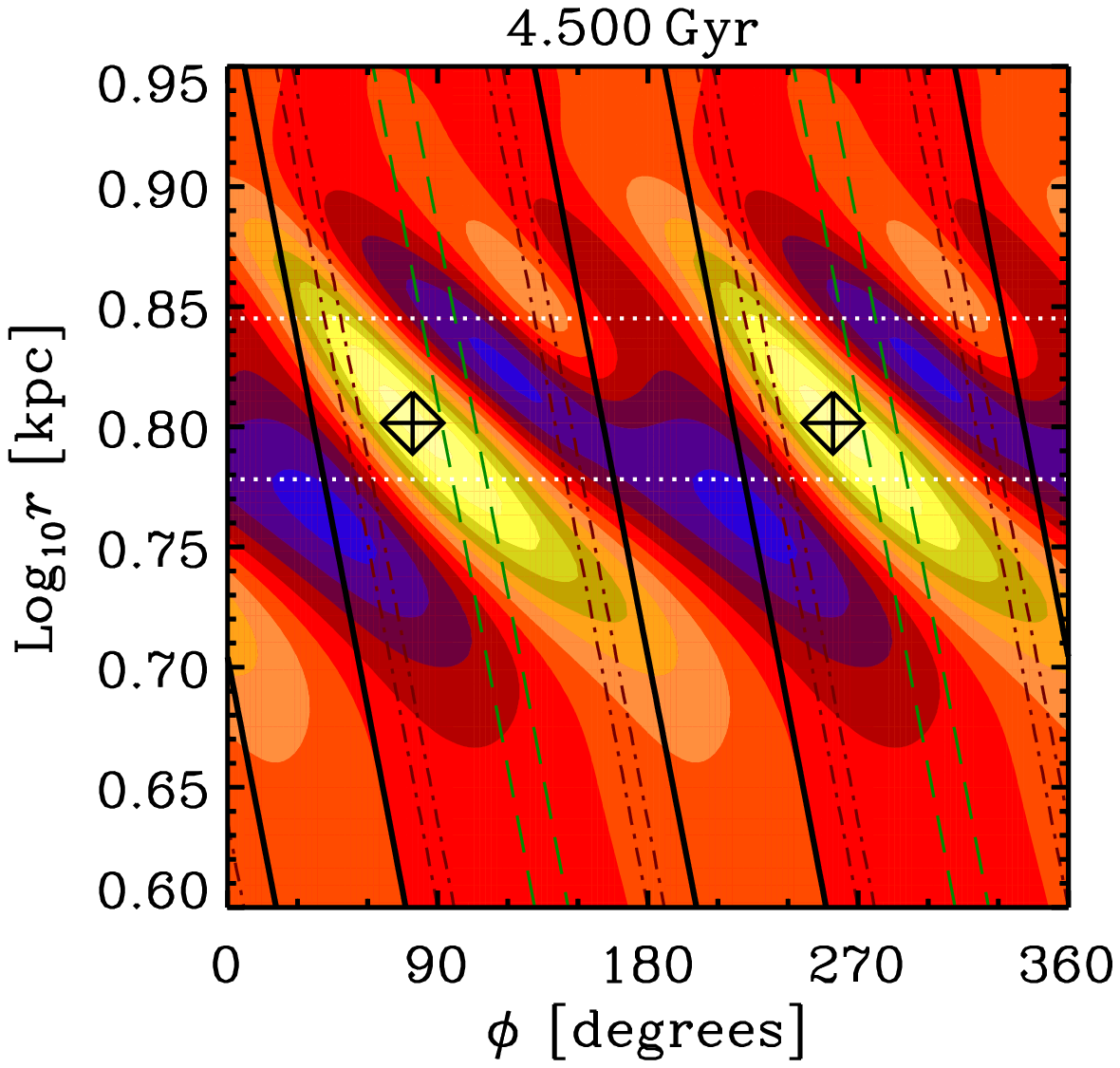}
    \includegraphics[width=42mm]{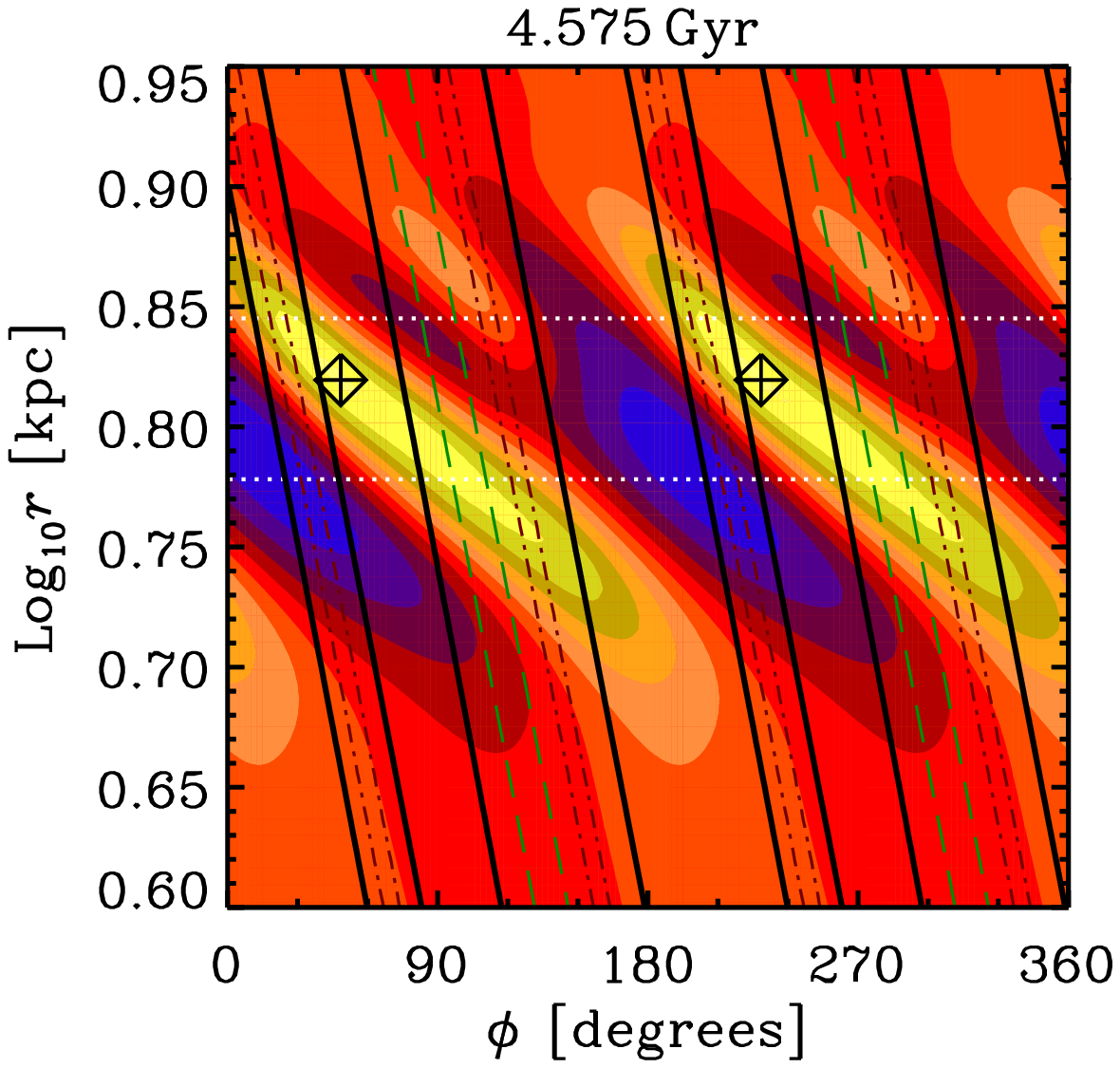}
    \includegraphics[width=42mm]{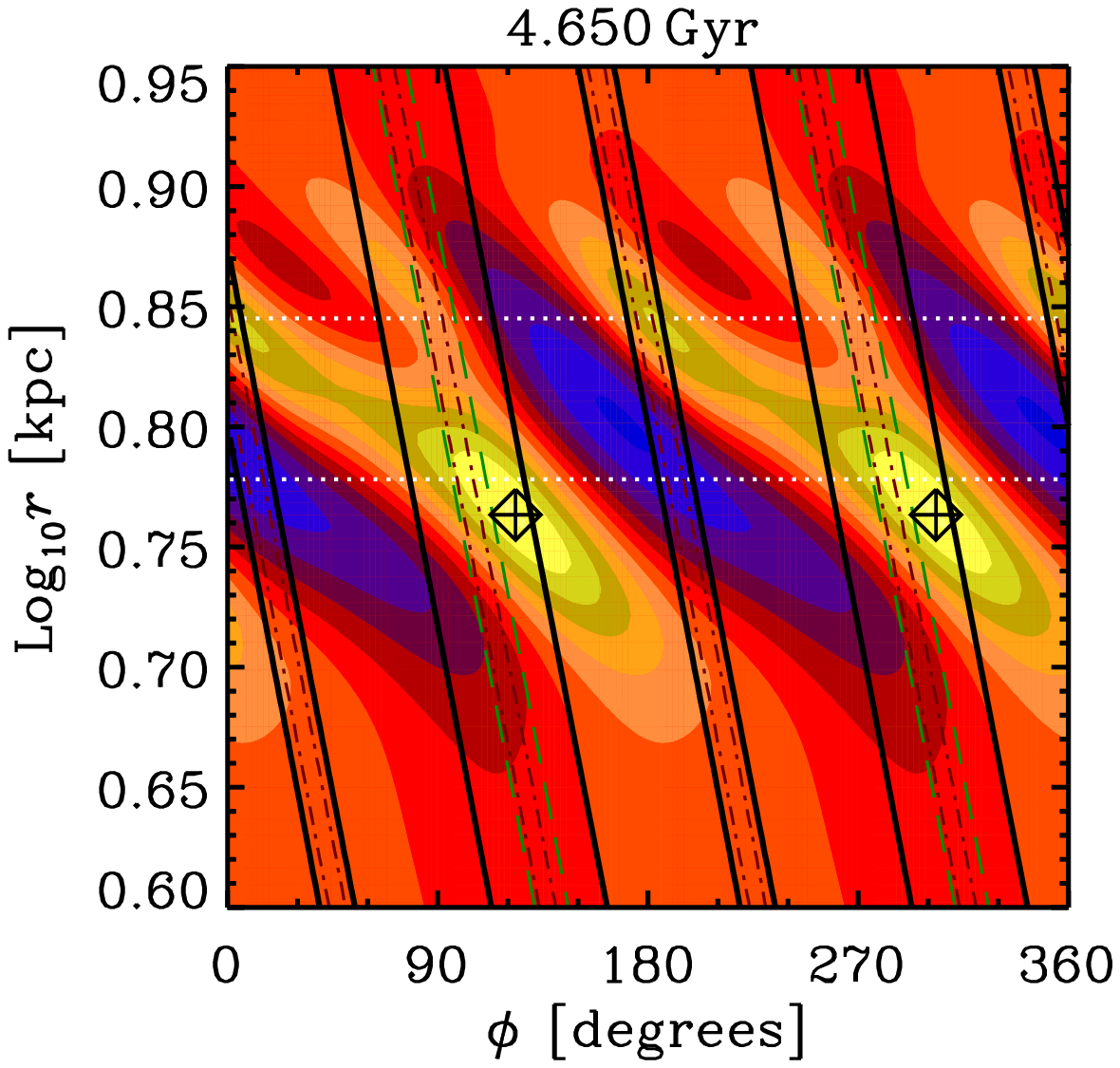}
    \includegraphics[width=42mm]{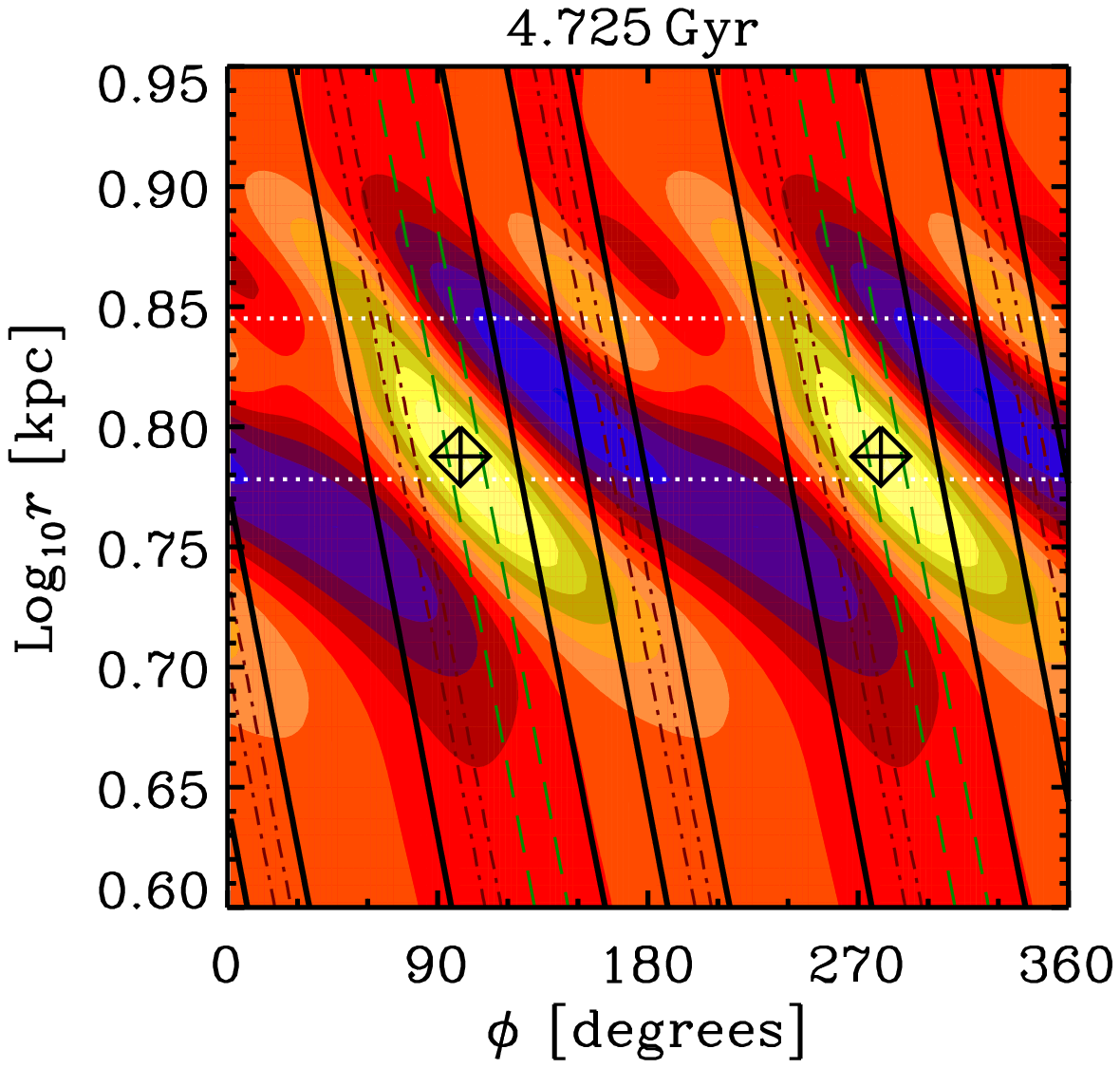}
  \end{array}                       
  $
  \caption{Same as figure \ref{fig:delta}, but now for Model~A24, and with spacing between successive snapshots increased.
           \label{fig:A24}}
\end{figure*}                       

\subsection{Evolution in time and transient morphological features}
\label{sec:evolution}
In the case of forcing by two patterns,
the morphology evolves with time along with the interference pattern of the $\alpha\kin$-spirals.
For ease of interpretation, all plots are shown in the frame corotating with the gas situated at $r=6\kpc$,
which corresponds to the frame of the two-arm pattern for, e.g., Models~A2 and A23.
In Figure~\ref{fig:delta} the evolution of the magnetic field is illustrated for Model~A23, for a total duration of $0.175\Gyr$,
which is roughly equal to half the period of the total $\alpha\kin$ pattern in this reference frame
$2\pi/[3(\Omega\1-\Omega\2)]=0.387\Gyr$.
[To save space, only half of a full period is plotted, 
but the full information is effectively recovered if, after reaching the end of the time sequence, 
one loops back from the beginning, while switching one's attention from the left (right) half of each panel,
to the right (left) half.]

Firstly, it may be noted that three magnetic arms (yellow/orange regions) are always visible, 
with one arm generally weaker than the other two, and also less extended, being confined to the outer disk.
Secondly, the spatial relations of the arms with one another evolve with time.
The third relatively weak arm usually appears to result from a bifurcation of one of the prominent arms,
but sometimes it would be better described as a disconnected armlet/filament.
As mentioned above, the sequence of plots may be loosely interpreted as effective interference patterns between two inner magnetic arm segments
(caused by the two-arm $\alpha\kin$-pattern) and three outer magnetic arm segments
(caused by the three-arm $\alpha\kin$-pattern).
When a bifurcation is present, two of the three outer magnetic arm segments are joined to the same inner segment, 
while the third outer segment is joined to the remaining inner segment
(`3a' and `3b' joined to `2a' and `3c' joined to `2b', say), e.g. at $t=4.55\Gyr$.
When a disconnected filament is instead present, two of the three outer segments are each joined to different inner segments, 
while the third outer segment is isolated from the inner segments 
(`3a' to `2a', `3b' to `2b' and `3c' isolated, say), e.g. at $t=4.65\Gyr$.

Since certain galaxies, such as IC~342, have been reported to contain an inner $n=2$ and outer $n=4$ pattern \citep{Crosthwaite+00}, 
Model~A24 is studied, which is identical to Model~A23 except that $n_2=4$ instead of $3$; 
a time sequence is shown in Figure \ref{fig:A24}.
Morphological features of magnetic arms discussed in Sect.~\ref{sec:morphology}, \ref{sec:evolution} and \ref{sec:pitch} below,
such as bifurcations, isolated armlets, and variation in winding angle and extent, are also seen for this model.
Note also that $\pi$-fold symmetry is always present in this model.

Although each individual spiral pattern has been modelled as steady and rigidly rotating,
it is known that spiral patterns may in fact be transient.
A brief discussion of the implications of spiral transience vis-\`{a}-vis the dynamo is presented in Appendix~\ref{sec:transient}.

\subsection{Pitch angles}
\label{sec:pitch}
The winding (or pitch) angle $p$ of the magnetic arms, as well as the pitch angle of the regular field $p_B$,
can be inferred from observations, and are thus important to explore in the models.
Model~A23 has variable winding (pitch) angles of the arms,
given by
\begin{equation}
  p=\cot^{-1}\left(\frac{\del\phi\ma}{\del\ln r}\right)\simeq\tan^{-1}\left[\ln(10)\frac{\Delta\log_{10}r}{\Delta\phi\ma}\right],
\end{equation}
where $\phi\ma$ is the location of the azimuthal maximum of $\delta$.
The quantity $p$ is seen to vary from arm to arm and also with time for any individual arm.
For example, whenever a bifurcation of an arm is seen, as discussed above, 
the winding angle of the outer branch is necessarily larger than that of the inner branch,
at least in the vicinity of the bifurcation.
Furthermore, the two main magnetic arms may have different winding angles from one another.
For example, this is true at the time $t=4.625\Gyr$, 
when the magnetic arm from $300^\circ$ to $360^\circ$ and $0^\circ$ to $210^\circ$ has a smaller winding angle, (about 6$^\circ$)
than the magnetic arm from $180^\circ$ to $360^\circ$ and $0^\circ$ to $30^\circ$ (about 9$^\circ$).
This can be thought of as simply a consequence of the variable azimuthal spacing between adjoining two- and three-arm segments 
that effectively combine to form a magnetic arm.
For smaller azimuthal separation, $p$ is larger and the arm thicker, 
while for larger azimuthal separation, $p$ is smaller and the arm thinner
because as the two- and three-arm segments separate, they effectively stretch out the magnetic arm between them.
The winding angles seen in Model~A23 are not very different from the (constant) winding angle
found for magnetic arms in Model~A2 ($p\simeq7^\circ$) or A3 ($p\simeq8^\circ$).
In all cases, $p$ is much smaller than the analogous quantity $p_\alpha$ for the $\alpha$ arms,
where $p_\alpha\simeq34^\circ$ for the two-arm $\alpha\kin$-spiral and $p_\alpha\simeq27^\circ$ for the three-arm $\alpha\kin$-spiral,
as also discussed in \citetalias{Chamandy+13a} for the single-pattern case.
Nevertheless, it is interesting that two interfering patterns can produce magnetic arms
with different winding angles and with winding angles that evolve with time.

The pitch angle of the magnetic field $p_B<0$ also shows small differences between the models, 
most notably in its extrema within the radial region of interest.
In Model~A23, $|p_B|$ has extrema of approximately ($3^\circ$, $13^\circ$), 
whereas for Models~A2 and A3 the extrema are approximately ($5^\circ$, $12^\circ$) and ($6^\circ$, $11^\circ$), respectively.
However, the mean value of the pitch angle of $-8^\circ$ is virtually the same for all three models.

\begin{figure*}                     
  $                                 
  \begin{array}{l l l l}              
    \includegraphics[width=42mm]{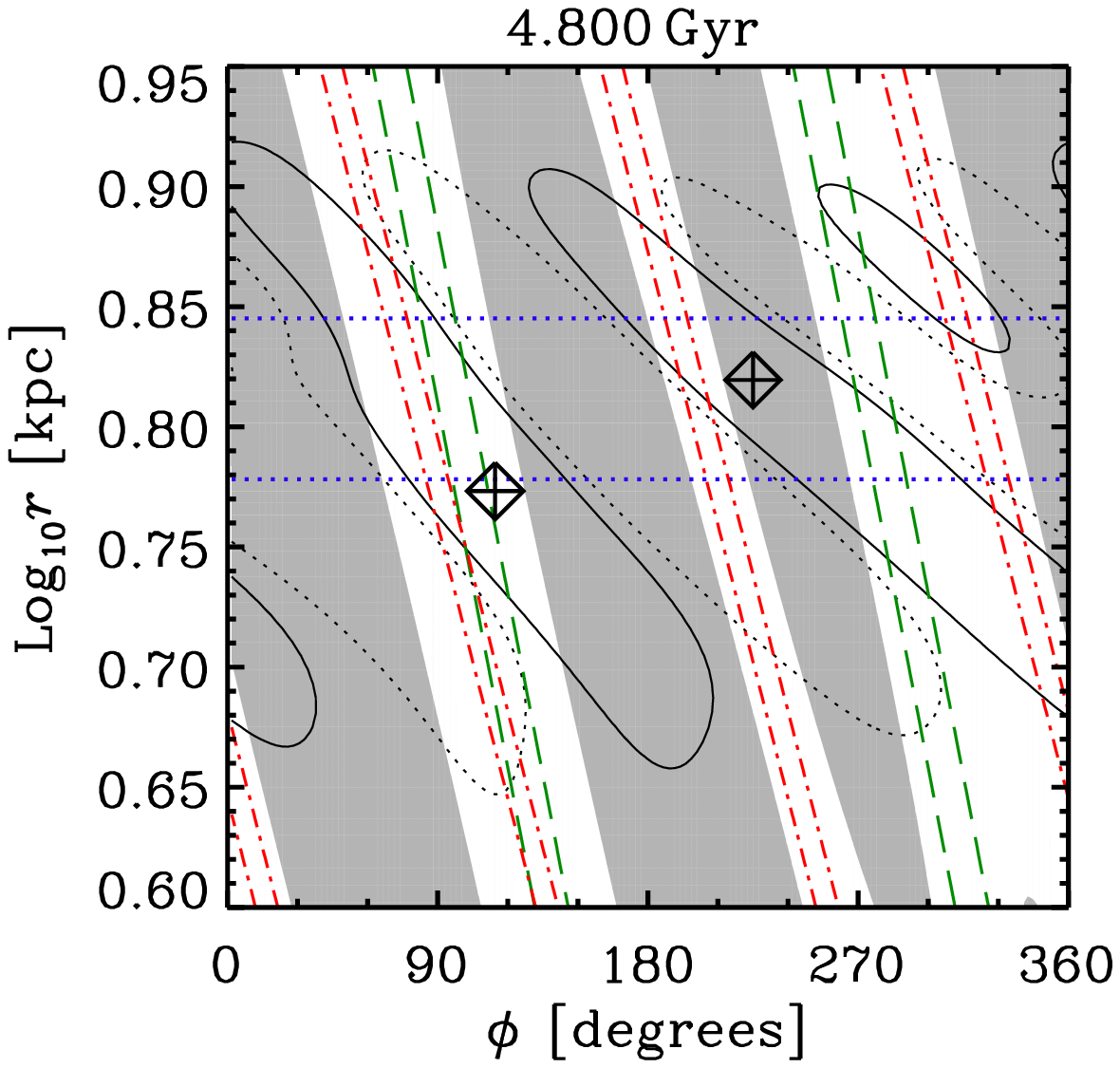}
    \includegraphics[width=42mm]{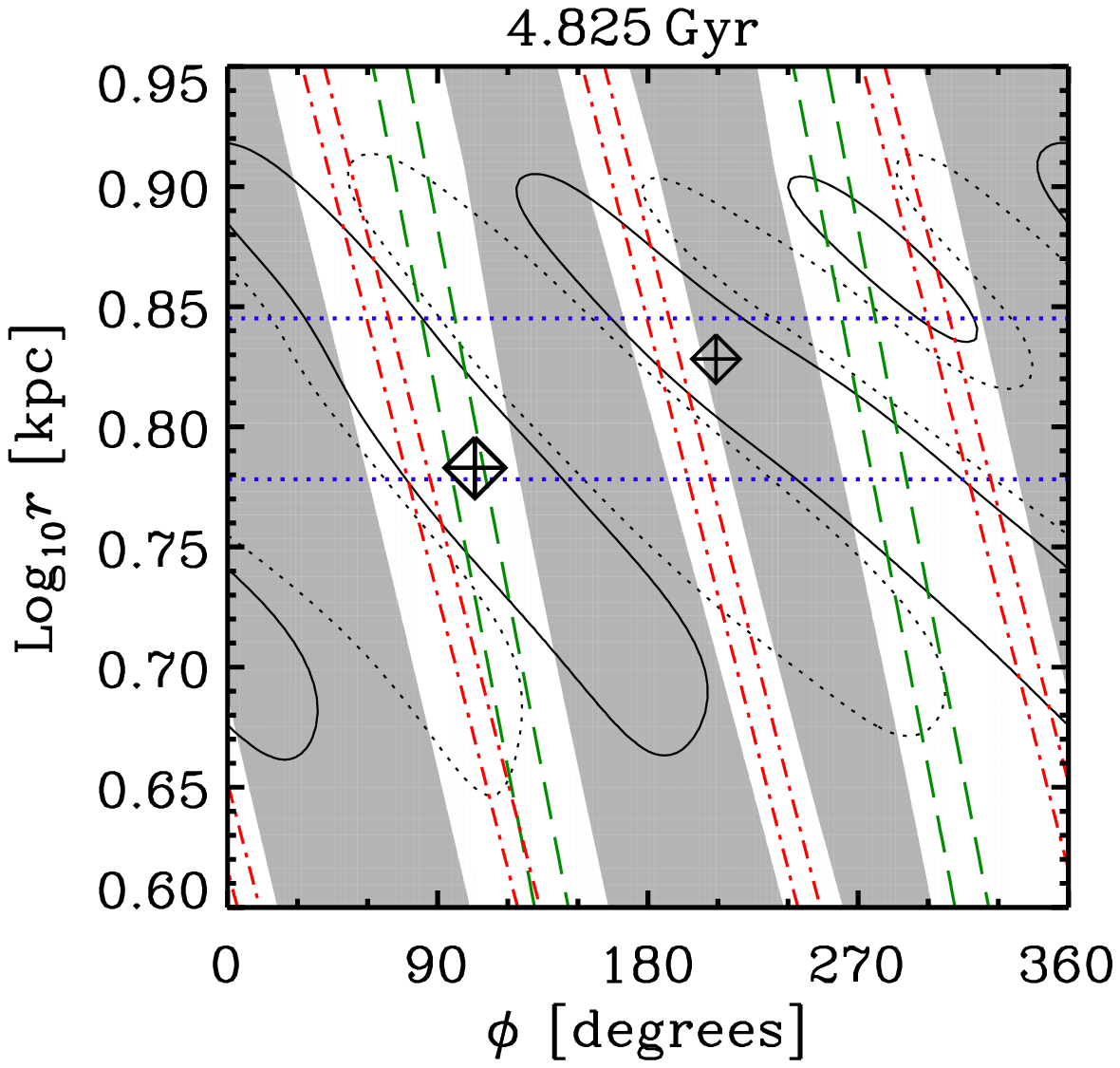}
    \includegraphics[width=42mm]{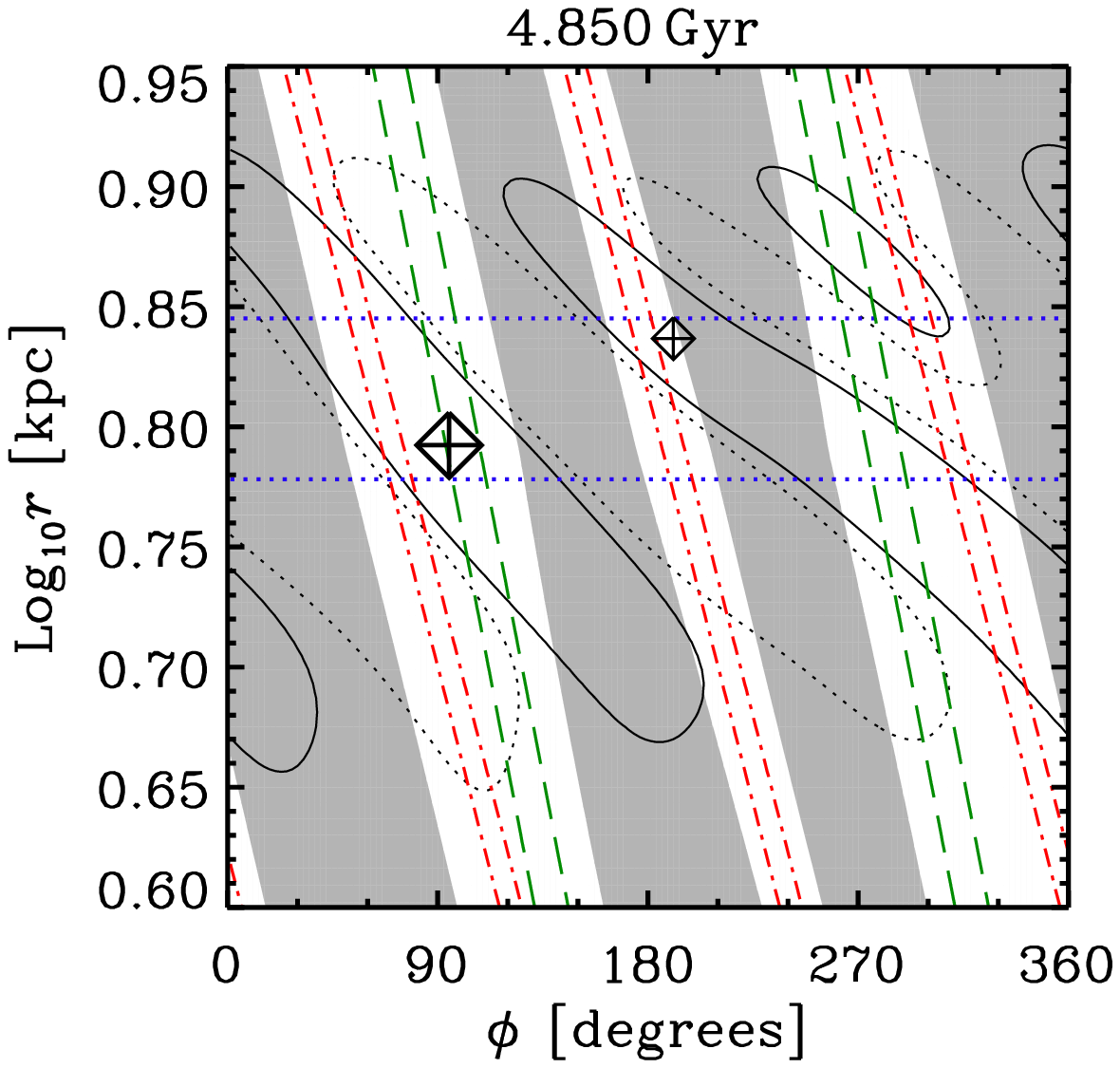}
    \includegraphics[width=42mm]{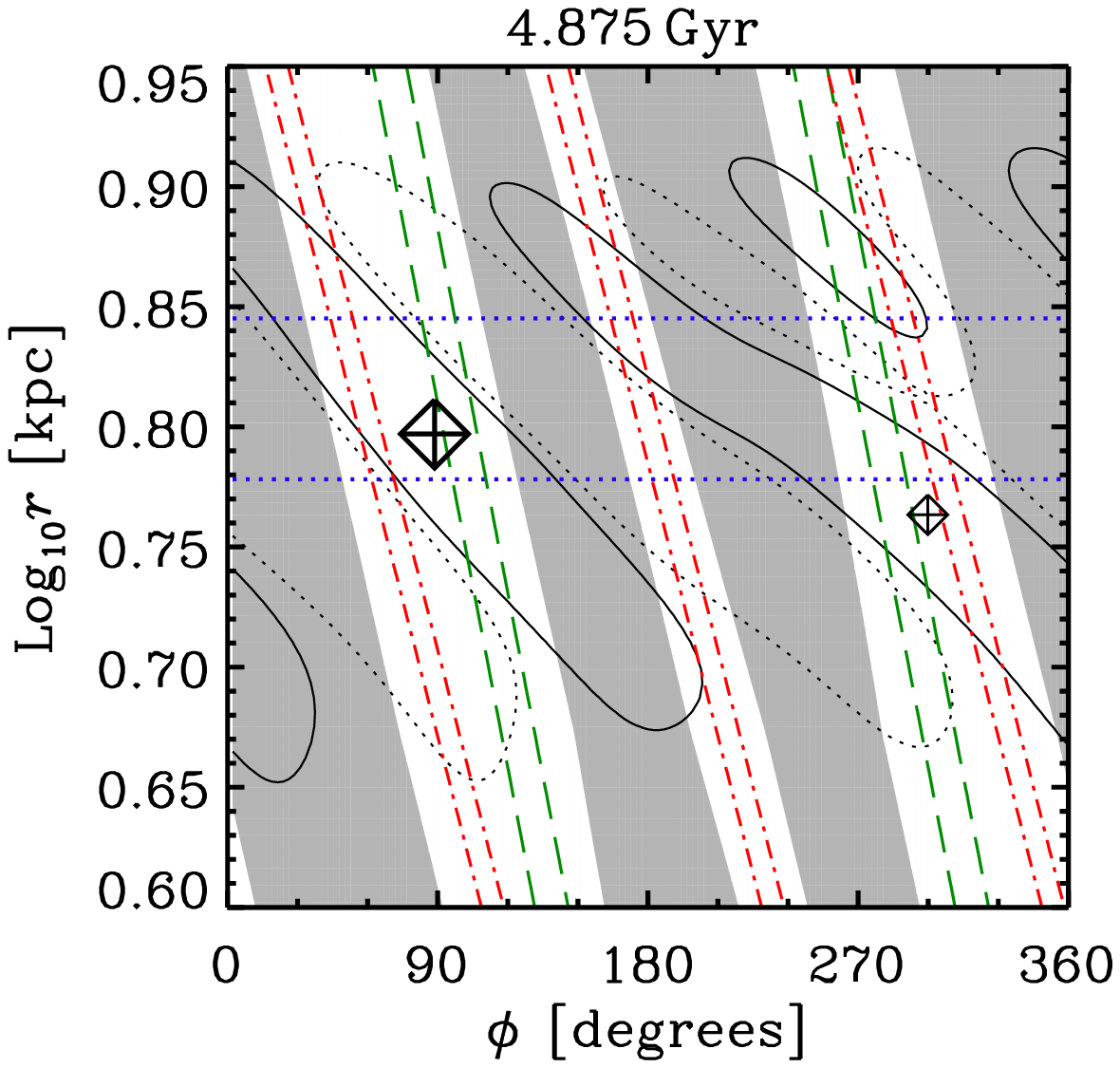}\\
    \includegraphics[width=42mm]{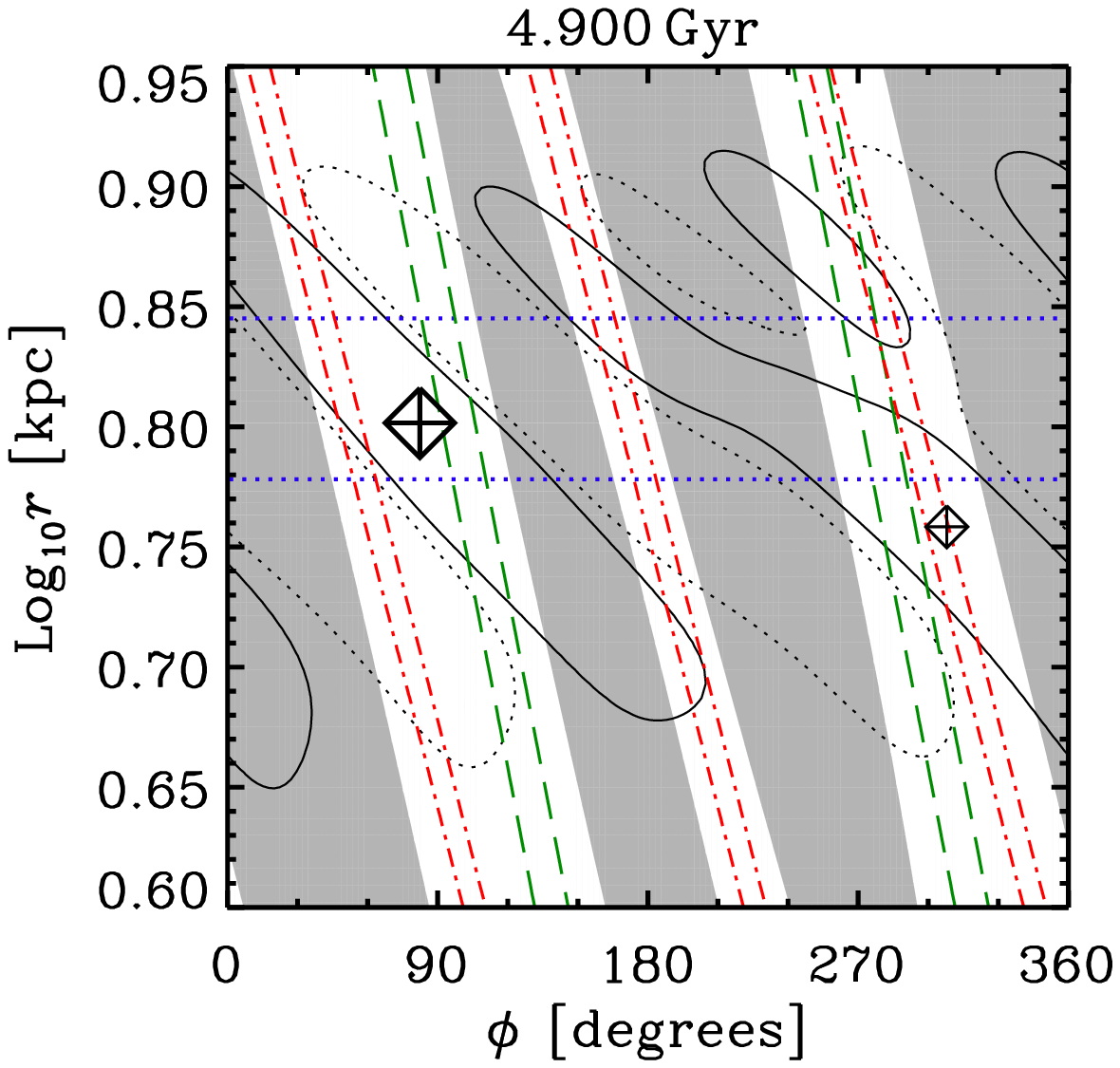}
    \includegraphics[width=42mm]{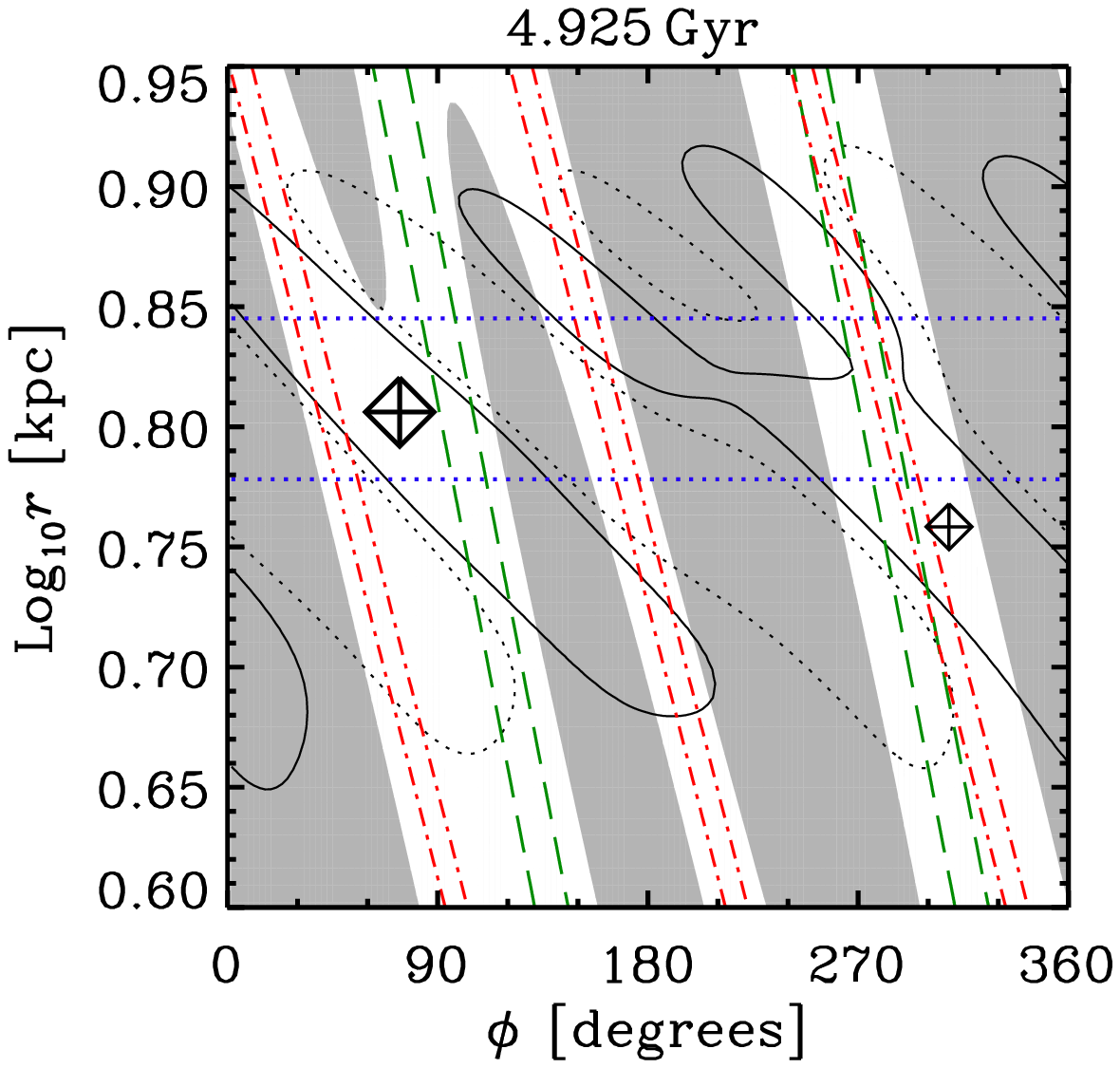}
    \includegraphics[width=42mm]{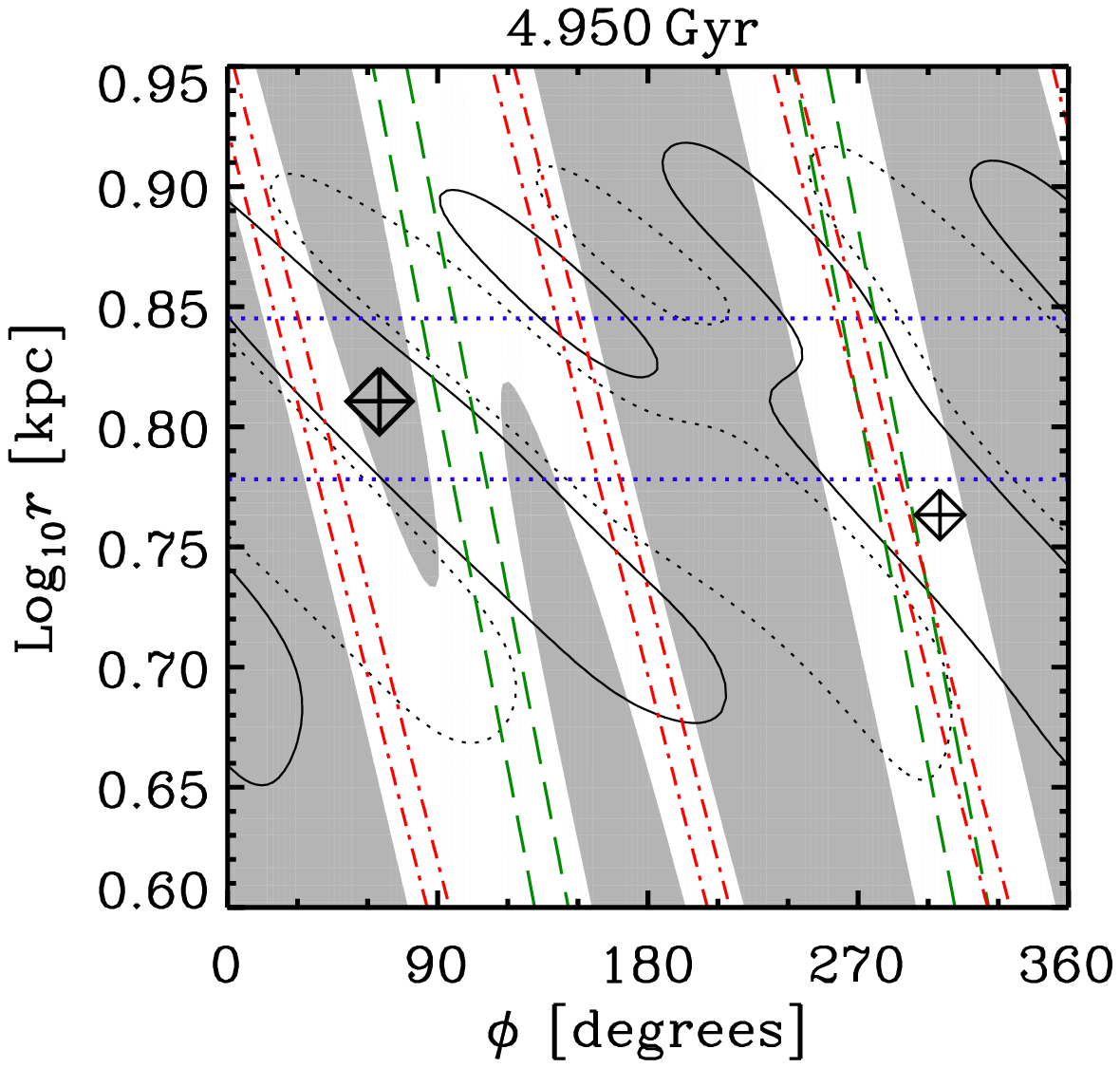}
    \includegraphics[width=42mm]{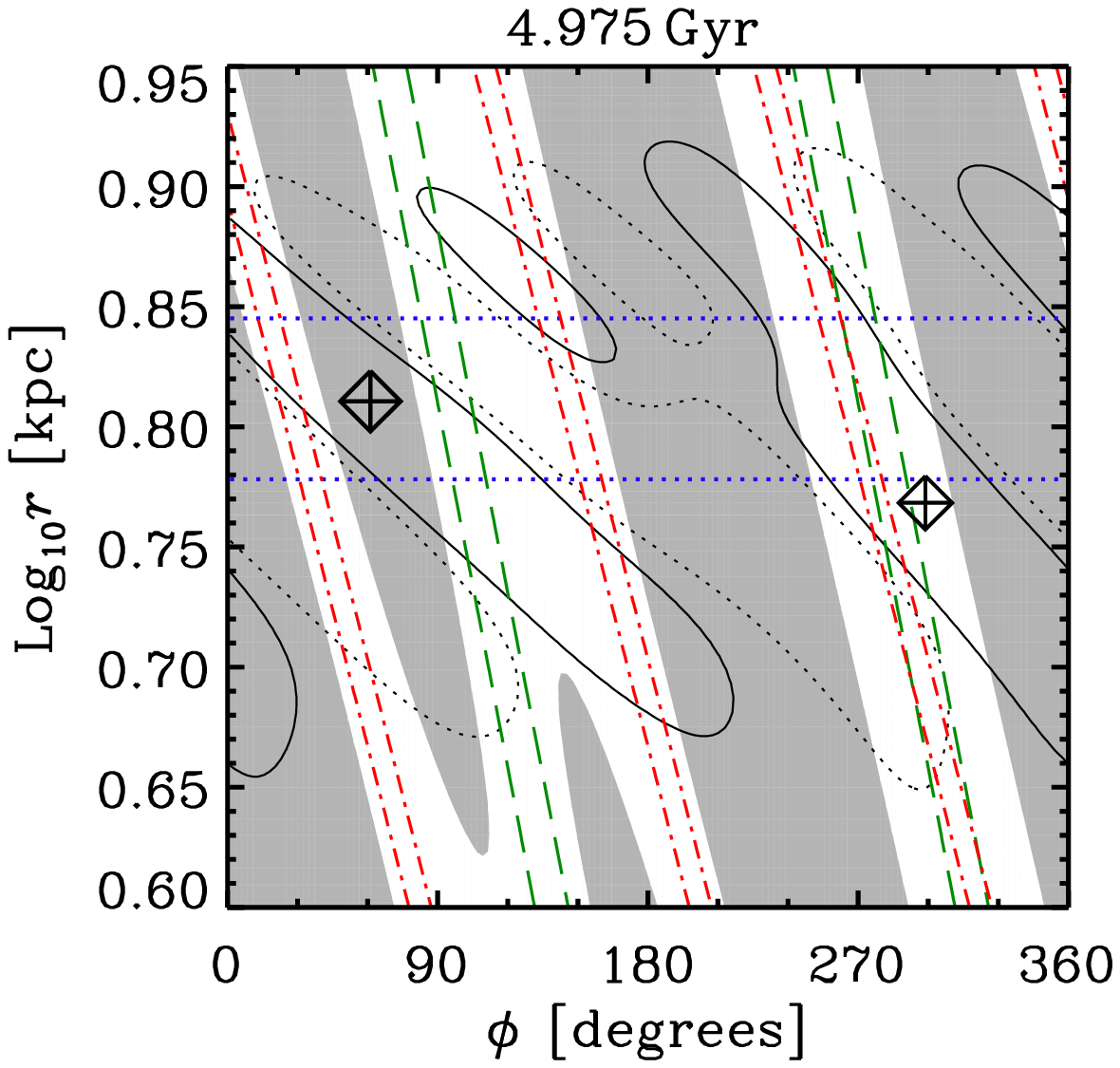}
  \end{array}                       
  $
  \caption{Time sequence showing the locations of the largest two local maxima of $\delta$ for Model A23,
           with respect to arm/interarm regions.
           White regions designate spiral arms [$\alpha\kin>0.5\max(\alpha\kin)$], while grey designates interarm regions.
           $\delta=0.1$ (-0.1) contours are shown with a solid (dotted) line.
           Other contours and symbols are the same as in previous figures.
           \label{fig:deltasimp}}
\end{figure*}

\subsection{Magnetic field in interarm regions}
As a consequence of magnetic arms being more tightly wound than $\alpha\kin$-arms, 
it can sometimes be difficult to decide, based on visual inspection, 
whether magnetic arms are stronger within $\alpha\kin$-arms or in between them.
This question is important because $\alpha\kin$-arms can be assumed to be correlated (or, possibly, anti-correlated)
with the gaseous spiral arms of the galaxy,
and, as mentioned in Sect.~\ref{sec:introduction},
magnetic arms have been reported to be present within the interarm regions of some galaxies like NGC~6946.
In spite of the small values of $p$ relative to $p_\alpha$, it is possible to answer this question by determining where $\delta$ peaks;
i.e. whether it peaks inside or outside one of the $\alpha\kin$-arms.
It is reasonable to adopt the definition that those regions for which $\alpha\kin>0.5\max(\alpha\kin)$ can be regarded as arms,
while those regions for which this condition is not satisfied are interarm.
(The arms of the original two- and three-arm patterns are now referred to as `constituent' arms.)
It can then be seen from Figure~\ref{fig:deltasimp} that the peak of $\delta$ is usually situated inside an $\alpha\kin$-arm (white regions), 
for example at $t=4.9\Gyr$.
However, at other times, the peak of $\delta$ is clearly located in interarm (grey) regions,
for example at $t=4.975\Gyr$, when it lies between constituent arms from the two- and three-arm patterns whose approximate separation is $70^\circ$.
Inspection of Figure~\ref{fig:deltasimp} shows that magnetic arms are strongest in the interarm regions about $1/4$ of the time 
(i.e. for four out of sixteen diamonds in the figure).
Note, however, that for this model, both diamonds (i.e. strongest local maxima) are never in the interarm regions at the same time.
                                    
\begin{figure}                    
  $                                 
  \begin{array}{c c c}           	   
    \includegraphics[width=42mm]{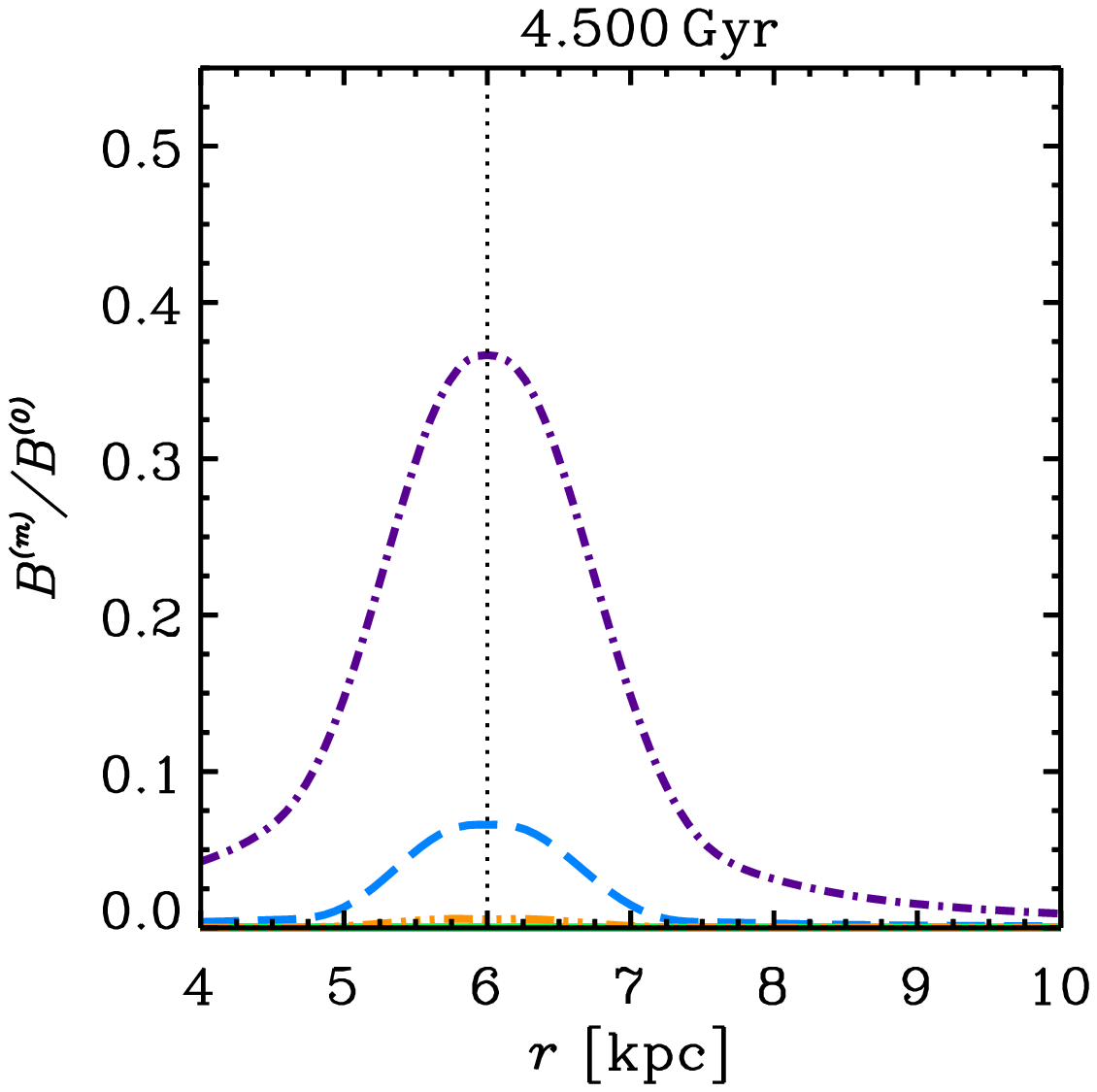}		
    \includegraphics[width=42mm]{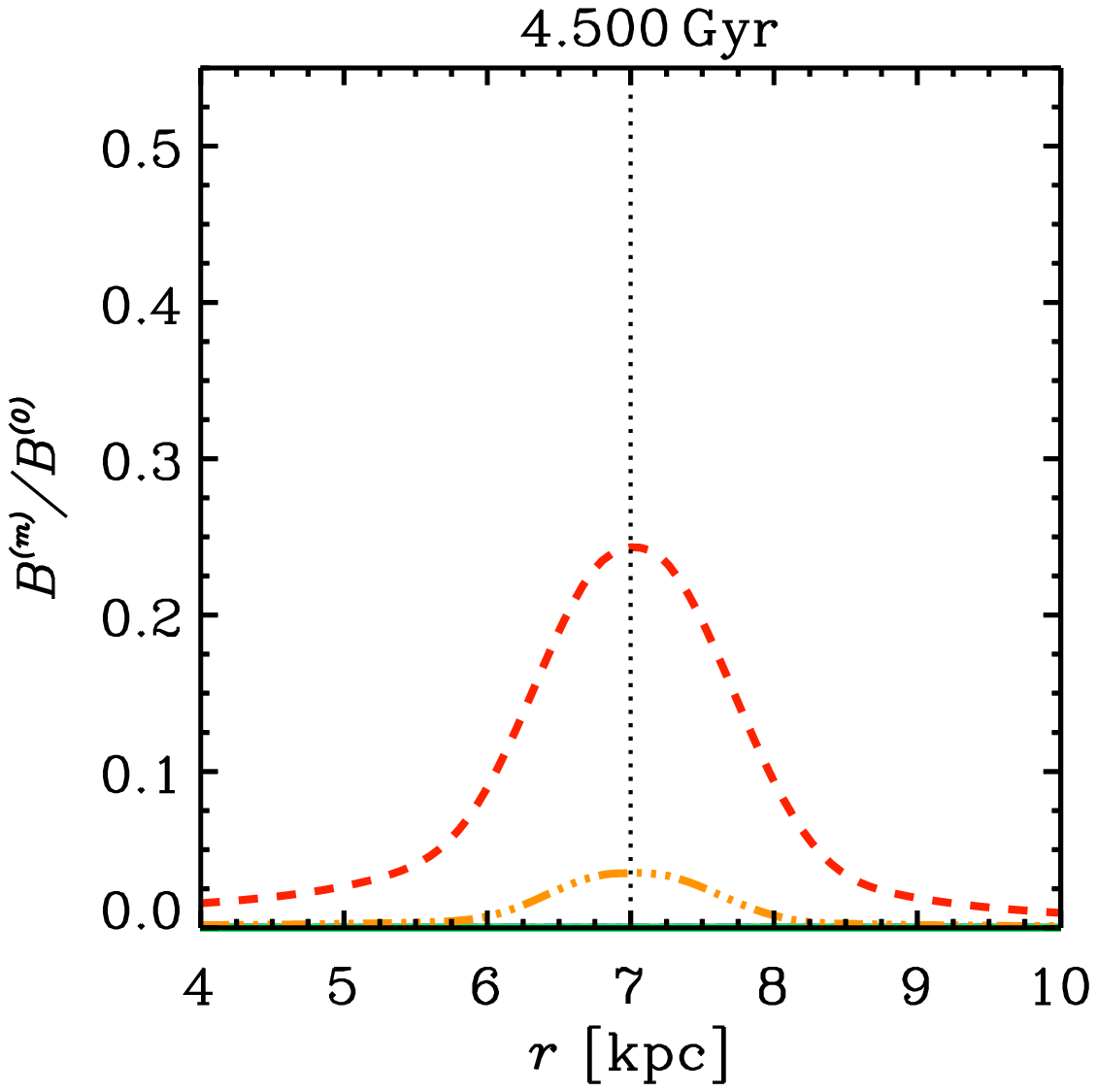}\\		
    \includegraphics[width=42mm]{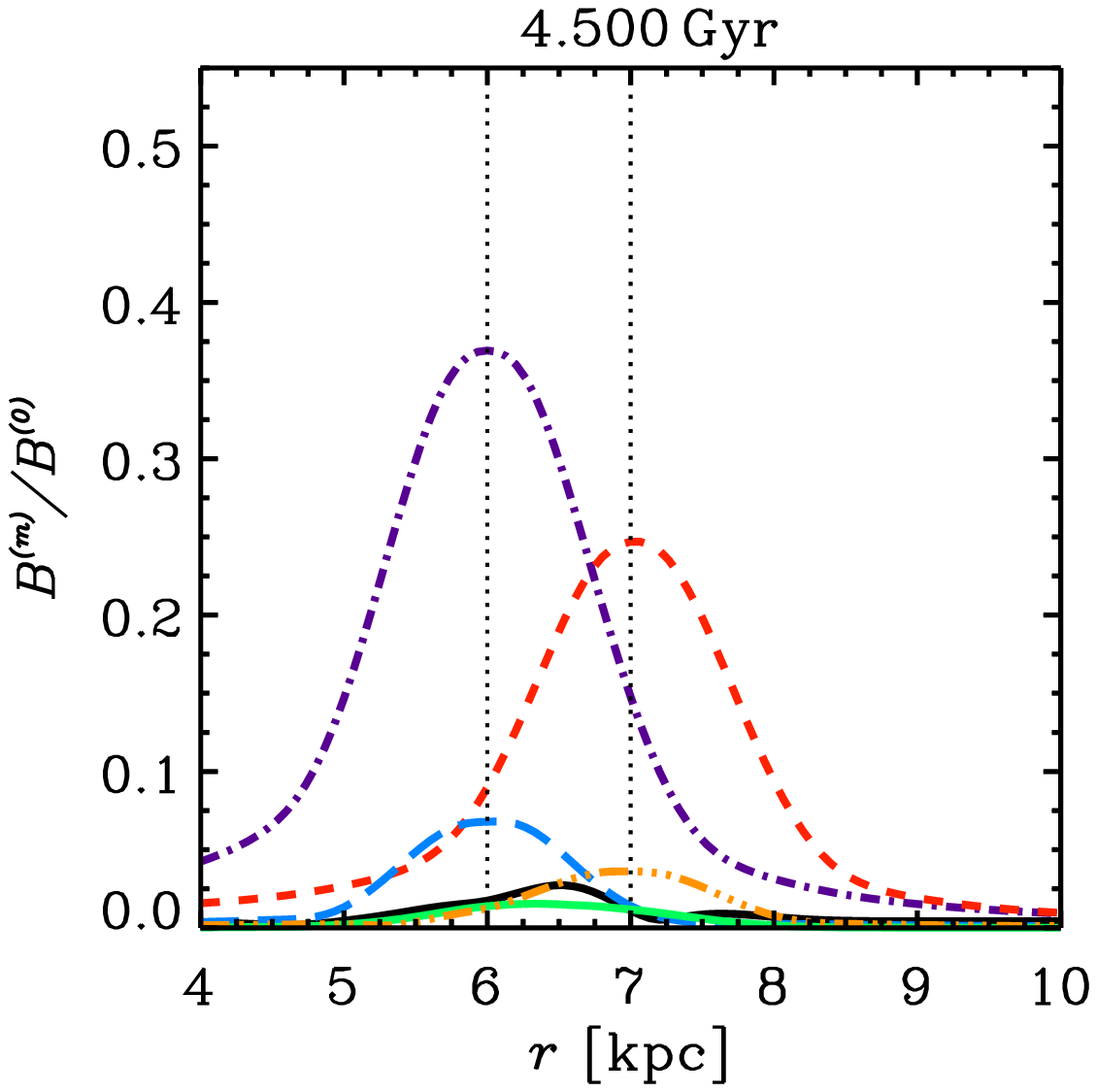}		
    \includegraphics[width=42mm]{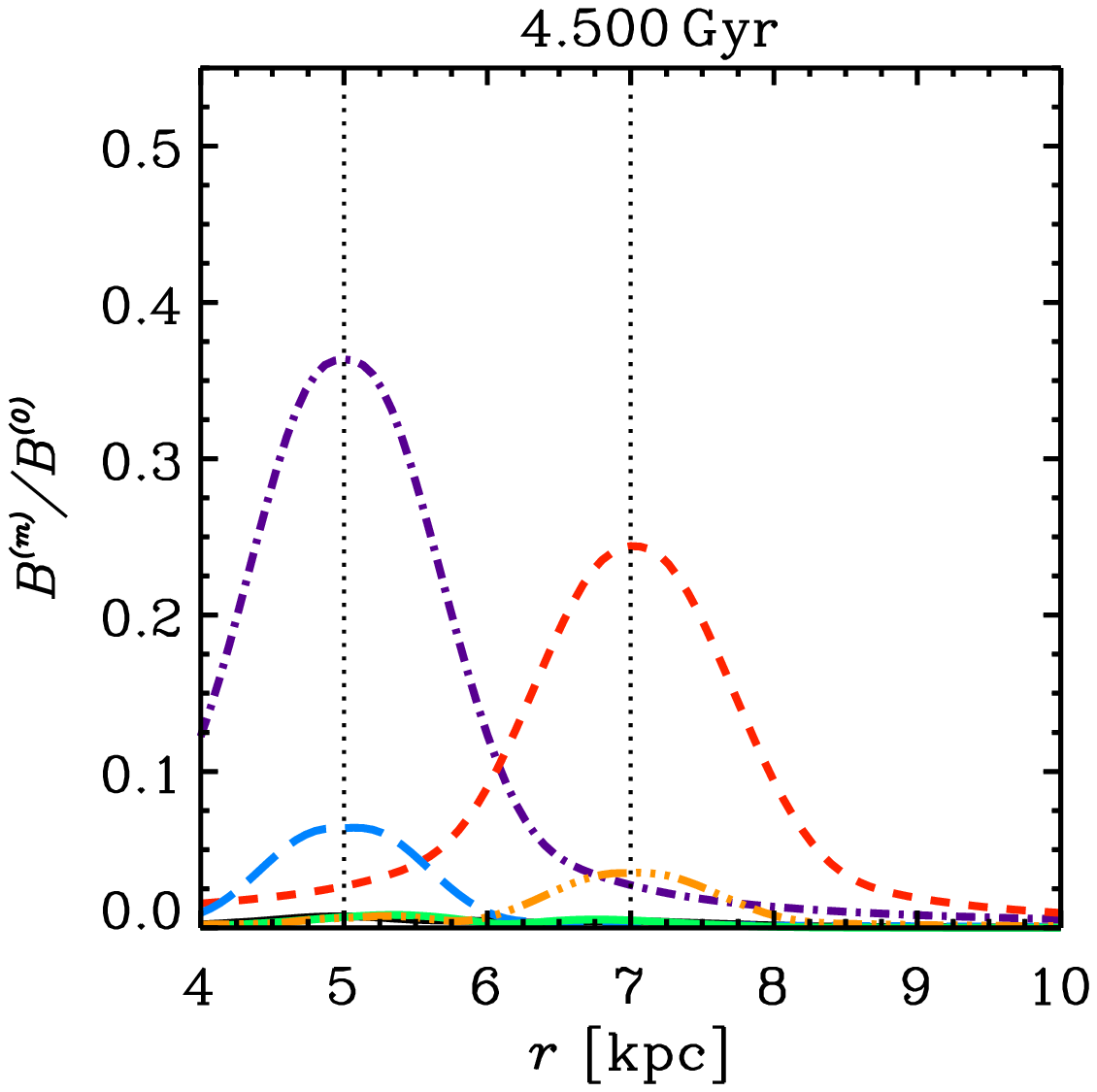}\\		
    \includegraphics[width=42mm]{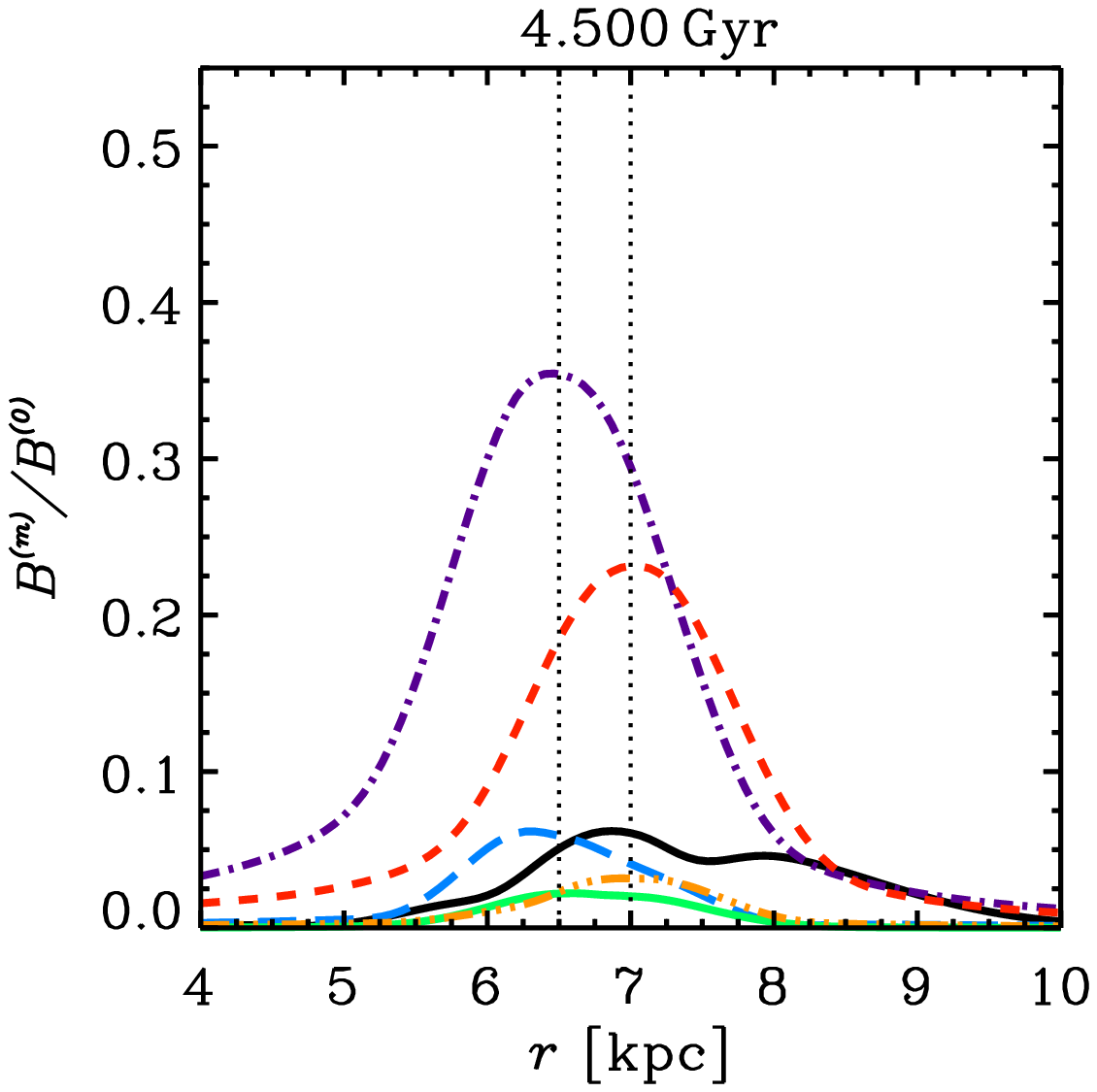}		
    \includegraphics[width=42mm]{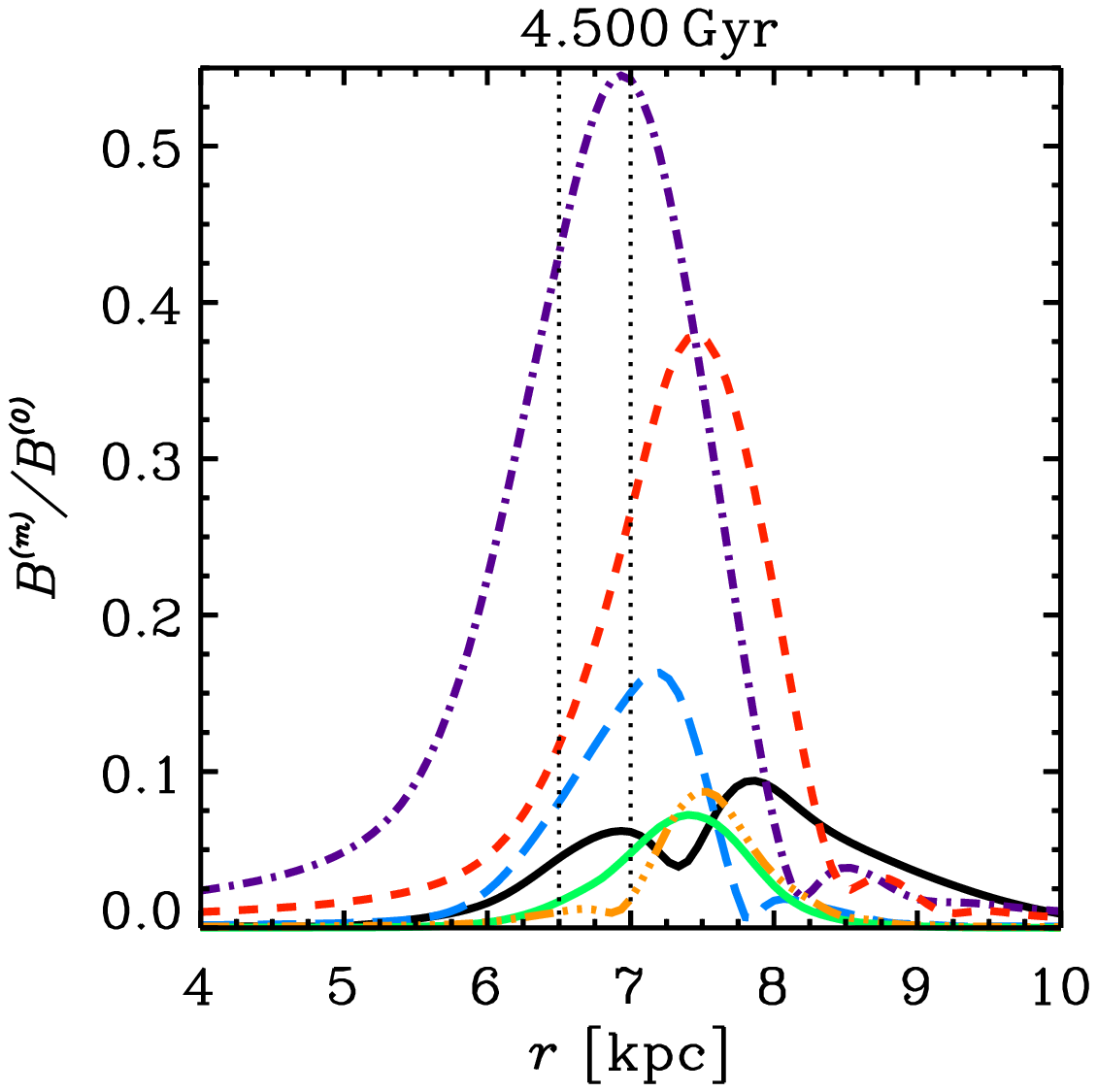}		
  \end{array}
  $
    \caption{The square root of the ratio of the magnetic energy in component $m$ relative to that in the axisymmetric component $[E^{(m)}/E^{(0)}]^{1/2}$, 
             as a function of radius, at $t=4.5\Gyr$: 
             $m=1$ (solid black), $m=2$ (dash-dotted purple), $m=3$ (short dashed red), $m=4$ (long dashed blue), 
             $m=5$ (solid green), $m=6$ (dash-tripple dotted orange).
             Modes with $m>6$ have much smaller energy and are hence not included in the plots.
             Corotation radii for each $\alpha\kin$-pattern are marked with vertical dotted lines.
             Top left: Model~A2. Top right: Model~A3. Middle left: Model~A23. 
             Middle right: Model~L23. Bottom left: Model~S23. Bottom right: Model~S23$\tau$.
             \label{fig:fourier}}        
\end{figure}

\subsection{Azimuthal {F}ourier components}
It is also interesting to ask to what degree various azimuthal components are present, as a function of radius.
To address this question, Figure~\ref{fig:fourier} has been plotted for Models~A2, A3, A23, L23, S23 and S23$\tau$ at the time $t=4.5\Gyr$.
For other times in the nonlinear regime, the plots do not show any important differences from $t=4.5\Gyr$.
Modes up to $m=6$ are illustrated, although it is worth noting that only modes up to $m=2$ have so far been detected observationally \citep{Fletcher10}.
In the single pattern model A2 (A3), $m=2$ ($m=3$) dominates, and is localized around the corotation radius.
In addition, the $m=4$ ($m=6$) enslaved component is also present, 
and is also localized around the $n_1=2$ ($n_2=3$) corotation, 
as expected (\citealt{Mestel+Subramanian91}, \citetalias{Chamandy+13a,Chamandy+13b}).\footnote{The $m=6$ component
is also present in Model~A2, since even modes are enslaved, though it is much weaker than $m=4$.}
In the dual pattern models, it can be seen, unsurprisingly, that the $m=2$ component is dominant, followed by the $m=3$ component, 
and that each is concentrated around the $n_1=2$ or $n_2=3$ corotation radius, respectively, as in the single pattern models.
More interestingly, the $m=1$ component (solid black), though relatively weak, is clearly present in the models with two patterns, 
whereas $m=1$ is negligible when only a single pattern is invoked.
This is important because $m=1$ symmetry has been observed in some galaxies \citep{Fletcher10}
and was reported to be the dominant azimuthal symmetry in the galaxy M81 \citep{Krause+89},
though the three-arm gaseous spiral symmetry of this galaxy has been found to be rather weak \citep{Elmegreen+92},
so it may not be well-suited to the model.
A small but finite $m=5$ component is also present in the dual arm models, 
which is natural given that there are in total five $\alpha\kin$ constituent arms forcing the dynamo.

If the pattern separation is increased from that in the standard model A23, 
this causes a decrease in $m=1$ and $m=5$ components,
as seen from Figure~\ref{fig:fourier} (middle right panel),
for Model~L23.
As the pattern separation is instead reduced from that in A23, 
as in Model~S23,
the $m=3$ symmetry becomes less obvious compared to $m=2$, 
and for S23 the field may best be described as having two asymmetric magnetic arms.
This can be seen from Figure~\ref{fig:fourier} (bottom left panel), which shows the Fourier decomposition for Model~S23.
The range of radii at which $m=3$ dominates is much smaller than in Model~A23 (middle left),
and moreover, the difference between $m=3$ and $m=2$ at those radii is also much smaller.
The $m=1$ component is also much more important in Model~S23 than in Model~A23
(resulting in the asymmetric appearance of the two main magnetic arms),
as is $m=5$.
This is not surprising, as stronger coupling between the two patterns, and hence larger $m=1$ ($=|n\1-n\2|$) and $m=5$ ($=n\1+n\2$) components,
would naively be expected to result from a reduction in pattern separation.
Interestingly, $m=1$ is comparable to $m=2$ and $m=3$ for $r\gtrsim8\kpc$.
The situation is similar for Model~S23$\tau$, which has $\tau=l/u$ rather than $\tau=0$.
However, in this case $\tau$ has enhanced and shifted each maximum away from the corresponding corotation radius, 
as discussed in Sect.~\ref{sec:tau} above.

\subsection{Optimizing the model}
The galactic disk model used thus far is meant to be rather typical, 
and has not been optimized to obtain any of the rather generic observational properties listed in Sect.~\ref{sec:introduction}, 
e.g. better alignment of magnetic and gaseous arms, larger radial extents of magnetic arms, 
and larger pitch angle $p_B$ as compared to the models presented above.
In Model X23, parameter values are chosen to make the model as conducive as possible 
to obtaining these properties,
while still being realistic.
In this model, the disk is flared with $h\D\simeq0.32\kpc$, giving $h=0.45\kpc$ at $r=10\kpc$.  
This smaller value of $h$ naturally leads to larger $|p_B|$ in the kinematic regime \citepalias[e.g.][]{Chamandy+13b},
and this seems to be true also in the non-linear regime.
$\alpha\kin$-arms are now taken to \textit{add} to the axisymmetric $\alpha\kin$, 
rather than just modulate it.
In other words, $l^2\omega/h$ is now the minimum value of $\alpha\kin$.
Furthermore, $\epsilon_1$ and $\epsilon_2$ are now taken to be equal to unity, but 
$\alpha\kin<u$ everywhere for all times, and $\alpha\kin<u/2$ for $r>4.7\kpc$ for all times.
Finally, the rotation velocity at $r=10\kpc$ is reduced to $\mup=180\kms$,
which happens to be similar to the value in NGC~6946 \citep{Fathi+07,Jalocha+10}.
This decreases the differential rotation in the disk.
Differential rotation acts to reduce the winding angle of the magnetic arms, as magnetic field gets advected along with the flow 
(though $\alpha\kin$ also depends on $\omega$ so the effect on winding angle is non-trivial).
By enhancing the strength of $\alpha\kin$-arms and reducing the differential rotation,
magnetic arms tend to become more aligned with $\alpha\kin$-arms.
More severe deviations from the Brandt profile, as would be appropriate for some galaxies, 
could make a substantial difference, but such a change is not explored here.
The vertical velocity $\muz$, which can be thought of as a mass-weighted velocity over all phases of the ISM due to a galactic fountain/wind,
 could take on values from $\sim0.2-2\kms$ \citep{Shukurov+06}.
Taking $\muz=2\kms$ further increases $|p_B|$ by $\sim5^\circ$ as compared to $0.3\kms$ \citep[see also][]{Sur+07}, 
but at the cost of the dynamo being too weak at large radius $r\sim10\kpc$.
The functional form
\[
\muz= U\f\Exp{-r^2/(2r_U^2)}
\]
is thus adopted, with $U\f=2\kms$ and $\muz$ made to equal $0.2\kms$ at $r=10\kpc$ so that $r_U\simeq4.66\kpc$.
Other parameters $r_\omega=2\kpc$, etc., are retained.\footnote{Deviations 
from Krause's law can also significantly affect the dynamo.
For example, putting the azimuthal mean value of $\alpha\kin$ constant with radius allows the dynamo to remain supercritical at large $r$, 
even if $\muz$ is large.
However, there is currently little theoretical justification 
for imposing an axisymmetric $\alpha\kin$ radial dependence different from the one used.}

\begin{figure*}                     
  $                                 
  \begin{array}{c c c c}              
    \includegraphics[width=42mm]{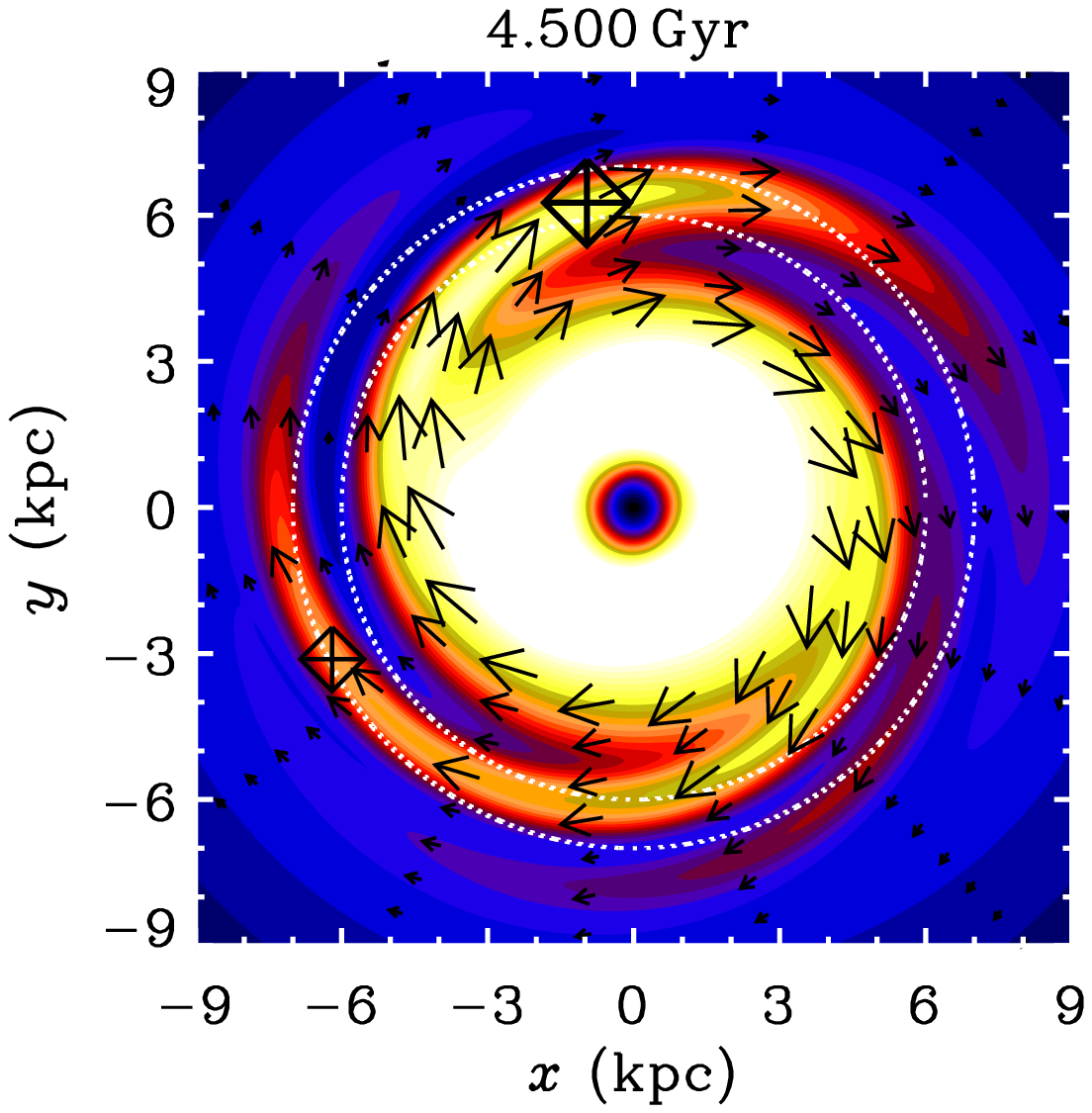}
    \includegraphics[width=42mm]{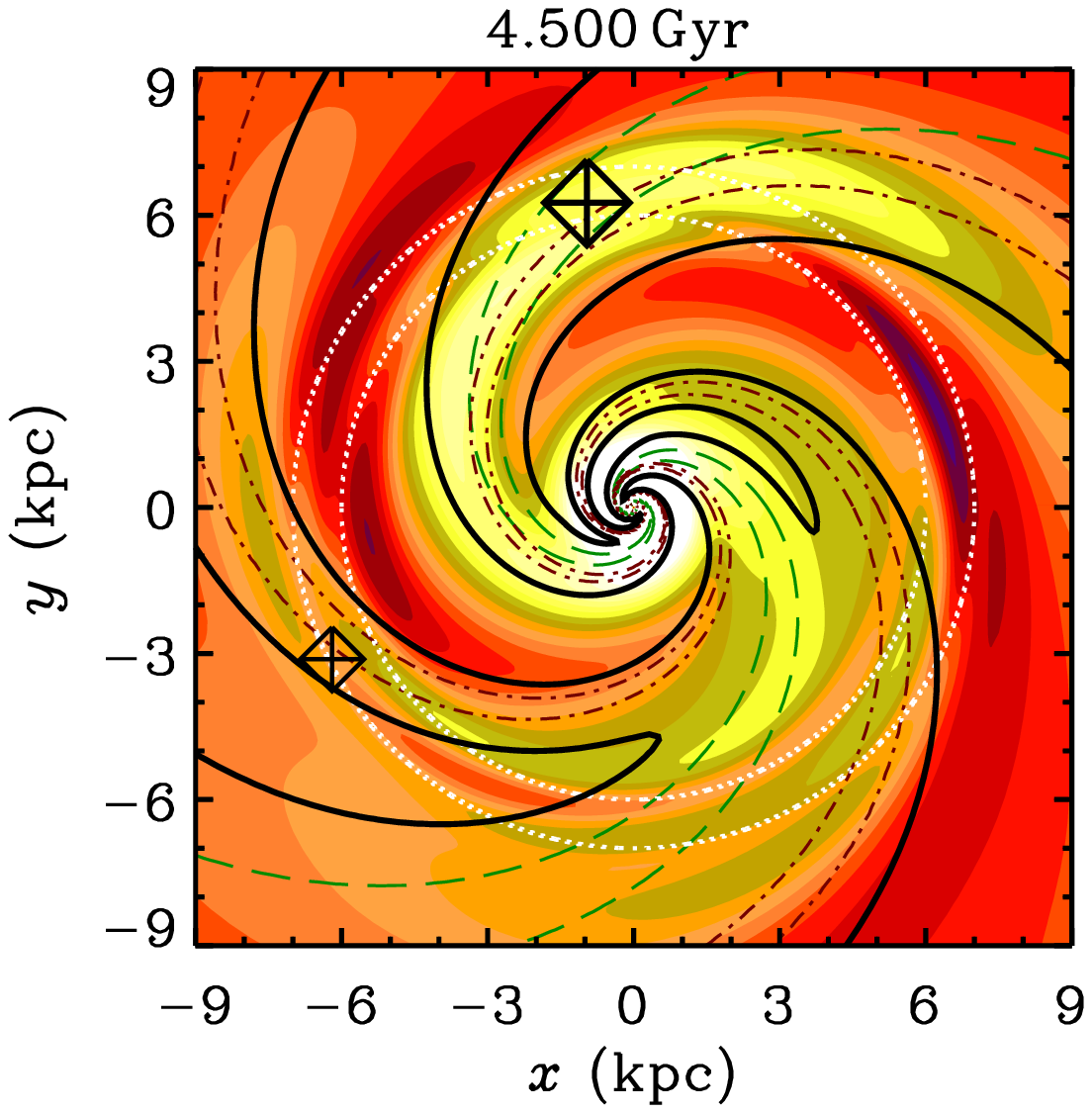}
    \includegraphics[width=42mm]{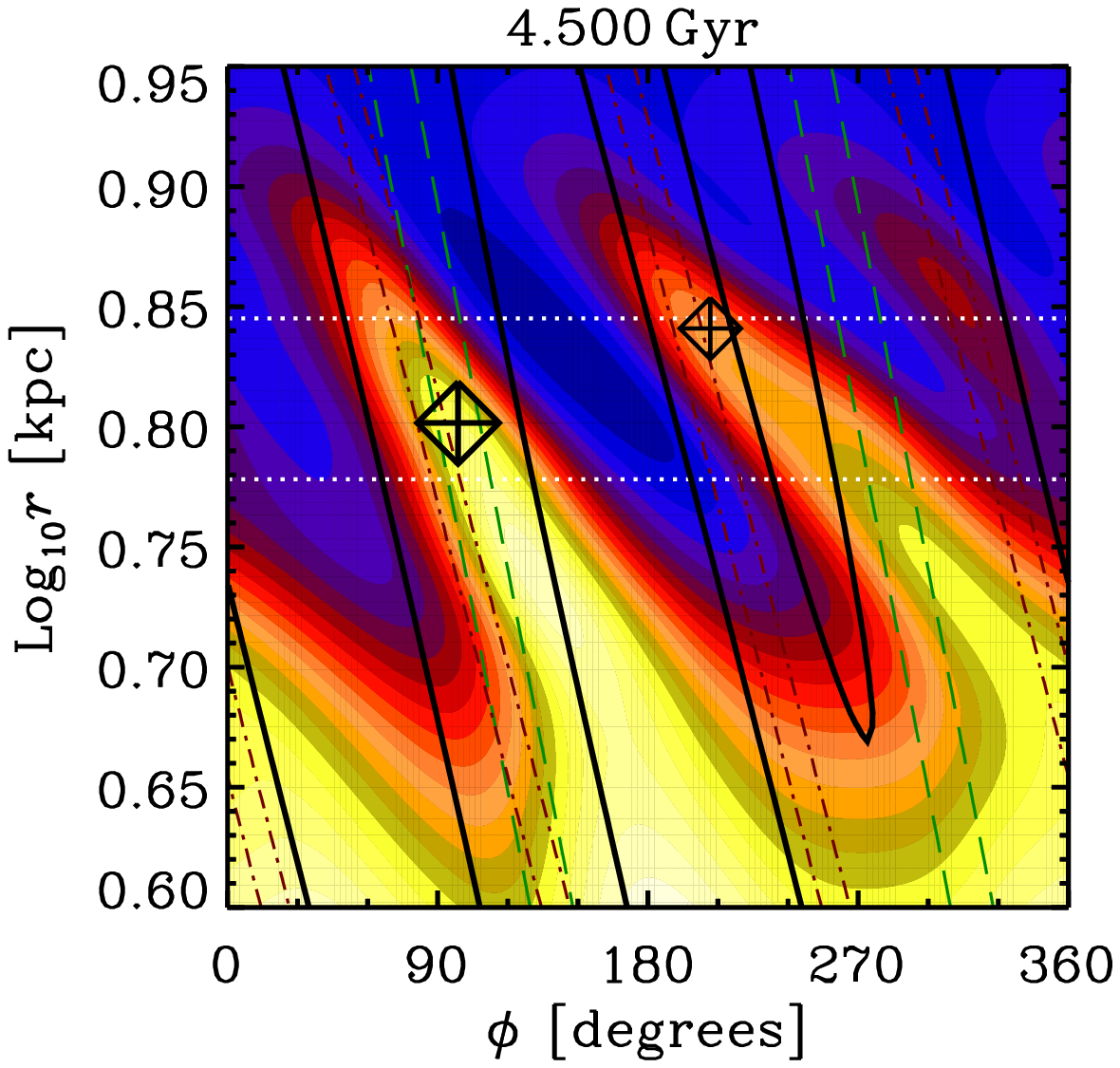}
    \includegraphics[width=42mm]{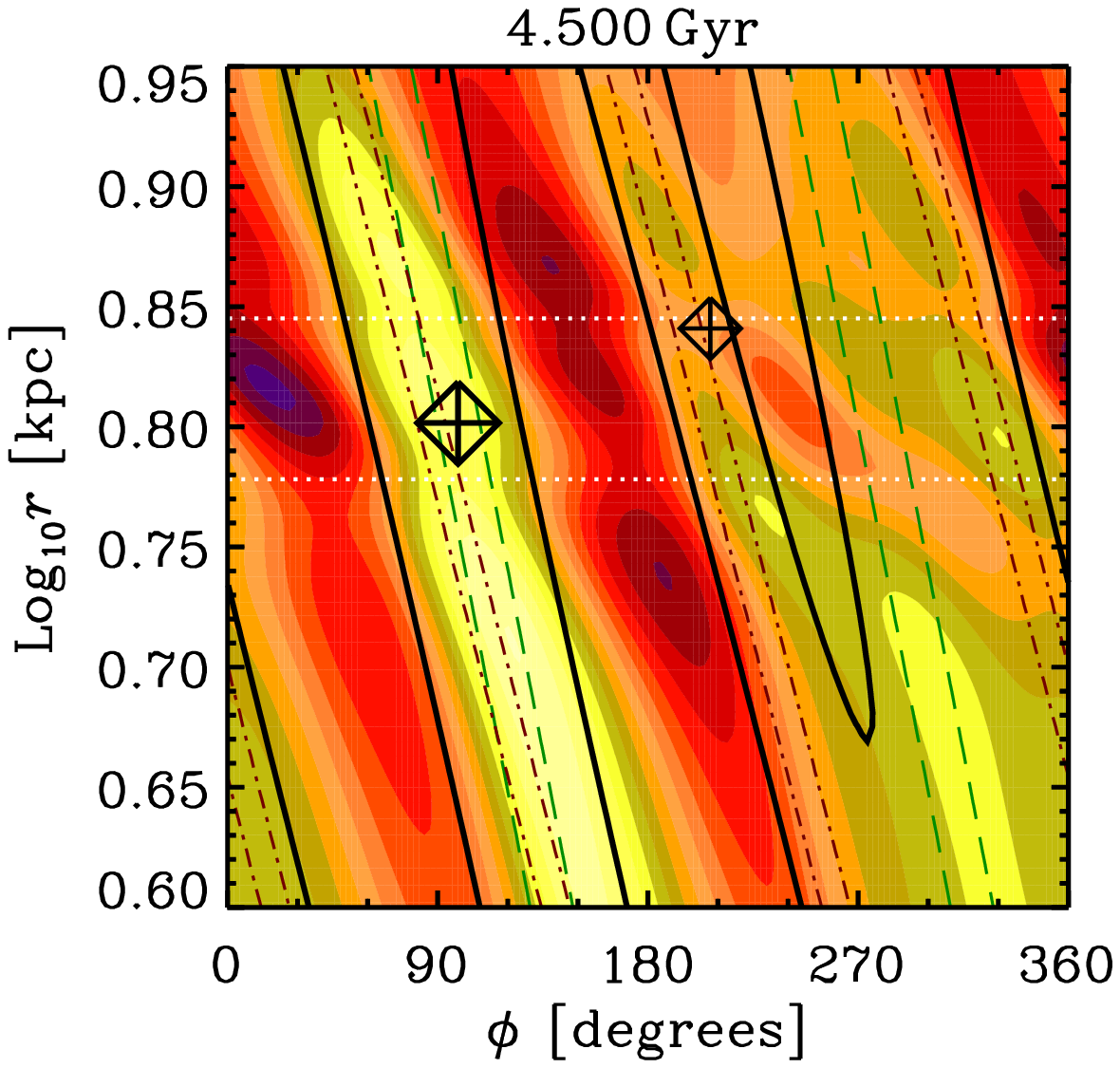}\\
    \includegraphics[width=42mm]{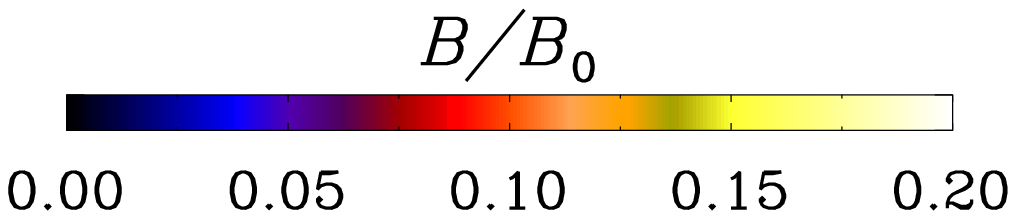}
    \includegraphics[width=42mm]{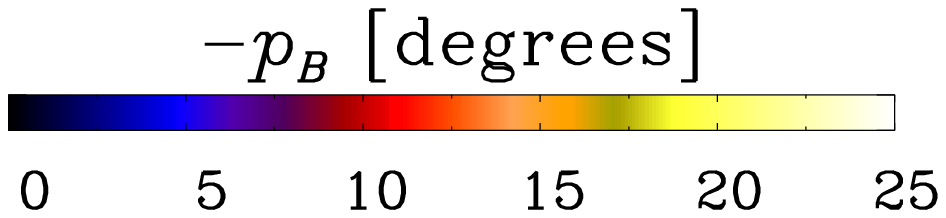}
    \includegraphics[width=42mm]{IS_205_colr_Bcart_nt180.eps}
    \includegraphics[width=42mm]{IS_205_colr_pitch_nt180.eps}\\
    \includegraphics[width=42mm]{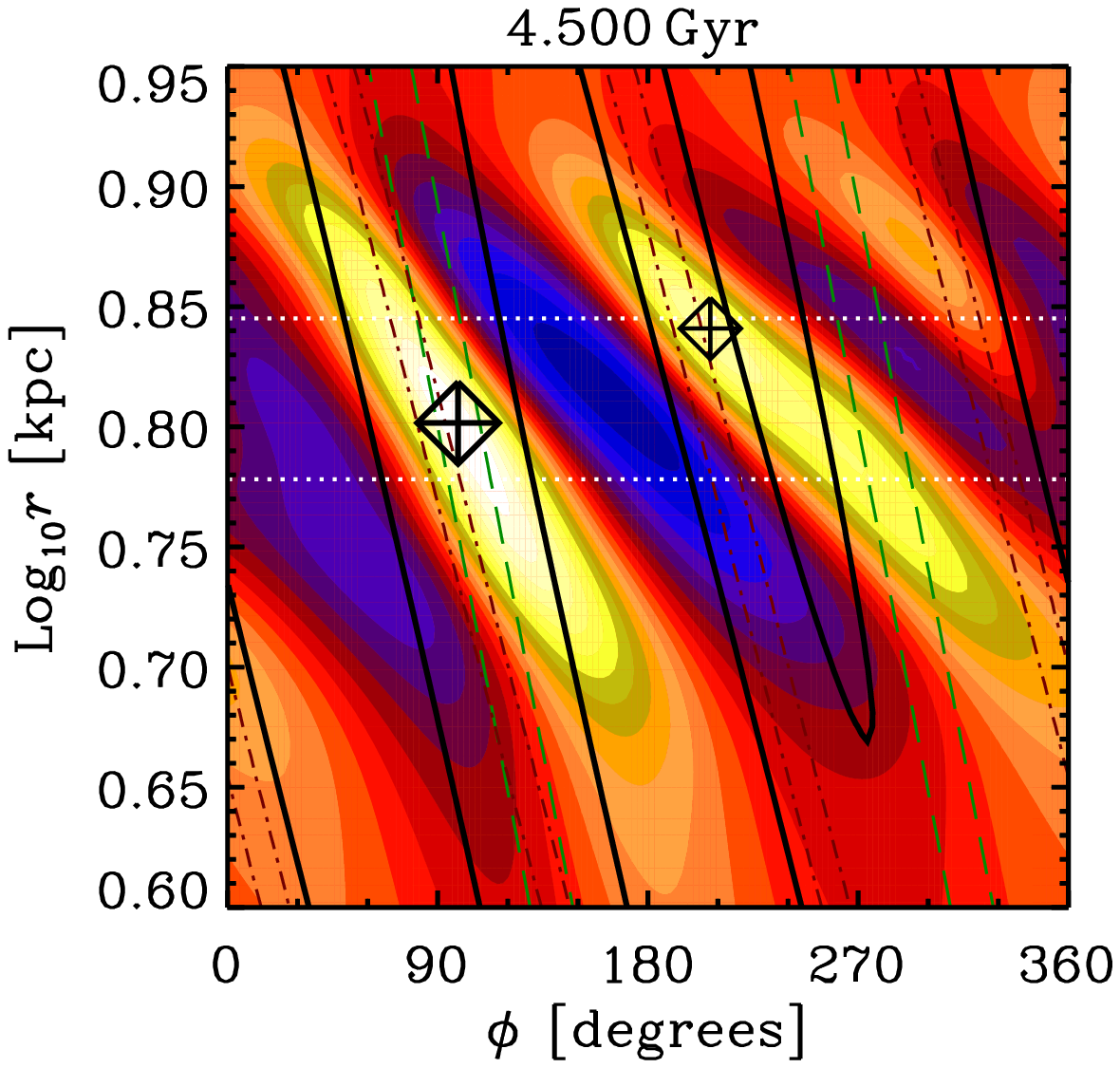}
    \includegraphics[width=42mm]{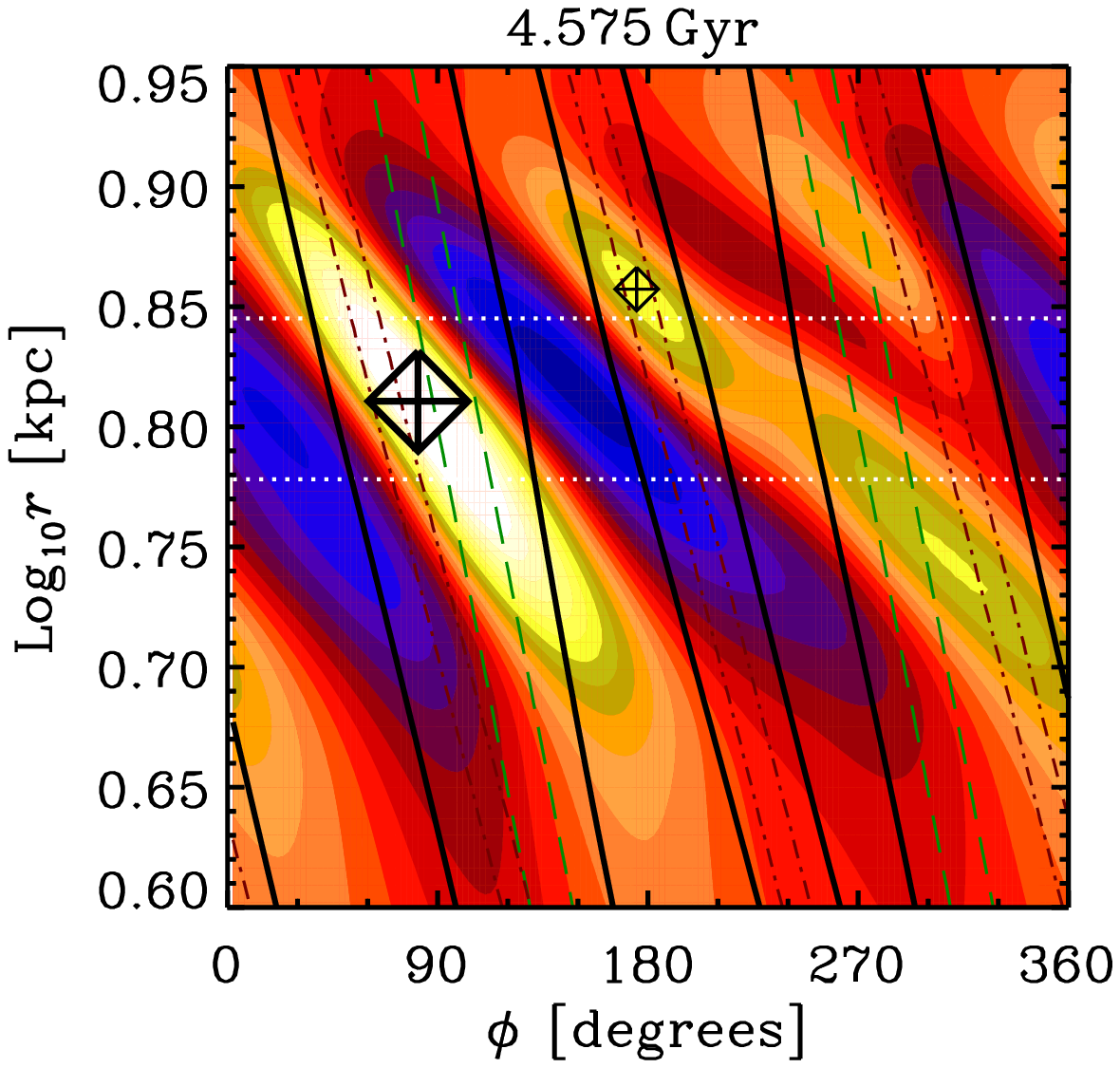}
    \includegraphics[width=42mm]{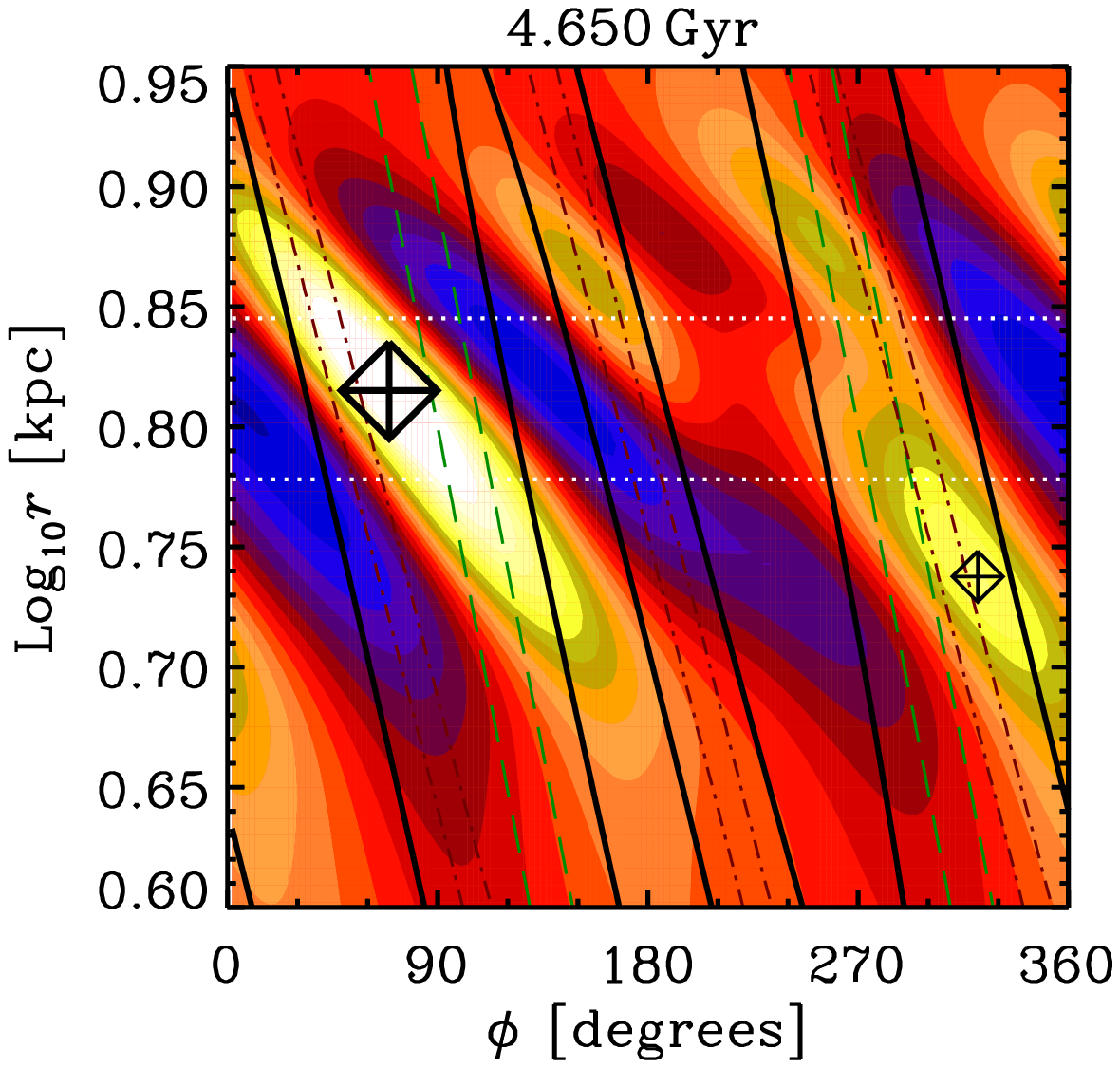}
    \includegraphics[width=42mm]{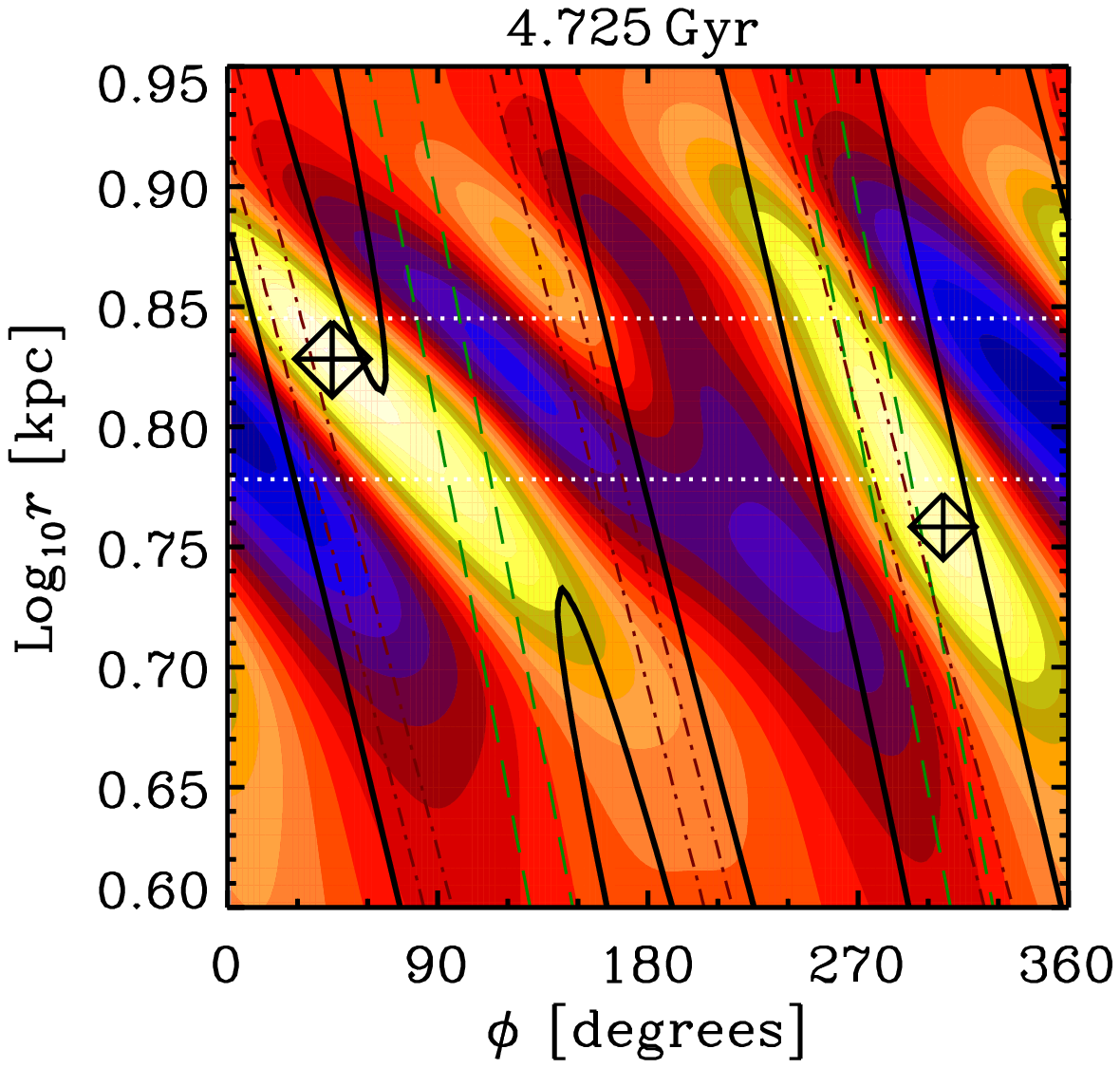}
  \end{array}
  $
  \caption{Results for Model~X23. Top panels show, from left to right, 
           magnetic field strength and negative of the pitch angle of the magnetic field in Cartesian coordinates,
           and then the same quantities in $\log r$-$\phi$ coordinates.
           Bottom panels show $\delta$ at various times.
           Color scheme for $\delta$ is the same as that in Fig.~\ref{fig:Bcompare}.\label{fig:X23}}
\end{figure*}

The results for Model~X23 are illustrated in Fig.~\ref{fig:X23}.
Top panels show the field strength and pitch angle, while bottom panels shown snapshots of $\delta$.
Magnetic arms are much more radially extended than in Model~A23. 
They are also more closely aligned with $\alpha\kin$-arms, 
especially at some times and places (e.g. for the strongest magnetic arm at $t=4.5\Gyr$).
Moreover, $|p_B|$ reaches up to $\sim23^\circ$ in the magnetic arms,
but is smaller in between magnetic arms.
Large-scale magnetic field is strongest within $\alpha\kin$ arms in this model,
though $\delta$ can be comparably large in arm and interarm regions.
This can be seen at $t=4.5\Gyr$, 
when the magnetic arm between about $150^\circ-360^\circ$ clearly stretches across an interarm region.
Qualitative features observed in Model~A23, such as isolated armlets and bifurcations, 
can also be identified in Model~X23.

Model~X23 comes closer to explaining the properties mentioned in Sect.~\ref{sec:introduction}.
However, extreme cases with $|p_B|$ as large as $\sim40^\circ$ as in the galaxy M33 and near-perfect alignment of magnetic and optical arms
as in NGC~6946 are clearly not possible to explain using such models.
The $\tau$-effect helps to shift the large-scale magnetic field into the interarm regions.
However, as seen in Sect.~\ref{sec:tau}, alignment of the two types of arms only gets worse, 
and the radial extent gets smaller, when the $\tau$ effect is included.
Allowing each individual spiral pattern to wind up, at least to some extent,
would help to further improve the alignment and increase the radial extent \citepalias{Chamandy+13a}.

\section{Discussion and conclusions}
\label{sec:conclusion}
Galactic mean-field dynamo solutions are obtained numerically 
for the case of non-axisymmetric forcing of the dynamo by multiple co-existing steady 
rigidly rotating spiral patterns.
Each pattern is modelled as a spiral modulation of the $\alpha$ effect with a fixed number $n$ of equally spaced arms.
Specifically, the cases of an inner two-arm and outer three-arm spiral pattern, 
with corotation radii separated by $0.5$, $1$ or $2\kpc$, are investigated.
A model with an inner two-arm and outer four-arm pattern is also studied.

The resulting magnetic field is found to exhibit magnetic arms that can be stronger, 
as well as more extended in radius and in azimuth, 
than those produced by any individual pattern acting alone.
This helps to explain the occurrence of pronounced magnetic arms with radial extents similar 
to those of the gaseous arms that have been observed in some galaxies.

The morphology of the magnetic arms in the models evolves with time on the scale of the beat period of the two patterns, 
as it depends on how the $\alpha\kin$-arms from the two-arm and three-arm patterns interfere to generate the magnetic arms.
For instance, the winding (pitch) angle of magnetic arms can vary significantly in time and from one arm to another,
and at times, magnetic arms can become quite well-aligned with the gaseous arms, 
though never as precisely aligned as sometimes seen in observations.
Furthermore, bifurcations of magnetic arms and isolated magnetic segments (filaments/armlets) are produced.

Interestingly, magnetic arms in some models 
are found to have maximum strength sometimes in between the $\alpha\kin$-arms, 
i.e. in the interarm regions, and at other times within an $\alpha\kin$-arm.
However, these models do not produce a systematic shift (for the whole galaxy and for all times) 
of large-scale magnetic field toward the inter-arm regions unless the $\tau$ effect is included \citepalias{Chamandy+13a, Chamandy+13b}.

It is also noteworthy that the magnitude of the pitch angle of the large-scale magnetic field 
can exceed $20^\circ$ for the most favourable (yet still realistic) parameters, 
but pitch angles $>25^\circ$ are not possible to explain as of yet.

Unsurprisingly, the $m=2$ magnetic azimuthal component is found to dominate near the $n=2$ corotation radius, 
while the $m=3$ component dominates near the $n=3$ corotation radius. 
However, other components, including $m=1$, are also present to some degree; 
this is important because $m=1$ is negligible for the cases of single two-arm or three-arm patterns, 
though it has been detected observationally in a number of galaxies \citep{Fletcher10}.\footnote{\citet{Mestel+Subramanian91} 
show, however, that $m=1$ has a growth rate comparable to that of $m=0,2$ for the different disk model used in that work.}
However, there is no reason why asymmetry in either the spiral arms or the coupling of spiral arms with the dynamo 
could not be present and produce significant power in $m=1$, e.g. as touched on in \citet{Moss98}.
This possibility should be explored in the future.

Some of the features found in the solutions are also found in the galaxy NGC~6946,
which is thought to contain an $n=2$ and an $n=3$ pattern \citep{Elmegreen+92},
and is famous for harbouring magnetic arms that are localized in the interarm regions \citep{Beck07}.
This galaxy is inferred to have five magnetic arms, 
with some extending radially for $\gtrsim10\kpc$ and others located only in the inner or outer regions \citep{Frick+00}.
This is reminiscent of the complex morphology seen in Model~A23 or X23, 
including magnetic `armlets' that are shorter than the magnetic arms and localized to the outer regions of the galaxy, 
as well as bifurcations of magnetic arms.
Another galaxy whose magnetic field has been studied in some detail, 
and that has also been shown to contain strong two- and three-arm spiral structure by \citet{Elmegreen+92} is IC~342. 
More recently, \citet{Crosthwaite+00} find a two-arm spiral pattern in the inner part of the galaxy, 
and four-arm spiral structure in the outer part.
The magnetic field literature for this galaxy is not as extensive as for NGC~6946,
but IC~342 is known to have a significant regular field, largely interarm,
extending $\sim8\kpc$ in radius, 
and apparently filamentary in structure in the outer regions \citep{Krause+89,Sokoloff+92,Beck+Wielebinski13}.
The large radial extent and outer filamentary structure are reminiscent of the magnetic field resulting from models such as A23 and A24.

Taken together, the studies of forcing by a corotating (maximally winding up) spiral and by a transient rigidly rotating spiral in \citetalias{Chamandy+13a},
and the present study represent initial steps in the direction of systematically exploring the effects of alternate (i.e. non-steady)
spiral morphology and dynamics on the galactic mean-field dynamo.
Such an exploration is beginning to lead to a better understanding of magnetic arms,
but there exists an additional motivation:
to invert the problem, using magnetic arms to learn about gaseous spirals.
This study and \citetalias{Chamandy+13a} already hint that the presence of radially extended magnetic arms 
that trace the optical arms in some galaxies implies the presence of spiral patterns that wind up (to some degree) with time.
However, more evidence needs to be gathered before such a claim is made.
For instance, it would be useful to study how the mean field dynamo responds to forcing by spiral patterns
whose pitch angles and envelope functions evolve with time, e.g., according to density wave theory \citep{Binney+Tremaine08}.
It would also be useful to determine how robust basic features of the solutions are when the spiral pattern and dynamo 
are coupled differently,
through the parameter $\eta\turb$ say, instead of through $\alpha\kin$,
or by imposing large-scale spiral streaming flows. 
Finally, given that spiral structure and dynamics involve non-linear effects,
it would also be useful to use data from an $N$-body simulation as direct input into the dynamo model,
as first attempted by \citet{Otmianowska-Mazur+02}.

\section*{Acknowledgments}

We thank Anvar Shukurov for his many useful inputs during the
course of our discussions on the origin of magnetic spirals 
and for useful comments on an early version of the manuscript.
We also thank the referee for helpful suggestions that improved the paper.
KS was visiting the University of Rochester when the idea for this work arose. 
He acknowledges partial support from NSF grant PHY-0903797 during this visit 
and the warm hospitality of Eric Blackman at Rochester.

\bibliographystyle{mn2e}
\bibliography{refs}

\appendix
\section{Effect of making the $\alpha$ spirals transient}
\label{sec:transient}
\begin{figure*}                     
  $                                 
  \begin{array}{l l l}              
    \includegraphics[width=42mm]{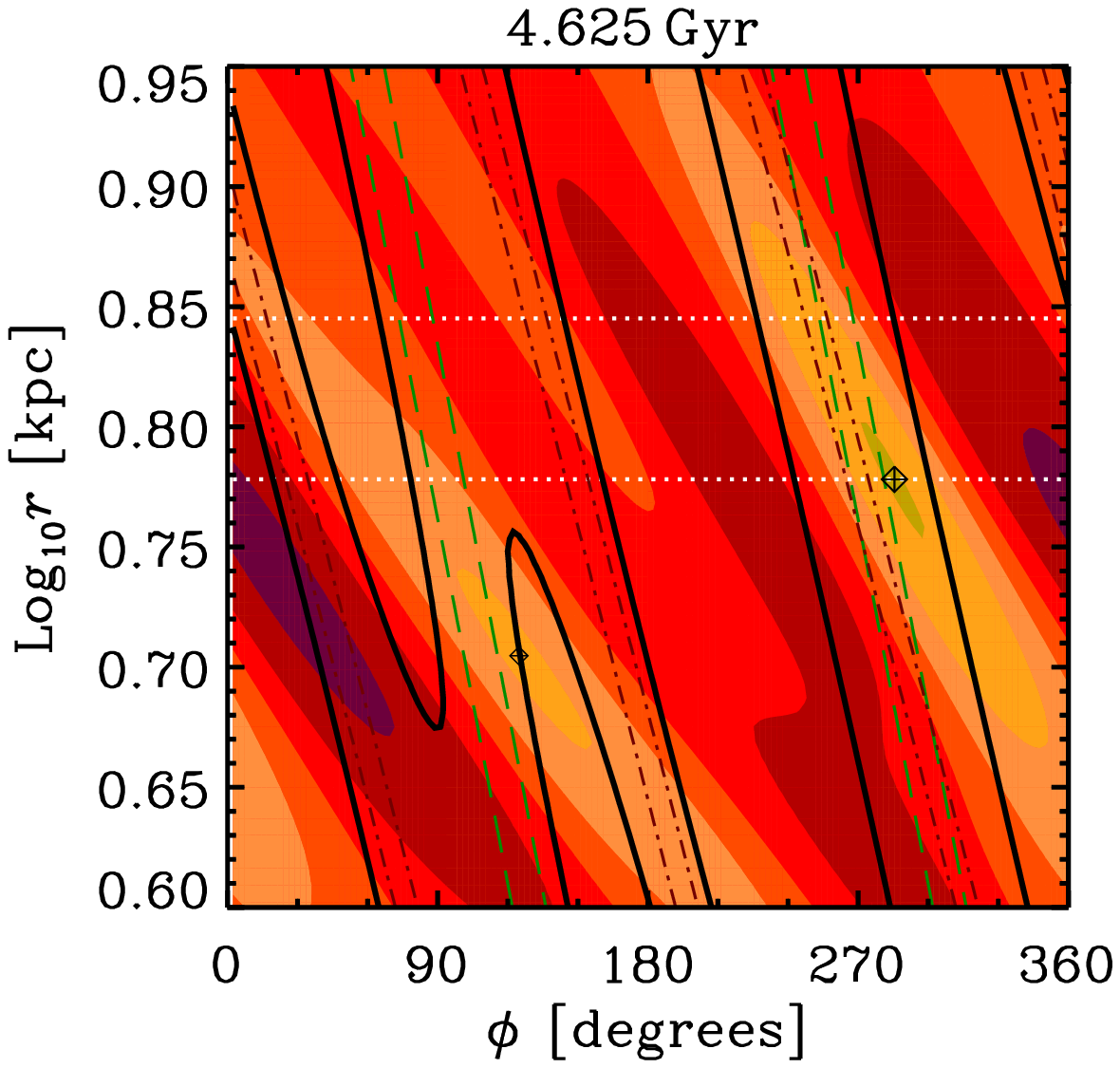}
    \includegraphics[width=42mm]{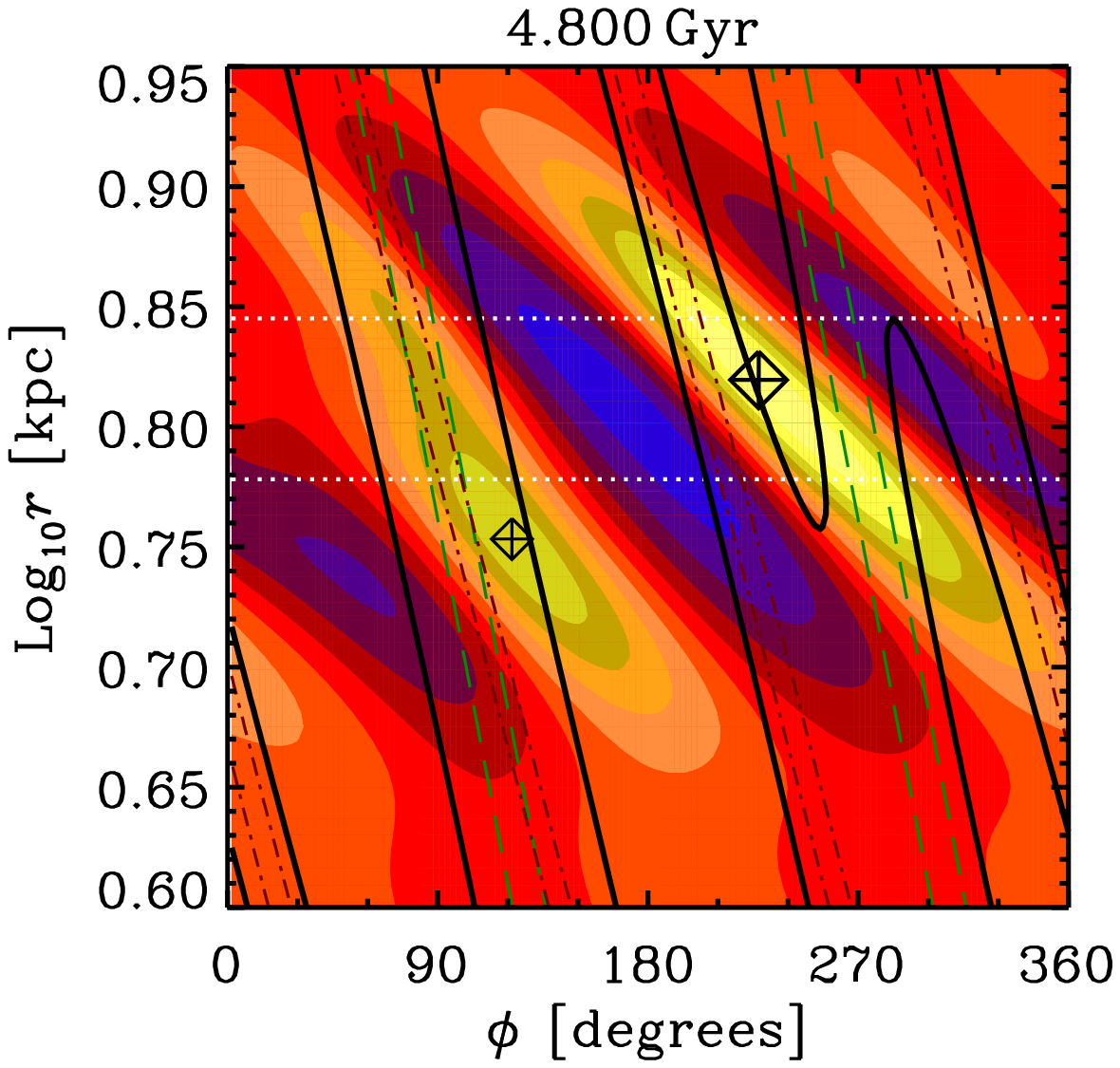}
    \includegraphics[width=42mm]{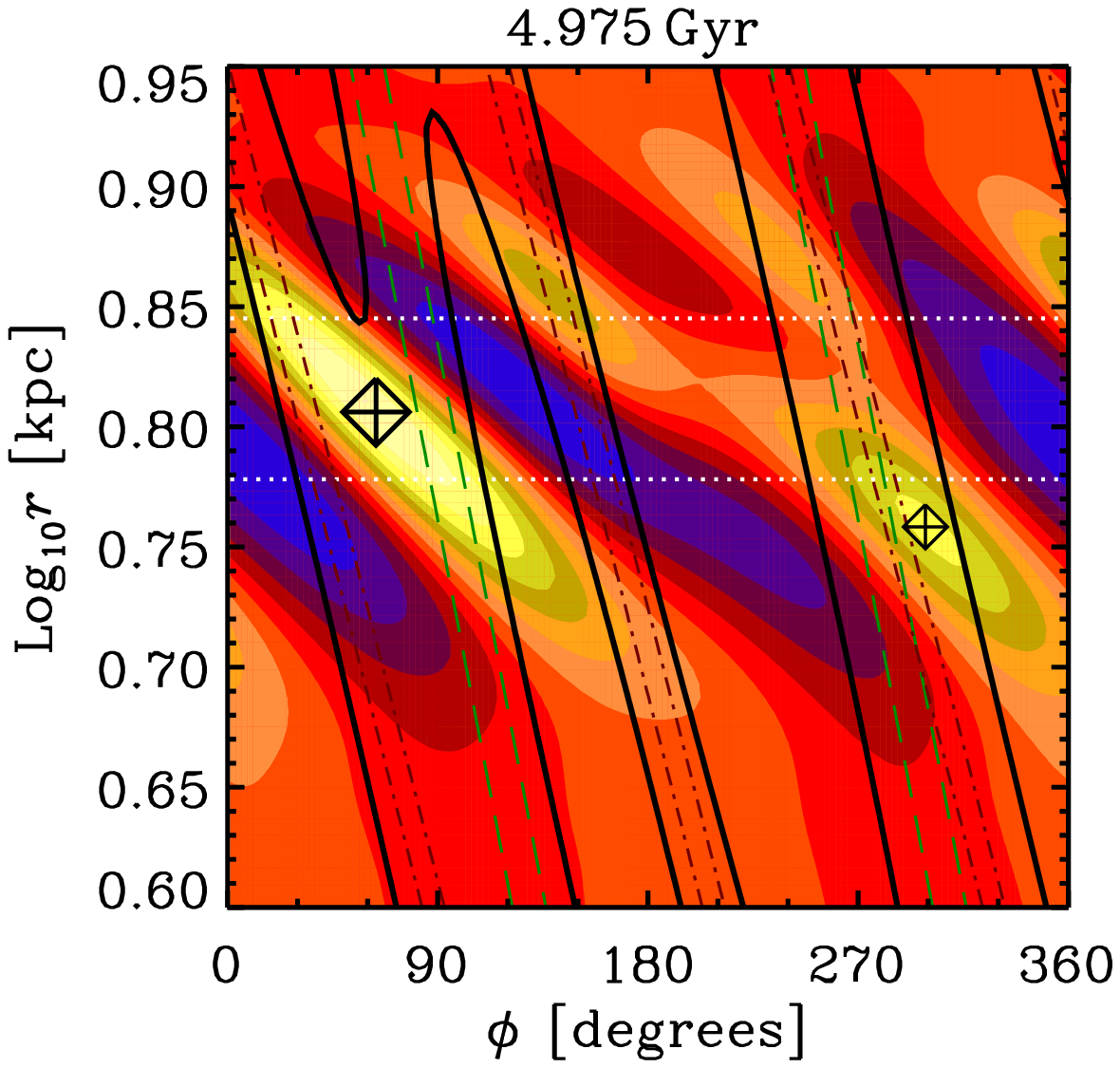}
    \includegraphics[width=42mm]{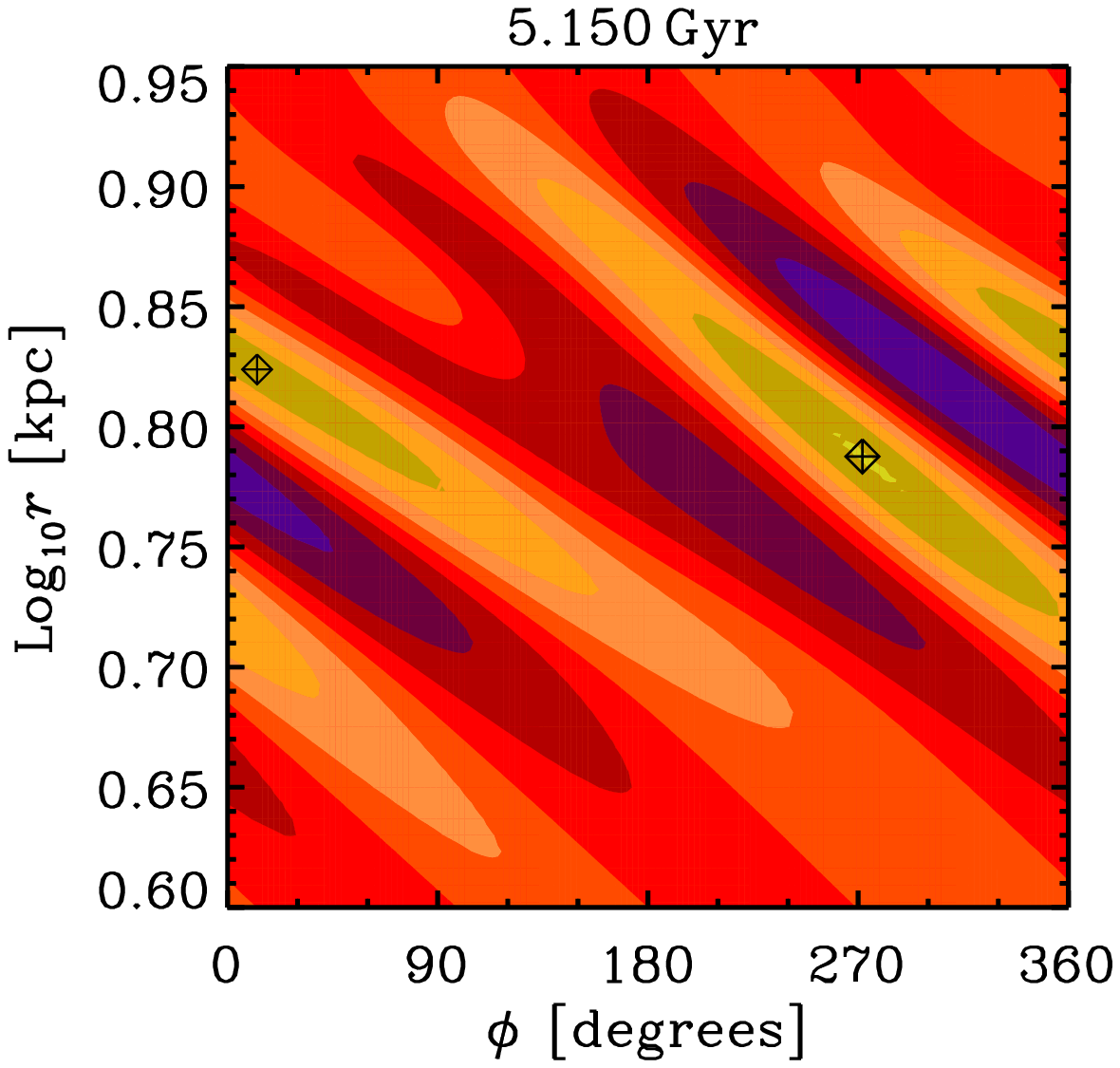}
  \end{array}                       
  $                                 
  \caption{Time sequence showing the evolution of $\delta$ for Model~T23 (transient spiral patterns). 
           The $\alpha\kin$-spirals are turned on at $t=4.5\Gyr$ and turned off at $t=5\Gyr$.
           Contours and symbols are the same as in previous figures.
           \label{fig:delta_transient}}
\end{figure*}

Galactic spiral patterns may typically last for only a few galactic rotation periods,
and so it is important to explore how magnetic arms develop and evolve when forced by transient patterns.
For this purpose, Model~T23 invokes both two- and three-armed patterns from the time $t=4.5\Gyr$, 
when the axisymmetric field is saturated.
These patterns are subsequently turned off at the time $t=5\Gyr$, 
i.e. $3.3$ rotation periods of the innermost pattern later.
Figure~\ref{fig:delta_transient} shows a sequence of snapshots of $\delta$, 
starting from $0.125\Gyr$ after the time when spiral forcing is turned on.
It can be seen that it takes only about $0.3\Gyr$ from the onset of spiral forcing (about $2$ rotation periods of the innermost pattern)
for the $\delta$ pattern to have amplitude and morphology comparable to that of Model~A23,
for which the spiral forcing was present from $t=0$.
After spiral forcing is stopped, magnetic arms remain at the level $\delta>0.1$ for a few hundred $\Myr$,
but their winding angle $p$ gets gradually smaller because of the galactic differential rotation.
For $\tau=l/u$, the magnetic arms last for a few hundred $\Myr$ longer still (not shown; see also \citetalias{Chamandy+13a}).
Of course, in reality, the interfering spiral patterns would probably turn on and off at different times, 
i.e. $t_{\mathrm{on},1}\ne t_{\mathrm{on},2}$ and $t_{\mathrm{off},1}\ne t_{\mathrm{off},2}$.
Having said that, the above discussion is not meant to cover all the possible scenarios, 
but only to study to what extent the dynamo can respond to the spiral forcing by multiple patterns
during the finite lifetimes of these patterns.
In summary, it is clear that even for transient spiral patterns, 
magnetic arms may have ample time to develop and then evolve along with the superposition of the $\alpha\kin$-spirals.
Therefore, the above results for other models generally continue to be valid even when the spiral patterns are transient.
However, the growth and decay times of magnetic arms may in fact be comparable to the lifetimes of the spiral patterns that produce them,
adding complexity to the problem.
To keep the problem simple, this complication has been left aside in other sections of the present work,
but any attempt to model a specific galaxy would have to take into account the possible transience of each individual pattern.

\label{lastpage}
\end{document}